\documentclass[english, openany]{book}
\usepackage[utf8]{inputenc}
\usepackage[english]{babel}
\usepackage{import}

\usepackage[]{graphicx}
\usepackage[]{color}
\usepackage{alltt}
\usepackage[T1]{fontenc}
\usepackage[utf8]{inputenc}
\usepackage{mathptmx}
\usepackage{adjustbox}
\usepackage{amssymb}
\usepackage[utf8]{inputenc}
\usepackage{amsmath}
\usepackage{mathtools}
\usepackage{tabularx}
\usepackage{array}\newcolumntype{C}[1]{>{\centering\arraybackslash}p{#1}}
\usepackage{xcolor}
\usepackage{marvosym}
\usepackage[top=60pt,bottom=60pt,left=78pt,right=78pt]{geometry}

\usepackage[amssymb]{SIunits}
\usepackage{cite}
\graphicspath{{figures/}}

\usepackage{subcaption}
\usepackage{hyperref}
\usepackage{listings}
\usepackage{mdframed}

\usepackage[official]{eurosym}

\usepackage{parskip}

\usepackage{subcaption}
\usepackage{lipsum}
\usepackage{booktabs}
\usepackage{multirow}
\usepackage{fancyhdr}
\usepackage{titlesec}
\titleformat{\chapter}
   {\normalfont\LARGE\bfseries}{\thechapter.}{1em}{}
\titlespacing{\chapter}{0pt}{40pt}{2\baselineskip}

\usepackage{float}
\floatstyle{plaintop}
\restylefloat{table}

\usepackage[tableposition=top]{caption}

\usepackage{authblk}
\usepackage{tcolorbox}

\usepackage{subfiles}

\begin{document}

\begin{titlepage}
	\clearpage\thispagestyle{empty}
	\centering
	\vspace{1cm}

	{
		\textsc{Latin American Strategy Forum for Research Infrastructure - LASF4RI}
	}
		\vspace{2.5cm}

\normalfont

	\rule{\linewidth}{2mm} \\[0.5cm]
	{ \Huge \bfseries Latin American Strategy for Research Infrastructures\\ for High Energy, Cosmology, Astroparticle Physics\\
	\vspace{0.2cm}LASF4RI for HECAP} \\[0.5cm]
	\rule{\linewidth}{2mm} \\[3.4cm]

    { \Large \bfseries LATIN AMERICAN HECAP PHYSICS BRIEFING BOOK}\\
    \vspace{1cm}
    { \Large \bfseries Preparatory Group}\\
    \vspace{0.5cm}
    \begin{table}[h]
        \centering
        \begin{tabular}{l|l}
            \textbf{Hiroaki Aihara - University of Tokyo}   &  \textbf{Alfredo Aranda - University of Colima} \\
          \textbf{Reina Camacho Toro- LPNHE/CNRS}   & \textbf{Mauro Cambiaso - Universidad Andrés Bello}\\ 
          \textbf{Marcela Carena - Fermilab/U. of Chicago } & \textbf{Edgar Carrera - Universidad San Francisco de Quito}\\
          \textbf{Juan Carlos D'Olivo -  UNAM} & \textbf{Alberto Gago - Pontifica Universidad Cat\'olica del Per\'u} \\
           \textbf{Thiago Goncalves - Valongo Observatory} & \textbf{Gerardo Herrera -  CINVESTAV}\\
           	\textbf{Diana López Nacir - DF/IFIBA UBA-CONICET} &\textbf{Marta Losada - NYUAD}\\
           	 \textbf{Jorge Molina - Universidad Nacional de Asunción} & \textbf{Martijn Mulders - CERN}\\
           	   \textbf{Diego Restrepo - Universidad de Antioquia}, &
		\textbf{Rogerio Rosenfeld - IFT-UNESP \& ICTP-SAIFR}\\
			\textbf{Arturo Sánchez- ICTP/INFN/ U. of Udine } &
        \textbf{Federico Sánchez - U. Nacional de San Martín} \\
        \textbf{Marcelle Soares-Santos - U. Michigan}& \textbf{Martin Subieta - U. Mayor de San Andr\'es}\\
        \textbf{Hern\'an Wahlberg - U. Nacional de la Plata}& \textbf{Harold Yepes Ramirez - YTU}\\
         \textbf{Alfonso Zerwekh - U. T\'ecnica Federico Santa Mar\'{i}a} & \\
        \end{tabular}
        \label{tab:my_label}
    \end{table}

\vspace{5cm}


\pagebreak

\end{titlepage}

{\hypersetup{linkcolor=black}
	\tableofcontents{}
}

\mainmatter


\normalfont

 
\chapter{Introduction}\label{chapt:intro}

\noindent For the first time the scientific community working at the forefront of research in high energy, cosmology and astroparticle physics (HECAP)  in Latin American have come together to discuss and provide  scientific input towards the development of a regional strategy for these fields of knowledge. The process for the creation of a long-term strategy is unfolded in two main avenues: the development of new
experiments/facilities in the region and the construction of synergies for a more compelling and 
coordinated 
participation in global projects.

\noindent The process derived from the October 2018 Iberoamerican Science and Technology Ministerial meeting and its corresponding declaration\footnote{https://www.segib.org/wp-content/uploads/Declaracion-III-Reunion-de-Ministros-y-Altas-Autoridades-en-Ciencia-Tecnologia-e-Innovacion\_ES.pdf}. This declaration was then ratified at the end of 2018 by the Heads of State\footnote{https://www.segib.org/wp-content/uploads/III-PROGRAMA-DE-ACCION.pdf}. 
The declaration gave the mandate to create
a Strategic Forum for Research Infrastructures, 
both to provide an incentive for participation in existing facilities
as well as 
to foster the development of new projects
and initiatives. 
It is important to note that in the area of HECAP Spain and Portugal take part in the European processes for the development of the corresponding strategy. Thus, a necessary first step was to pursue the HECAP strategy at the Latin American level that would then interface with 
those of 
Spain and Portugal.

\noindent The Latin American Strategy Forum for Research Infrastructures (LASF4RI \footnote{\tt lasf4ri.org}) took shape as a result of the first workshop organized at ICTP-SAIFR in Sao Paulo, April 30-May 1st 2019 \footnote{\tt www.ictp-saifr.org/workshop-on-the-latin-american-strategy-forum-for-research-infrastructure/}. The decision was then made to move ahead to pioneer  the first LASF4RI initiative with HECAP.
\noindent The LASF4RI-HECAP process began with the establishment of the Preparatory Group 
(PG) 
and the High Level Strategy Group
(HLSG). The purview of the Preparatory Group is to develop the community input process and related discussions to produce the Physics Briefing Book. 
Consequently,  a broad invitation was made to the Latin American HECAP community to submit their proposals.  By December 2019 40 proposals were submitted, see full list of submissions in section\ref{wp}. 
The submissions   were considered by the members of the Preparatory Group and organized by topic into  six working groups as defined below.
The next step in the process was to hold an Open Symposium to 
present
the submitted proposals and 
allow for additional input and discussions among the community members 
on the outlook for HECAP in the region\footnote{\tt www.ictp-saifr.org/lasf4ri2020/}. This took place in July 2020 with 
broad
participation of the community albeit the distress induced by the pandemic. 
At the online Open Symposium, there were  brief presentations of the submissions, together with a
compiled summary per topical working group presented by PG members, in their role a group convenors. 
Extensive discussion sessions were held for each 
topic and a final  general discussion closed the event. Additional input was sought from the community, when necessary, in order to 
strengthen and improve on
the content and output of the process. 

\noindent The present document, the Latin American HECAP Physics Briefing Book, is the result of this ambitious grass roots effort. This report contains the work performed by the Preparatory Group to synthesize the main contributions and discussions for each of the 
topical
working groups: 

\begin{itemize}
\item Astronomy, Astrophysics and Astroparticle Physics
\item Cosmology
\item Dark Matter
\item Neutrinos
\item Electroweak and Strong Interactions, Higgs physics, Flavour and CP Physics and Beyond the Standard Model Physics
\item Instrumentation and Computing
\end{itemize}

\noindent It is of interest to note that recent work done in Europe to update the European Strategy for Particle Physics summarized in the corresponding Physics Briefing Book 
(\url{https://arxiv.org/abs/1910.11775})
dives into a broader and more in depth discussion on the global physics landscape and future,
that complements the Latin American regional perspective presented here.
Similarly the European Astroparticle Physics Strategy\footnote{(\url{https://www.appec.org/wp-content/uploads/2017/08/APPEC-Strategy-Book-Proof-23-Nov-2.pdf}}
devised the specific strategy for astroparticle physics for the timeframe 2017-2026. In addition, the US has also started its community based Snowmass decadal process that will inform the corresponding strategy in that nation.
\noindent This Latin American HECAP Physics Briefing Book 
is fundamental to inform 
 the Strategy Document Committee 
 in the preparation of 
 the
 Strategy Document 
 that will provide 
 essential 
 input to the 
 LASF4RI-HECAP HLSG. The  main role of the HLSG is to analyze the resulting work of the process and in particular to provide feedback, assess, and validate the Strategy Document. The expectation is that the Strategy Document should be revisited on a 3-5 years timeframe. Nevertheless, the extent of the impact of the COVID-19 pandemic on the timescale is still not clear at the time of writing this report.
 
\noindent As can be seen in the subsequent chapters of this 
briefing book,
for the first time a landscape of 
the Latin American
experimental activities in HECAP 
is presented in a succinct and combined form.  The focus is on the scientific questions addressed by experiments that are either located in Latin America or that have a significant participation of Latin American research groups. 

\noindent This briefing book  discusses the relevant emerging projects developing in the region and 
considers potentially impactful future initiatives and 
participation of the Latin American HECAP community in international flagship projects.
Identifying synergistic contributions 
to these
endeavors is the main outcome of this work. 

\noindent The comparative advantages for experimental projects for astroparticle physics and cosmology located in the Latin American region, the increased growth and participation of research groups in HECAP, the strengths and expertise that have been consolidated 
as well as the remaining shortfalls in capabilities,  infraestructure and sustained support
are some of the main takeaways of this process. For example, accelerator physics is identified to be an important gap in the existing scientific capabilities in the region.

\noindent The second, third and fourth chapters
of this briefing book
are devoted to the confluence of astronomy, particle physics and cosmology experiments that cover multi messengers of the Universe (visible light,
cosmic rays, gamma rays, neutrinos, gravitational waves, etc), precision measurements of the CMB (baryon acoustic oscillations in radio, studies of CMB polarization at large and small angular scales), next generation gravitational waves projects, including also  the nature of dark matter/dark energy from galaxy surveys and direct/indirect dark matter 
detection
experiments. 

\noindent 
Current and prospective 
astroparticle and cosmology 
probes
build on the demonstrated ability to host and operate such  type of large experiments in the Latin American region
and significantly 
 benefit from favorable geographic and atmospheric conditions. The projects 
 under consideration
aim to ensure that a  leading role on the global scale in this field is maintained in the region. The Latin American participation in direct detection dark matter experiments 
has a unique potential searching for 
dark matter candidates in the low mass (sub-GeV) mass region.
Impactful participation in liquid Argon detector based experiments and, most recently, demonstrated capabilities based on Skipper-CCD technology  offer a clear path ahead.
A game-changer 
opportunity is presented to design and construct experiments based on Skipper-CCD detectors for dark matter and neutrinos in the prospective ANDES facility.

\noindent The fifth chapter focuses on neutrino physics in which there is a significant participation of Latin American scientists both in experiments trying to pin down the fundamental nature and properties of neutrinos, as well as in experiments using neutrinos as a probe. There is a wealth of physics domains including precision oscillation measurements, neutrino cross sections, non-standard interactions, CP violation, new neutrino states, astrophysical neutrinos, etc. 
As examples of forefront technological advancements of great impact for  current and future neutrino experiments with Latin American participation are the aforementioned Skipper-CCD detectors as well as the ARAPUCA photon detection
system. The latter is a new technology developed in Latin America, that has a central role in  detecting the light emitted in detectors  at current Fermilab neutrino experiments, and will be instrumental to the prospective Deep Underground Neutrino Experiment (DUNE) physics program. As well as, the planned small photo-multipliers subsystem of JUNO and the optical calibration system for KM3NeT.

\noindent Chapter six is devoted to experiments at colliders that probe the fundamental understanding of electroweak and strong interactions at the energy and luminosity frontiers. One of the main features of the last two decades has been the significant 
growth 
of the high energy physics community in the Latin American region.
This is
exemplified by the 
increased
participation 
of Latin American researchers 
in all Large Hadron Collier (LHC) experiments, having thus contributed to some of the most outstanding discoveries and measurements related to Higgs physics, the quark gluon plasma, precision electroweak physics, flavour and CP physics. This last topic is also complemented by the experimental efforts in Belle 2.  At the same time these energy frontier experiments have the potential for new discoveries beyond Standard Model physics that could explain some of the most fundamental unknowns in our understanding of the microscopic world. As a result of these activities clear areas of expertise and excellence emerge that  encompasses detector design and construction, triggers, Data Acquisition (DAQ) and readout systems, software and computing, as well as leading certain physics analyses.


\noindent The final chapter addresses the area of instrumentation and computing. Future experimental activities with their high specification requirements across HECAP drive the focus of the instrumentation R\&D efforts as well as the evolving needs of computing resources. The Latin American expertise and development of technologies and devices that can be utilized across HECAP is presented including specific chip and board design, firmware and construction, electronic readout systems, photonics, charged coupled devices, 
Resistive Plate Chamberss and water Cherenkov detectors. The computing infrastructure 
requirements are also indicated and the case is made for a strategic R\&D computing approach to benefit HECAP in Latin America, including the advanced training, development and retention of expertise.

\noindent Throughout the chapters there are important references to advanced training and capacity building which is a strength in the region and is key  to increase  participation and the  visibility of the community in  the amazing projects that are considered here.

\noindent It is worth mentioning that in all of the physics topics mentioned above the fruitful exchanges between theoretical and experimental physicists have been of great importance to
support activities and capability growth,
motivate new searches, physics analysis, or the development of new techniques. 
The main objective of this report 
is to evaluate the research infrastructures, rather than concentrate on the theoretical aspects of HECAP, although a significant participation of Latin American theorists was noticeable throughout this process. 
\noindent 
Throughout the previous sections,
ample consideration is given to
the overlap of the scientific drivers 
when appropriate, e.g. direct searches for dark matter or new neutrino states at the LHC.

\noindent 
The present 
Physics Briefing Book of HECAP activities in Latin America 
aims at providing
a landscape of existing expertise
and capabilities. It
will hopefully 
instruct some recommendations for further 
development of strategic research infrastructures relevant for future  HECAP efforts in the region as well as for a continuous and  successful involvement of Latin American HECAP  researchers in the global arena.

\chapter{Astronomy, Astrophysics and Astroparticle Physics}\label{chapt:astro}

\vspace{1cm}
\noindent{Juan Carlos D'Olivo (UNAM, Mexico) \\
		Federico Sánchez (U. Nacional de San Mart\'{i}n, Argentina) \\
		Martín Subieta (U. Mayor de San Andr\'es, Bolivia)}


\selectlanguage{english}


		
	


	




\section{Introduction}\label{sec:intro}

The subtle connection between the macro- and micro-cosmos is probably the most prominent
feature of the astroparticle physics field, which is a research area that in the last decades  had an explosive development 
as demonstrated by the
six Nobel Prizes that have been awarded in the last two decades: 2002, 2006, 2011, 2015, 2017, and 2019.
Being at the interface of astronomy, particle physics and cosmology, this field
will certainly play a major role in pushing our knowledge of the Universe to the edge and,
undoubtedly, its prospects for the near future are very exciting\footnote{It worth noting that the international community is recently organising itself around these research axes. For example, in 2012, the European community formed the Astroparticle Physics European Consortium (APPEC) as a consortium operated on the basis of a Memorandum
of Understanding with the overarching aim of
strengthening European astroparticle physics and the
community engaged in this field. For more information: \href{https://www.appec.org/}{https://www.appec.org/}.}.

\noindent The domain of elementary particles with their fundamental
interactions, on one side, and the world of the celestial objects with their evolution, on the other, 
have established two highly successful standard scenarios:
the {\it Standard Model} of particle physics (SM)
and the {\it Big Bang Model} (BBM) of cosmology . Both are playing a crucial role in the
current understanding as well as in the unsolved enigmas of our Universe. 

\noindent The main open questions that can be addressed from astroparticle physics might be summarized in two broad sets. Namely,
\begin{itemize}
	\item {\bf the non-thermal universe}: what can be learned from the most energetic messengers (gamma rays, neutrinos, cosmic rays, and gravitational waves) of our universe?
		What is the link among them? What are the acceleration sites of these high-energy particle and what are the mechanisms involved?

Particles with energy above $10^{9}$\,eV are the messengers of the extreme Universe. 
In addition to the radiation emitted by stars and produced predominantly
by thermal or collisional emission, we know that some forms of radiation are
independent of gas temperature and require the action of a collective process of
concentration of energy on a small population of particles. The energy spectrum of gamma and cosmic
rays, which cannot be reduced to a thermal process, are a good example. We also know that both are indeed 
an important piece of information in our task to describe the Universe
and its fundamental laws since their energy density is of the
same order of magnitude as the energy density of stars light and the galactic
magnetic fields \cite{hecr1989}.

Several observatories around the world have inaugurated a new way of observing the Universe by means of
cosmic rays, gamma-rays, cosmic neutrinos and gravitational waves. 
As an example of the connection of all these observations, it has been argued that mergers of black holes,
both of stellar mass and of extreme masses, are a key ingredient for
producing ultra high energy cosmic rays (UHECRs), high energy neutrinos, high energy photons and gravitational waves. The key recognition
is that, during a merger, the individual spins and the accompanying relativistic jets swing around,
at full power, until they are aligned with the orbital spin \cite{gergely}
This
swerving around of a full power jet system would be directly visible in many
observations.

The simultaneous observation of cataclysmic events in our Universe
with multiple messengers provides, therefore, new insights into the astrophysical properties of compact objects 
and also stringent tests of physical laws and particle properties, which additionally
contributes to the search for dark matter (DM). 

	\item {\bf the dark universe}: what is the nature of dark matter? What is the role of tiny mass neutrinos? Do axion-like particles exist?

Unlike normal matter, well explained and understood within the context of the SM, DM does not seems to interact 
electromagnetically: it does not absorb, reflect or emit light. The existence of DM is inferred only from its gravitational
effects, particular at large scales. It may be comprised of 
particles whose masses span orders of magnitude from $10^{-22}$\,eV to 1\,PeV and the most frequently proposed candidates 
arise in theories that suggest physics beyond the SM, such as supersymmetry and extra-dimensions.

The currently dominating paradigm describes the DM as a relic density of weakly interacting massive particles (WIMPs) in the range $10^{2}$\,GeV to $10^{4}$\,GeV.
The products of the self-annihilation of WIMPs 
would interact with the environment at the place of annihilation.
The self-annihilation would therefore
be more sizable in astrophysical environments, such as galaxies and clusters, richest in dark matter.
In fact, the products of DM annihilation would show up, in such objects, as an additional,
{\it exotic}, gamma component on the top of the astrophysical one. This extra component
depends on the
nature of the dark matter through its annihilation cross-section, its mass,
its couplings to the Standard Model, all eventually affecting the spectrum of
gamma–rays emitted. This indirect searches of DM annihilation (or decay) have
therefore become one important tool of all Gamma-Ray observatories.

On the other hand, models of superheavy DM (SHDM) that have been proposed since the early 90s \cite{shdm}
are being revived due to the null detection (both direct or indirect) of WIMPs.
If super-heavy particles decay into SM particles, a flux of ultra-high energy photons could
be observed preferentially from regions of denser DM density such as the center of our Galaxy.
So far, no photons with energies above 1\,EeV have
been unambiguously identified \cite{auger1}. This can
translate into constraints on the properties of
the SHDM particles like their mass or lifetime.
Ultra-high-energy cosmic rays observatories might, therefore, also contribute to solve the DM mystery. 

\end{itemize}

\noindent The Latin American region has an unique-opportunity window to host many large international projects in the field of
astroparticle physics. In observational astronomy Latin America is strategically important given its geographical position. The present, the mid- and the long-term future are covered by different proposals.
Nowadays, the landscape with the ongoing observatories is bright, but the prospects for the future might be undoubtedly stunning with the proposed projects.

\noindent The Pierre Auger Observatory ({\bf PAO}) [I-17], installed in Argentina since 20 years ago, 
is the world's largest and most sensitive facility to detect giant particle showers generated by the 
interaction of high-energy cosmic rays. It has certainly paved the way for other collaborations to
come to Latin America. The next generation ground-based gamma-ray observatory, the Cherenkov Telescope Array ({\bf CTA}) [I-6, I-24],
that will be hosted in Chile, is a clear example in this sense. The construction of CTA, which is foreseen to finalize in 2025, will
improve the sensitivity, angular resolution, field of view and energy coverage of the current experiments by around one order
of magnitude, providing an impressive boost to gamma-ray observations.
Another example is the High-Altitude Water Cherenkov Gamma-ray Observatory ({\bf HAWC}) located
in Mexico since 2007, which is the basis for the next-generation Southern Wide-field-of-view Gamma-ray
Observatory ({\bf SWGO}) [I-23]
to be hosted in the region. Also important for the region, is the Andes Large-area PArticle detector for Cosmic-ray physics and Astronomy 
({\bf ALPACA}), another gamma ray observatory that will be located very near to Mt. Chacaltaya, in Bolivia.

\noindent Cosmology and radio-astronomy experiments have benefited from the crystal-clear skies in the Latin America region: 
the Q\&U Bolometric Interferometer for Cosmology ({\bf QUBIC}) [I-37] 
and the
Large Latin American Millimetre Array ({\bf LLAMA}) are under construction in the north-west of Argentina.

\noindent A spin-off of all these efforts is the Latin American Giant Observatory ({\bf LAGO}) [I-29], a network of detectors
distributed all over Latin America. It represents a non-centralized and almost completely regional effort 
whose purpose is the study of high-energy phenomena, space weather, and atmospheric radiation.

\noindent Finally, the prospects for the region to be host of a Giant Radio Array for Neutrino Detection ({\bf GRAND}) [I-25]
 show the strong commitment of Latin America to the astroparticle physics field both in developing research facilities and
leading them with highly trained human resources.

\noindent In short, the infrastructures already deployed, as well as those being constructed, planned or in their design phase, will 
exploit all confirmed high-energy astroparticle messengers from the Universe and
are essential for enabling new, or strengthening the already existing, Latin American scientific 
capabilities.

In this chapter the various fields and, particularly, the different observatories in astroparticles most relevant for the Latin American region
are briefly reviewed.  
Before the concluding remarks, some considerations on the synergies with
other chapters of the briefing book are also addressed.

\begin{figure}[h]
	\centering
	\includegraphics[width=0.9\textwidth]{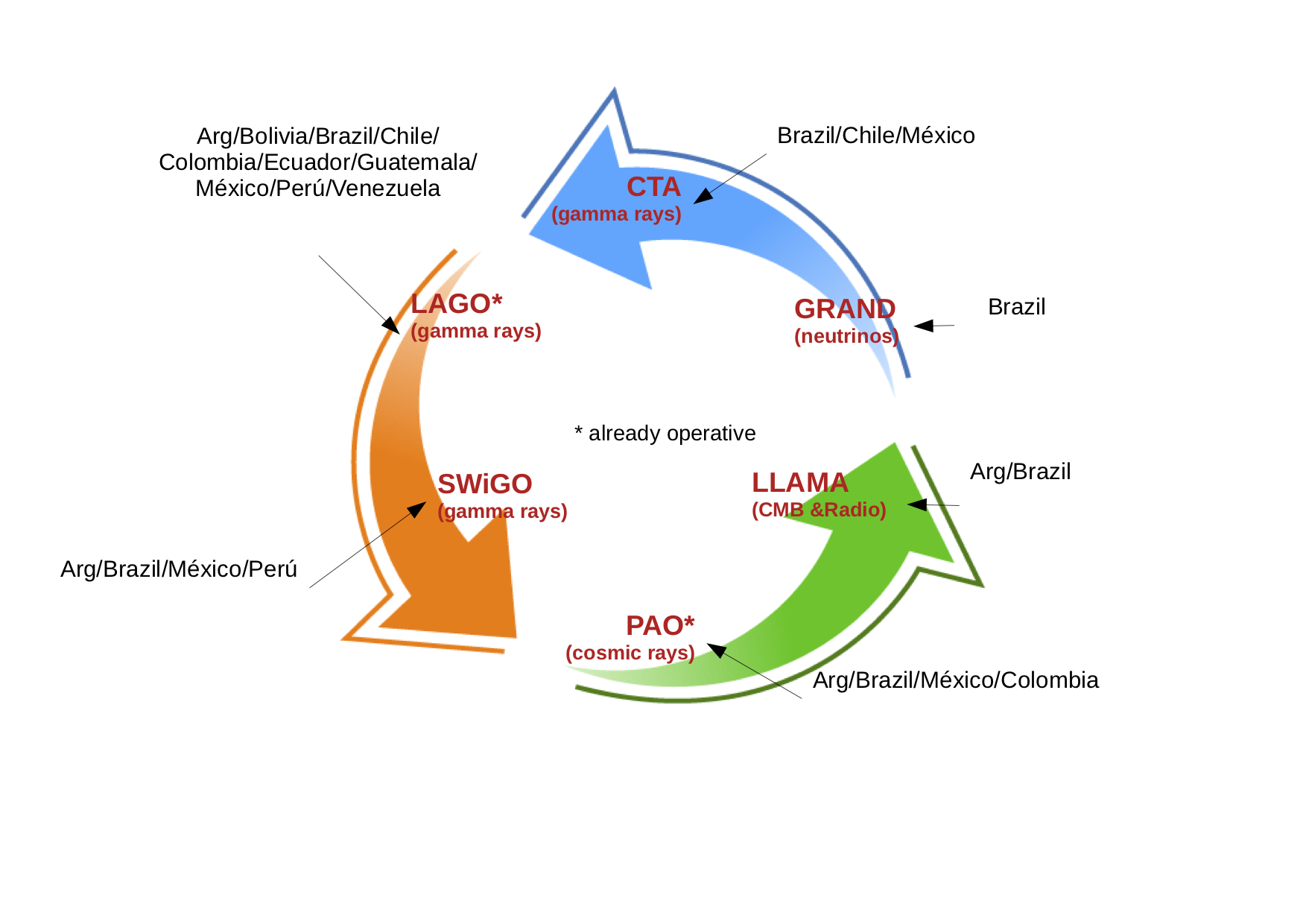}
	\caption{Latin American prospect in astroparticle physics. Five out of six large international projects 
	have already selected the Latin American region as host sites. Participation of
	Latin American countries is also indicated.} \label{fig:LA_projects}
\end{figure}


\section{Involvement of Latin American Countries}\label{sec:projects}
Latin America has a strong commitment in astronomy, astrophysics and astroparticle physics field as demonstrated by the 
several intercontinental collaborations that decided to place their projects in the region (see Figure~\ref{fig:LA_projects}). There are several 
reasons behind these choices but most of them rely on
the well-established comparative advantages with respect to other areas in the world.
The access to crystal-clear skies as well as  the support of local scientific communities are certainly among
the key advantages provided by the  countries in the region.
The following presentation follows the white papers of the proposals submitted to LASF4RI-HECAP.

\subsection{Pierre Auger Observatory}
PAO's main physics goals are:
\begin{enumerate}
  \item To elucidate the mass composition and the origin of the flux suppression at the highest energies.
  \item The search for a flux contribution of protons up to the highest energies ( sensitive up to 10\% fraction of p in the suppression region).
  \item Understanding extensive air showers and hadronic interactions.
  \item To probe the origin and characteristics of primary cosmic rays from below  $10^{17}$\,eV up to $10^{20}$\,eV.
\end{enumerate}

The PAO corresponds to a 3000\,km$^2$ area in Mendoza Province (Arg.) embedded with water-Cherenkov detectors, flourescence telescopes, 
plastic-scintillation detectors, radio-antennas detectors, and resistive-plate chambers.

The PAO collaboration consists of 80 institutes from 17 countries (Argentina, Australia, Belgium, Brazil, Colombia, Checz Rep., France, Germany, Italy, México, 
Poland, Portugal, Romania, Slovenia, Spain, Netherlands, USA). 
The construction finalized in 2008 and since 2015 is constructing the upgrade. The completion of the 
upgrade is foreseen to be in 2022.

\subsection{Latin American Giant Observatory}

LAGO's mains physics goals are:
\begin{enumerate}
  \item Search for Gamma Ray Bursts (GRBs) events.
  \item Study the flux modulation of galactic cosmic rays at different locations on Earth.
  \item Study of variation of secondary particles at detection level.
  \item Study of the global magnetic structure cloud reaching the space environment surrounding the Earth.
  \item Space weather studies.
\end{enumerate}

This widespread network of observatories is deployed throughout Latin America, in sites located at different latitudes, from Mexico to Antarctica, and at different altitudes.
 LAGO is based on
single or small arrays of water-Cherenkov detectors.
The LAGO collaboration consists of 29 institutes from 11 countries (Argentina, Bolivia, Brazil, Chile, Colombia, Ecuador, Guatemala, México, Perú, Venezuela, Spain).
This network started activities in 2006. The current detector upgrade plan is expected to take place in 2020-2025.


\subsection{Cherenkov Telescope Array}
CTA's main physics goals are:
\begin{enumerate}
  \item Understanding the origin and role of relativistic cosmic particles (sites and mechanism of high-energy particle acceleration).
  \item Probing extreme environments (physical processes at work close to neutron stars and black holes, characteristics of relativistic jets, winds and explosions, radiation fields and magnetic fields in cosmic voids).
  \item Exploring frontiers in physics 
(the nature and distribution of dark matter, quantum gravitational effects on photon propagation).
\end{enumerate}

The CTA observatory consists of two array site locations, one in the Southern (Paranal, Chile) and one in the Northern hemisphere (Canary Island, Spain). It will use imaging atmospheric Cherenkov telescopes of three-different kinds: the Small Size telescopes (SSTs), the Middle Size Telescopes (MSTs), and the Large Size Telescopes (LSTs).

The CTA collaboration consists of 200 institutes from 31 countries (Armenia, Australia, Austria, Brazil, Bulgaria, Canada, Chile, Croatia, Checz Rep., 
Finland, France, Germany, Greece, India, Ireland, Italy, Japan, México, Namibia, Netherlands, Norway, Poland, Slovenia, S. Africa, Spain, Sweden, Switzerland, Thailand, Ukraine, UK, USA).
The current timeline foresees completion of construction by 2025.

\subsection{Southern Wide-field-of-view Gamma-Ray Observatory}
SWGO's main physics goals are:
\begin{enumerate}
  \item Measure TeV halos around nearby Pulsar Wind Nebulas.
  \item Identify sources of PeV galactic cosmic rays. 
  \item Measure the galactic center and Fermi Bubbles morphology.
  \item Measure the galactic diffuse emission, the local galactic cosmic ray anisotropy
and solar cosmic rays interactions.
  \item Search for new VHE gamma ray galactic sources and neutrino VHE gamma rays counterparts,
detect AGN flares and issue alerts, and search for periodicity and long term emission in AGNs.
  \item Search for counterparts to GW events and nearby bright GRBs.
  \item Search for dark matter annihilation and decay, Lorentz Invariance violation, primordial black-holes or axion-like particles.
\end{enumerate}

SWGO's site candidates are: Atacama dessert (Chile), Cerro Vecar (Argentina), ALPACA site and Mt. Chacaltaya (Bolivia), Laguna Sibinacocha (Perú).
It will be based on water-Cherenkov technique graded in
4000 detectors deployed in 0.08\,km$^2$ and 1000 detectors an extended area of 0.22\,km$^2$. The SWGO is foreseen to be completed in 2026.
The SWGO collaboration consists of 43 institutes from 10 countries (Argentina, Australia, Brazil, Germany, Italy, México, Perú, Portugal, UK, USA).

\subsection{Large Latin American Millimeter Array}
LLAMA's main physics goals are:
\begin{enumerate}
  \item Astrophysics in multiple wavelengths (solar physics, cosmology, galactic structure, astrochemistry, interstellar Medium, star formation).
  \item Observations of atomic molecular spectral lines coming from astronomical objects located in a wide range of distances and
surveys of the southern sky for different molecular lines.
  \item Very Long Baseline Interferometry (VLBI) observations. 
  \item Study of the polarization of Cosmic Microwave Background Radiation (CMBR) at small angular scale (complement to QUBIC). 
\end{enumerate}

It is under construction in North-Western Argentina (Alto de Chorrillos, Salta Province) at 4800\,m above sea level. The LLAMA is a radio telescope with a single dish of 12\,m diameter operated in the frequency range of 95 to 950\,GHz.
LLAMA is a joint project between Argentinian and Brazilian institutions. 
Construction planning started in July 2014 after the formal signature of an agreement between the main institutions involved. First light is expected in 3-4 years. 

\subsection{Giant Radio Array for Neutrino Detection}
GRAND's main physics goals are:
\begin{enumerate}
  \item Study the origin of the Ultra High Energy Cosmic Rays (UHECRs).
  \item Perform neutrino astronomy, study GZK neutrinos and neutrino physics.
  \item Detect Ultra High Energy (UHE) neutrinos and gamma rays and discover UHE neutrino point sources.
  \item Probe millisecond astrophysical transients (fast radio bursts and giant radio pulses).
  \item Map the sky temperature with mK precision and measure the global signature of the epoch of re-ionization and study the Cosmic Dawn.
\end{enumerate}

The site for the prototype is in China but future sub-arrays might be located in Latin America. GRAND will be a network of air shower ground-detectors based on radio-antennas.
8 institutes from 6 countries (Brazil, China, France, Germany, Netherlands, USA) are working in this project. GRANDProto35 (35 antennas and 24 scintillators) was in commissioning phase and several upgrades are planned spanning a timescale up to 2030s.

\section{Leadership Areas}\label{sec:leadership}
The area of {\bf Ultra-High Energy Cosmic Rays} observation is undoubtedly one of the most broadly spanned over in the Latin American region.
As one of the first and largest collaborations settled in Latin America, the Pierre Auger Observatory
has highly boosted the Latin American astroparticle physics community. 
In the past two decades, utmost contributions from the Latin American scientist were made. 
Along the years, many Latin American scientists have been in charge of managing roles in the governance of the International Collaboration
or leading scientific analyses.
For the current upgrade of the Observatory and its related R\&D program,  Latin American countries
are leading the construction of the Underground Muon Detector (AMIGA) and the Resistive Plate Chamber array (MARTA).
Both playing a major role towards solving the
problem referred to as the {\it muon puzzle} in extensive air showers:
none of the current hadronic interaction models tuned after  the last run of the Large Hadron Collider (LHC)
is able to correctly describe muon production in atmospheric showers induced by cosmic rays.
So far, the most unambiguous experimental evidence of this disagreement
was revealed in the analysis of PAO data.
Thus, this is certainly one of the most compelling issues to be solved in the upcoming years.
It is a key contribution to understand the composition of high-energy cosmic rays and, ultimately, the origin
of the most energetic particles in the Universe.

\begin{figure}[h]
	\centering
	\includegraphics[width=0.8\textwidth]{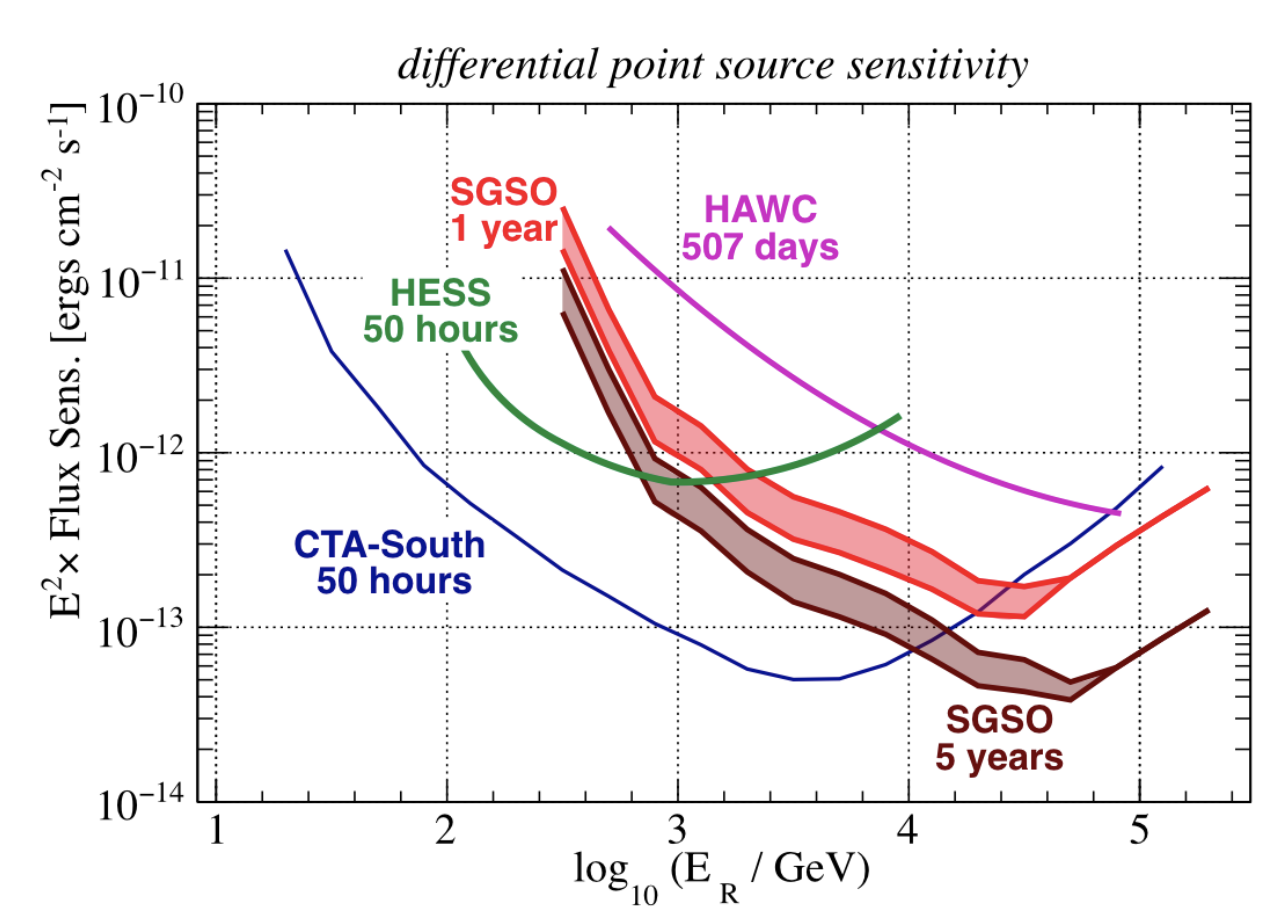}
	\caption{CTA and SWGO differential  point source sensitivity~\cite{2019arXiv190208429A}. Initially SWGO was known as the Southern Gamma-Ray Survey Observatory (SGSO).} \label{fig:CTA-SWGO-flux}
\end{figure}

{\bf Gamma-Ray} observation is also a highly developed area in Latin America. No doubt that
the construction and the scientific results of HAWC have greatly contributed to this aim.
But it is also undeniable that the construction of CTA and SWGO will foster and strengthen it even further. As an example, 
in the past 10 years, the number of scientists involved in CTA only in Brazil, has grown from three to sixty.
The science program that will be covered by CTA and SWGO is very broad
and will certainly transform our understanding of the high-energy universe exploring questions 
of fundamental importance: the origin and role of relativistic cosmic particles,
the physics in extreme environments, the physics beyond the standard models. Figure \ref{fig:CTA-SWGO-flux} shows the differential energy flux sensitivity of CTA and SWGO,  in comparison with other experiments.

{\bf Space-Weather} studies will benefit from the LAGO project, a spin-off of the Pierre Auger Observatory and
a seeder for the development of astroparticle physics in the Andean region. LAGO is rooted in the Latin American region both, 
for the sites equipped with detectors and for the conformation of the collaboration. LAGO will allow simultaneous 
measurements of the modulation of the galactic cosmic ray flux at different locations on Earth that can provide
important information on the properties and global structure of magnetic clouds reaching the terrestrial environment 
during interplanetary Coronal Mass Ejections. By combining all the data measured at different locations in the 
detector network, the LAGO project will provide very detailed and simultaneous information on the temporal 
evolution and small- and large-scale characteristics of the disturbances produced by different transient and long-term 
space weather phenomena.

\begin{figure}[h]
	\centering
	\includegraphics[width=0.8\textwidth,height=8cm]{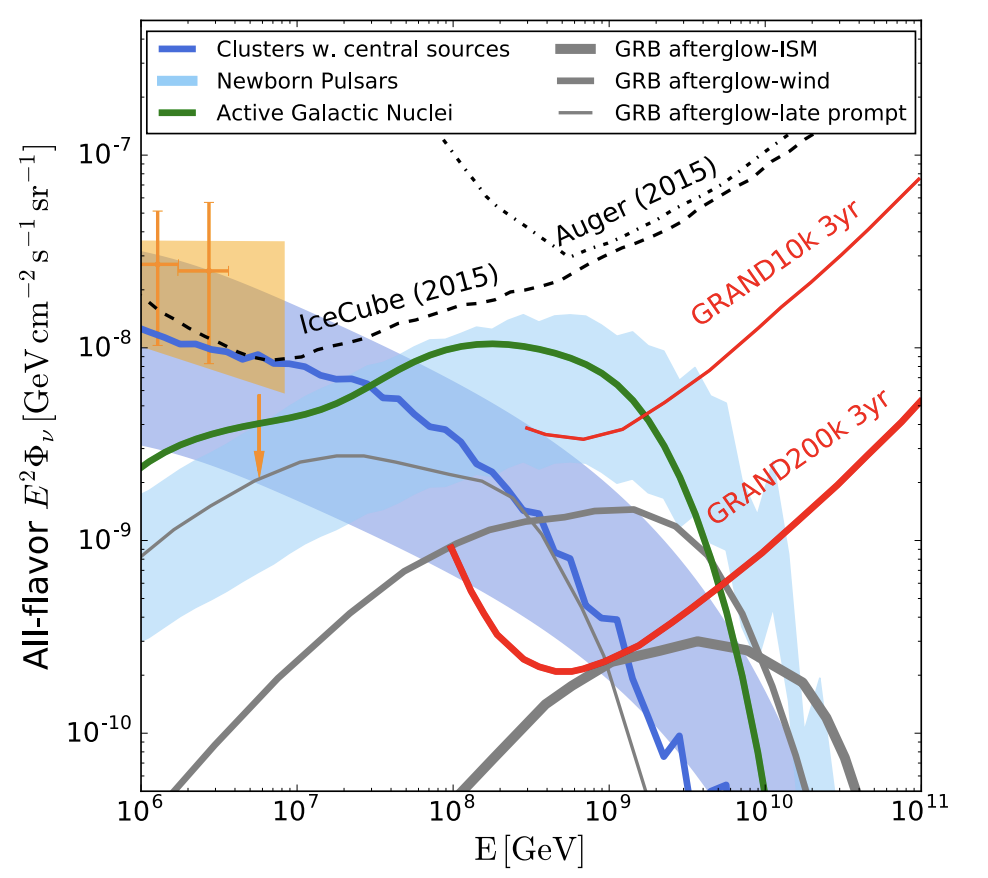}
	\caption{GRAND sensitivity to source produced neutrinos (all flavors)~\cite{2019AlvesBatista}.} \label{fig:GRAND}
\end{figure}

{\bf High-Energy Neutrino} observations with GRAND benefit
from the experience gained, among others efforts in the world, from the PAO engineering array for
radio detection (AERA). 
Since the early 1960s it was suggested that extensive air showers induced by cosmic particles 
may emit radiation at radio wavelengths.
In recent years the interest in this technique has been strongly growing. Besides the already mentioned AERA in the PAO,
the activities
around the world include LOFAR in the Netherlands, ARIANNA in Antarctica and Tunka-Rex in Siberia. 
Significant efforts
are made to calibrate the response of those detection systems with high precision, both for the signal
strength and the signal arrival time.
It is worth mentioning thay the cost of deploying radio antennas is lower than for other types of detectors. Moreover, their duty cycle
is extremely high and therefore may be a key technique in the near future for observing highly-energetic cosmic radiation.
The participation of the Latin American community in GRAND is therefore very important in this line of research. Figure\ref{fig:GRAND} illustrates the sensitivity to source produced neutrinos at GRAND.

\noindent Finally, {\bf Experimental Cosmology and Radio Astronomy} are also areas in which the Latin American scientific 
community will play an important role in the coming years. Two experiments, QUBIC and LLAMA, will be conducted 
in the region. LLAMA will be a multipurpose observatory that will allow observations for long periods of time 
and explorations of large portions of the sky, impossible to carry out at present through telescopes with 
international competition, providing excellent opportunities for groups with interests as diverse as solar physics, 
cosmology, galactic structure, astrochemistry, interstellar medium, star formation, polarization and magnetism, 
very large base interferometry, and more. For its part, QUBIC is designed to measure the polarization of the cosmic 
microwave background and may reveal the presence of gravitational waves with primordial
origin. Such measurements can probe inflationary cosmological models that predict that
quantum effects during an accelerated expansion at the earliest stages after the Big-Bang
produced gravitational waves along with the density fluctuations that later seeded galaxy
formation. QUBIC will join other international efforts currently pursuing this goal using a
novel approach, which combines the sensitivity of bolometric detectors with the control of
systematic effects provided by interferometry\footnote{Recent information on a new projet was submitted, the Next Generation Very
Large Array (ngVLA)
will be constituted by 256 antennas of 18-m each distributed across a
region 300 lm wide. It will have one order of magnitude better angular
resolution and sensitivity than the VLA. The ngVLA will be distributed in
the south of the
USA and the north of Mexico. This distribution will provide optimal images
of southern sources such as the center of our Galaxy. The radio
astronomical community of Mexico is very interested in participating in
this project that will have as
major partners the USA, Canada, Germany and Japan. It will be very positive
to involve other Latin American countries in this collaboration. Even a few
percent of guaranteed telescope time will allow the Latin American
groups to undertake major projects, given the speed of the instrument.}. 

\section{Drivers for Multiple Approaches}\label{sec:dirvers_cluster}

We have recently witnessed the dawn of the era of multi-messenger astrophysics. It was marked by the
detection of gravitational waves from a binary neutron star in coincidence with electromagnetic
counterparts, and by the observation of high-energy neutrinos from a blazar flare together with
radiation across the whole electromagnetic spectrum~\cite{Abbott_2016,2018Aartsen}.
Evidently multi-messenger exploration of the extreme Universe is in full swing alongside the quest for Dark
Matter and the search for the true nature of tiny mass neutrinos.

\noindent Cosmic ray detectors, gamma ray and high-energy neutrino observatories will benefit each other.
The Latin American scientific community has the unique opportunity to host several projects
covering all these ways of exploring the Universe and, therefore, significantly contributing to
push forward our knowledge of the Cosmos and its evolution.

\section{Synergies}\label{sec:synergies}

As already demonstrated by previous experiences around the world, 
there are multiple synergies between particle and astroparticle physics at various levels. From the
point of view of infrastructures,
large underground facilities or the engineering and management of large projects are good examples. 
From the point of view of detectors and enabling technologies, several R\&D programs including photosensors, 
electronics or computing are also
a taste of how basic astrophysical research strengthens the link between science and technology (see Chapter~\ref{chapt:instru}).
From  both the theoretical and phenomenological side and the fundamental unknowns, 
there is a large overlap between particle and astroparticle physics: 
\begin{itemize}
	\item The excess of muons with respect to simulated air showers, first observed in 2000, is a good
example.
While the experimental measurements are consistent within their uncertainties with predictions up to energies of
approximately 100\,PeV, corresponding to proton-air collisions at nucleon-nucleon centre-of-mass energies
of $\sqrt{s_{NN}}$<14\,TeV, at higher energies a muon excess is observed that systematically increases with the
shower energy for all hadronic
interaction models.
The main challenge in the description of muon production in the atmosphere is the treatment of 
hadron-ion interactions in the forward region over many orders of magnitude in energy. 
Conventional options
to improve the description of muon production in hadronic models are the enhancement of the production of
baryon-pairs, neutral rho mesons, or heavy quarks, as well as possibly missing physics in the soft-QCD
regime of hadron-ion collisions, or other new phenomena.
Ongoing studies on the muon discrepancy at the LHC include detailed measurements of prompt hadron
production cross-sections in the forward region in proton-proton and proton-lead collisions and measurements 
of the ratio of electromagnetic and hadronic energy flows, to which the production of GeV muons is
particularly sensitive. 
Similar studies at central rapidities allowed to improve the hadronic models
with the first LHC data. Running the LHC with oxygen beams is planned for 2023, which was largely
motivated by the muon excess and the need to study the nuclear modification of hadron production in the
first collisions in EAS. Also ongoing is an upgrade of the SMOG system in LHCb, which will extend the
unique capability of LHCb to perform fixed target experiments with LHC beams to study late interactions in
extensive air showers. 

	\item Another aspect is the contribution of gamma ray observations to solving the question of the nature of 
the dark matter. In particular, SWGO will scrutinize nearby extended sources of dark matter 
for evidence of gamma rays produced in annihilation or decay processes. Weakly
interacting massive particles (WIMPs) that were once in thermal equilibrium in the early universe
remain one of the most promising explanations for dark matter. Thermal WIMPs with masses
between $\sim$2 TeV and $\sim$100 TeV are largely still viable, and only astrophysical experiments
can probe heavy dark matter (>1 TeV). CTA will probe thermal WIMPs from
100 GeV to 10 TeV, leaving the heaviest of thermal WIMP masses unconstrained. SWGO, on the other
hand, will be able to probe almost the entire thermal mass range of WIMPs 
by extending the sensitivity up to 100 TeV. Also, both SWGO and CTA will be sensitive to
thermal WIMPs with masses from about 1 to 10 TeV. If there is a gamma-ray signal in that mass
range, both experiments should see it, leading to independent confirmation of the signal. See Chapter~\ref{chapt:DM} for further discussion.
Concomitantly to the exploration of the complete WIMP parameter space with
gamma ray observatories, the search might be also extended to cosmic ray observatories like PAO by means
of the limits (if not detected) of ultra-high energy photons that test the SHDM models. 

	\item The ultra-high-energy neutrinos that might be observed by ground-based experiments like GRAND, are also a probe of
fundamental physics. Specific signatures in neutrino-related observables, namely energy spectrum, 
angular distribution, and flavor ratios, are sensitive to the search for new physics.
In addition, neutrino observatories can extend the measurements of the cross section for neutrino-nucleon interactions, see Chapter \ref{chapt:neutrinos} for further discussion.

\end{itemize}

\noindent Finally, beyond the link with particle physics of most of the projects presented in this chapter, 
there is also a synergy with cosmology as LLAMA will contribute to this field by means of measuring the polarization 
of the CMB (see Chapter~\ref{chapt:cosmo}). The specific contribution will be on smaller scales (arcmin or smaller) for the same part of the sky 
scanned by QUBIC.

\section{Conclusions}\label{sec:conclusions}


The fields of astronomy, astrophysics and astroparticle physics have shown, in the past decades, that the Latin American community 
has the capacity to conceive, design and execute large scientific projects that reinforce 
the connection between basic science and technological applications,
favoring the progress of the region both in research and innovation.
Local support, together with the geographical features of the Latin American region, have played
a decisive role for the selection of Latin America as a hub for hosting project managed
by large intercontinental collaborations. In this chapter, we have revised the already established experiments, those being constructed and finally those projected in astroparticle
physics. The number of projects and the people involved in them, place the Latin American community
in a leading position in the area beyond any doubt.
We are aware that to foster the regional academic community, strengthen  Latin American science and 
technology capabilities, and consolidate the research-industry link, 
it is ideal to host such large projects. 
The development of capacities and enabling technologies are also key ingredients for 
social progress. The revenue generated by hosting this kind of efforts highly exceeds the investment by individual countries. 
The Latin American region has both an enthusiastic scientific community as well as geographic advantages 
(clear skies, large flat areas with low anthropogenic noise, etc.) to support these kinds of endeavors.




\chapter{Cosmology}\label{chapt:cosmo}

\vspace{1cm}
\noindent Diana López Nacir (DF/IFIBA UBA-CONICET, Argentina) \\
Marcelle Soares-Santos (U Michigan, USA)\\
Thiago Gonçalves (Valongo Observatory, Brazil)





\section{Introduction}

\noindent Cosmology is a branch of physics and astronomy, 
 dedicated to understanding the global properties and evolution of our Universe. It is a rapidly evolving field that stands at the interface between theoretical physics, particle physics, and astrophysics. Recent advances in both observational capabilities  and theoretical understanding  has lead to exhilarating deep connections between fundamental physics and cosmology. 
  
\noindent As a result of a worldwide community effort, a standard cosmological model has been built.  
The main pillars of the model are the standard theory of gravity, general relativity (GR), the Standard Model (SM) of particle physics, and the standard physical laws.  
In standard cosmology, the  evolution of the Universe is described  starting from a homogeneous and isotropic background Universe with small inhomogeneities characterized by certain  statistical distribution.
The base model, known as $\Lambda$CDM,  has only six free parameters that can be determined from several different observations.

\noindent Remarkably, the success  of the  $\Lambda$CDM model requires the existence of  Dark Matter (DM), a mysterious component for which there is no fundamental particle inside the SM of particle physics.  Therefore, to explain the evolution of the Universe new physics is required.  The DM  species constitutes roughly  27$\%$ of the current energy budget of the Universe. In contrast the ordinary matter and radiation in the SM corresponds to just the $5\%$  of the budget. The rest of the energy is described with the so-called cosmological constant $\Lambda$. 

\noindent The main reason for the introduction of  $\Lambda$, as a free parameter in the theory, is that its value can be adjusted to describe the  acceleration of the late time expansion of the Universe. There is no fundamental explanation nor understanding for the inferred particular value.  
  While this provides a  simple model that     is consistent with current observational data, other models provide alternative explanations of this acceleration. For example, some models attribute the acceleration to the presence of a dynamical component known as Dark Energy (DE), and others to a modification of the gravitational laws on cosmological distances.
A   scientific driver of the community is  to understand to what extent it is possible to discriminate among the different models from observations, and whether any of the models are better at fitting the data than what is currently the most accepted explanation, $\Lambda$.

\noindent Another mystery of the current understanding of the evolution of the Universe, is the origin of the  tiny inhomogeneities in the matter density.
Standard theory can explain the formation of structures (such as galaxies, galaxy clusters, etc.) as a consequence of the growth of smaller inhomogeneities in the matter density, but it does not explain the origin of the inhomogeneities.
  It has turned out to be  necessary to assume certain  (very particular) properties of the small perturbations that describe the initial departures of a spatially homogeneous and isotropic Universe, that is, of the primordial cosmological perturbations, that are inexplicable within the strict framework of the model.   Indeed, the model does not explain why the Universe we observe is so homogeneous and isotropic at ``large'' distance scales (as inferred from observations of the cosmic background radiation, see below), why large abundances of topological defects such as magnetic monopoles or cosmic strings are not observed, and why the Universe is practically spatially flat. 
Currently, the most widely accepted solution to these problems is given by the inflationary mechanism. Basically, this consists of adding a period of exponential or quasi-exponential expansion of the Universe, prior to what is known as the hot big-bang Universe. The challenge of explaining the physics of inflation is considerable. Inflation is assumed to have occurred  at a huge energy scale (perhaps as high as $\sim 10^{15}$ GeV), well out of reach of ground particle accelerators. Any description of the inflationary epoch, therefore, requires a huge extrapolation from the known laws of physics. 

\noindent The Cosmic Microwave Background (CMB) radiation,  provides very valuable information about the different parameters that describe the Universe on  large scales  (among which are the ones that determine the current expansion rate of the Universe $H_0$,  the relative abundance of matter, radiation, DM and the constant $\Lambda$, as well as those that characterize the primordial perturbations mentioned above and give us information about the inflationary model).  The next frontier of CMB research involves measuring its polarization.  After the first detection in 2002, measurements of the CMB polarization  have further confirmed the standard cosmological model and have increased the precision on the determination of cosmological parameters.  

\noindent Observational cosmology has undergone extraordinary advances in the past decade.   Nowadays,   there are  surveys characterizing the large scale structure (LSS) of the Universe by observing galaxies at different frequencies, taking their spectrum, creating catalogs with different properties.  
For instance, numerous experiments have measured with  high precision the  so called Baryon Acoustic Oscillations (BAO)  (which is a well understood feature in the
correlation function of a tracer of the LSS),  using both galaxies and the Lyman-$\alpha$ forest as tracers. 
Current data from  (Stage III)   spectroscopic surveys   already provide constraints  of comparable precision to those from the CMB, for some parameters (see for instance \cite{Philcox:2020vvt}).

\noindent With the increase of precision, some  evidence for discrepancies in  the values of the basic parameters have emerged. Primarily  when the inferred values for the  Hubble constant $H_0$ are compared with more direct measurements such as from supernovae catalogs or the strong lensing effect.  Depending on the data sets and  their combinations, the discrepancy  ranges between  $4\sigma$ and $5\sigma$ significance \cite{Verde:2019ivm}. 
 Other challenges of the standard  paradigm concern the description of some observables on ``small'' scales, such as the ``missing'' DM sub-halos or the so called core-cusp discrepancy \cite{Bullock17}.

\noindent From an observational perspective, the current status  highlights the importance of having different probes, studying cross correlations between different observables. 
 
\noindent Recently,  radio surveys are starting to implement a technique called Intensity Mapping (IM). This takes the integrated radio emission from unresolved gas clouds, and uses the redshifted $21$ cm line of Neutral Hydrogen gas (HI) (observed at radio frequencies)  as a tracer of the distribution of the LSS.  They are also detecting  mysterious Fast Radio Bursts (FRBs) which may be useful for cosmology or for some discovery.

\noindent The cosmological gravitational waves (GWs) are another important prediction of inflationary models and  can also leave their imprint in the CMB polarization. 
Whether there is a fraction of the polarization that was caused by GWs or not could be disentangled by using a  decomposition of the polarization properties into E-modes and B-modes.
 Scalar (density or temperature) inhomogeneities, at first order, generate only E-modes. Tensor (GWs) perturbations can produce both. The amplitude of the E-mode signals are much larger than the expected ones for B-modes.  Beyond the first order,    B-modes can be generated  from scalar  inhomogeneities, by an effect known as gravitational lensing, which rotates primordial E-modes into B-modes.
These  B-modes (called ``secondary B-modes") have been observed in the CMB polarization at relatively small angular scales, on the order of a few arc minutes. 
These successful measurements provide  an accurate mapping of the gravitational lensing effect which  can be used to study  the distribution of dark matter on large scales, and may even allow an indirect measurement of the mass of the neutrinos.
 In contrast, the so called primordial B-modes,  imprinted by  GWs have not been observed yet. They are expected to peak at an angular scale on the order of one degree. Therefore, a degree scale B-mode polarization signal could  be an indirect evidence for the existence of GWs with cosmological wavelengths.  
 From a specific inflationary model, together with a model for the subsequent cosmological history, one can  predict the expected level  for the amplitude of the signal in B-modes. To characterize the different predictions, a parameter  $r$,  known as the ``tensor-to-scalar ratio'', is defined. Therefore,  measurements and constraints on  $r$ can be used to probe  inflationary scenarios.
 The study of the generation of primordial GWs  during inflation has received increasing attention in recent times. 
 Beyond the context of inflation,  GWs can be produced in the early Universe by other mechanisms, such as by strong phase transitions after inflation or in non-inflationary alternative scenarios.

\noindent By itself,  the detection of  GWs is an event of spectacular historical importance.
GWs are a central prediction of General Relativity, and we have had indirect experimental evidence of their existence for many years: in particular, the Hulse-Taylor binary pulsar PSR B1913 + 16, discovered in 1974, is gradually losing  orbital energy at a rate that can be predicted based on gravitational radiation losses. A direct evidence has been the observation of GWs coming from the coalescence of a binary black hole system, carried out in 2016 by LIGO. What this  represents for physics is of enormous importance, since it allows the exploration of astrophysical objects that would otherwise be invisible, as well as the study of the physical phenomena associated with production of the waves and the corresponding theories.

\noindent The beginning of the multi-messenger astronomy era including GWs started with the detection of the  GW  together with the coincident  Gamma Ray Burst (GRB) and other electromagnetic signals,  associated to the coalescence of a binary of two neutron stars \cite{2017ApJ...848L..12A}. 
The combination of GW detections with electromagnetic-wave astronomy and the study of high-energy particles opened an unprecedented window  to probe the Universe.  This allows new bounds on the violation of Lorentz invariance, and tests the equivalence principle by constraining the Shapiro delay between gravitational and electromagnetic radiation, as well as constrains on  alternative theories of gravity and on models of the dark sector.

\noindent From the theoretical perspective, the current status highlights the need of  analyzing data to obtain physical quantities, test robustness, to deepen the understanding of data and how they are analyzed, to connect results from cosmological, astrophysical and laboratory experiments, to study alternative theories, their consistency, robustness, connections with particle physics and with the fundamental physics laws, to evaluate the validity of the standard assumptions, etc.

\noindent The main science drivers (SD) of the topic,   can be summarized in the following questions:

\noindent {\bf  What are the nature, properties and origin of the dark components?}
Thanks to many observations, not only  cosmological  but also on astrophysical  scales,  it has been possible to infer some  properties  of the dark components.
All observational evidence for the dark components is due to its direct or indirect gravitational effects. Their nature is still a mystery, and so far there has been no convincing detection of their possible non-gravitational interactions.

\noindent {\bf What is the origin of matter-antimatter asymmetry? }
All visible matter in the Universe is made of fundamental building blocks, the elementary particles. For each elementary particle, there is an antiparticle that has the same properties but opposite charge. In principle,  the Big Bang produced equal numbers of particles and antiparticles. However, the Universe today is considered to consist almost entirely of matter (particles) rather than antimatter (antiparticles).
The reason why there is matter, and no antimatter, in the energy budget of the Universe is still a mystery.  Where have all the antiparticles gone?    Currently there is no acceptable understanding of this asymmetry problem.

\noindent {\bf What are the nature and the properties of neutrinos?}
As overviewed in \cite{Strategy:2019vxc-1}, one of the great mysteries in particle physics is the determination of  the neutrino mass and its fundamental nature.
The possibility  that neutrinos could be their own antiparticles may be linked to the matter-antimatter asymmetry problem. 
 A cosmological neutrino background  is a   predicted relic of the standard cosmological model. Although it has not been directly  detected  yet, it has been indirectly verified by its role in the prediction of the observed primordial abundance of light elements (known as Big Bang Nucleosynthesis process) and by its imprints on  the CMB and  the LSS.
Cosmological data such as CMB and the distribution of LSS are sensitive to the total mass of  the neutrinos.   
From cosmological observations it is also possible to infer the number of relativistic neutrino states in thermal equilibrium in the early Universe, which in standard cosmology it is consistent with three neutrino species, and in general sets further constrains on alternative models and on light sterile neutrino scenarios.

\noindent {\bf What are the nature and properties of Black Holes (BHs)?}
This addresses  questions such as: Do primordial BHs exist? Are they (part of) the DM?  How can BH catalogs be used to address these questions and/or to extract useful cosmological information?

\noindent {\bf What is the origin and nature of the primordial perturbations?}
 Observations have made it possible to infer many of the statistical properties of primordial perturbations; for example, they are currently compatible with a primordial gaussian distribution of only adiabatic density perturbations and, furthermore, it is possible to set bounds on the level of non-Gaussianities and on the level of primordial GWs. As a result, many of the inflationary models have been ruled out.

\noindent {\bf Are the standard assumptions wrong? }
For instance,  one can replace GR by another theory of gravity, consider theories were  Lorentz  symmetry is violated, non-standard thermal or non-thermal history, etc.

\section{Experiments and infrastructure with cosmological impact with LA participation} 
In this Section we briefly describe the LA participation in some existing and planned experiments as well as some infrastructure and training that have an impact in the area of cosmology.

\subsection{{\bf B}AO from {\bf I}ntegrated {\bf N}eutral {\bf G}as {\bf O}bservations (BINGO)}

The BINGO telescope is a new instrument designed specifically for observing BAO in the frequency band $960 - 1260$ MHz and to provide a new insight into the Universe at $0.13\leq z \leq 0.45$ with a dedicated instrument. The optical configuration consists of a compact, two 40m diameter, static dishes with an exceptionally wide field-of-view ($15^\circ$ x $7^\circ$) and 28 feed horns in the focal plane. 

\noindent BAO observations will be carried through IM HI surveys and its main scientific goal is to claim the first BAO detection in radio in this redshift range, mapping the 3-D distribution of HI, yielding a fundamental contribution to the study of DE, with observations spanning several years. No detection of BAOs in radio has been claimed so far, making BINGO an interesting and appealing instrument in the years to come. Moreover, in view of the observation strategy and with some adjustment in its digital backend, BINGO will also be capable to detect transient phenomena at very short time scales ($\lesssim 1$ ms), such as pulsars and  FRBs.
 
{\bf Current LA  Involvement:} BINGO has currently about 50 participants in more than 20 institutions from Brazil (with the most participants, approximately 30), China, France, Germany, Portugal, Spain, South Africa, Switzerland and England.  The project is led by Brazilian researchers, with most of the investment from Brazilian funding agencies (see below), and will be hosted in Brazilian territory.

{\bf Facilities and resources:}
BINGO is being built in Para\'iba, Northeastern Brazil, with a strong support of state authorities. There is also support from the Ministry of Science and Technology and FINEP, a related funding agency. 

{\bf Technological advances}: Receiver construction, operating at room temperature, but with a careful front-end design, delivering a very low system temperature and very good stabilization. This subsystem is under development and can be considered a technological advance.

The optical design was mostly done in Brazil, with a complex focal plane distribution of the horns. BINGO is probably the largest example of a “single dish, many horns” telescope. 
The fabrication of the two 40m dishes are being carried by a Brazilian company, with more than 30 years of experience in telecommunications.

{\bf Advanced training}: There are currently three Brazilian postdocs and a number of M.Sc. and Ph.D. students deeply involved with aspects of the mission preparation, including development of the necessary instrumentation for the telescope. They are either leading or participating in the first series of papers of the instrument, to be submitted to Astronomy \& Astrophysics not later than October 2020.

{\bf Connection with Industries}: As mentioned before, the construction of the instrument is led by Brazilian institutions and in being carried by Brazilian companies. As of the time of this writing, there are no spin-offs or start-ups from BINGO.

 {\bf Timeline and Major Milestones:}
BINGO will start operations in late 2021. Commissioning of a few receivers operating in the telescope should start by early 2022. Currently the engineering projects for telescope construction are being completed and the terrain is being prepared for construction.

The project has undergone a major review in July 2019, with the some recommendations that include the receiver configuration, final optical design and an "end-to-end" pipeline to simulate the mission behaviour prior to the beginning of operations. The project is within schedule to commence operations as planned.

\subsection{Macon Ridge Astronomical Site:
The ABRAS and TOROS projects}
The goal is to  develop a new astronomical site at the Macon Ridge, in Salta,  Argentina (in the Atacama Plateau region) starting with  two complementary projects:    The Transient Optical Robotic Observatory of the South (TOROS)  and The Argentina-Brasil Astronomical Center (ABRAS) projects.

\noindent The main scientific goal concerns GW transient events with a main focus on those arising from the coalescence of BNS  systems, for which an electromagnetic  (EM)  follow-up was shown to be promising. For the coalescence of a  binary of two Black Holes (BBHs), the most frequently detected events thus far,  no  EM  emission is expected. Nevertheless,    exceptional  scenarios  are being considered  for which an EM signal could be detected.  
There are still clear uncertainties as to the frequency of these events and the improvements in search methods and the progress in the commissioning and final operation of the detectors undergoing through the planned upgrades. 
 
\begin{figure}[htbp]
\begin{center}
\includegraphics[width=15cm]{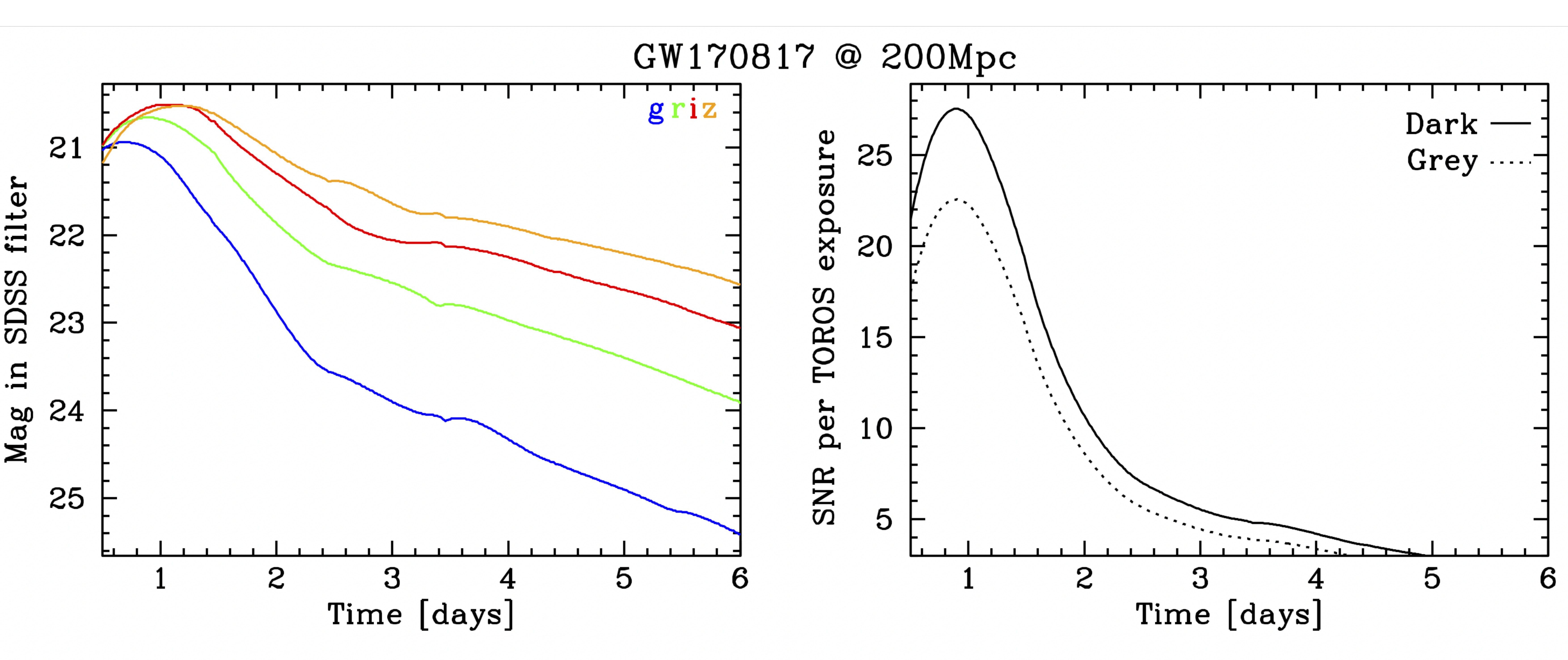}
\caption{Left: light curves of GW170817 \cite{2017ApJ...848L..17C}    shifted to a distance of 200 Mpc. Right: expected SNR of TOROS photometry of such an event. ``Dark'' and ``grey'' refer to typical sky brightness values associated with the new and quarter Moon, respectively. Even in the most pessimistic case, TOROS remains sensitive (SNR$>3$) to kilonova events for at least 4 days. Figure reproduced from \cite{MACON:WP}. }
\label{TOROSFOM}
\end{center}
\end{figure}

\noindent The TOROS collaboration was involved in the EM follow up  of the  first GW detection  of  several BBH mergers discovered during the O2 observational campaign (for which there was no significant EM counterpart found). 
 For the  first GW transient associated to  a BNS coalescence,  GW170817, the  collaboration obtained photometry for two nights \cite{Diaz2017}. 
As for   GW170817, the localization of gravitational waves
by the network of three detectors is expected to be of a few tens of square degrees. Thus the EM observations need to deal with a rather large positional uncertainty. Figure\ref{TOROSFOM} shows the expected light curve of the EM counterpart of a BNS merger at a distance of 200 Mpc, as well as the signal-to-noise ratio of the resulting TOROS photometry.

\noindent The  TOROS  project will be a fully robotic facility which   will   offer an  extended coverage of the southern skies. 
   The  envisioned    operation modes, in decreasing order of priority, are: 1) Follow up of gravitational-wave triggers, releasing coordinates of potential transients through the GCN circular methods \footnote{Details about the GCN system are available from \url{http://gcn.gsfc.nasa.gov}}, and stacked images to the entire astronomical community; 2) Follow up of short-duration $\gamma$-ray bursts  (GRBs) events that will serve as GW  triggers  by performing a similar strategy used to detect the GW counterparts.     In the northern hemisphere,  this search has started in the last few years by the Palomar Transient Factory and the Zwicky Transient; 3)  Baseline imaging of the entire surveyable area (expected to be completed within the first year of operation) to have  previous imaging of any region of interest as well as   previous knowledge of the rates of transients and variable sources with similar temporal characteristics,  useful  for estimating the foreground/background contamination rate from other astrophysical events.

\noindent The  ABRAS telescope will   complement   TOROS detector at design sensitivity, by carrying out both near-infrared imaging and optical spectroscopic follow-up characterization of transients identified by TOROS.

{\bf Current LA involvement}: 
TOROS    involves participation from 3  Latin American countries, USA and Poland.
In the submitted white paper \cite{MACON:WP} there are   9 researchers  in Argentina (8 members from IATE and 1 interested - not a member -  from  CASLEO),   4 members in
 Chile (3 from  Universidad de La Serena and 1 from   Pontificia Universidad Catolica), 2 members in  Mexico (1 from 
Instituto Nacional de Astrofisica  and 1 from Optica y Electronica) and  8 members from non-LA countries (7 from USA and 1 from In Poland).

ABRAS  involves  Argentina and Brazil. The main institutions behind  are  IATE and the IAG (Instituto de Astronomía, Geofísica e Ciencias Atmosféricas), from Sao Paulo, Brasil, while the funding institutions are the  MINCyT (Ministerio de Ciencia, Tecnolog\'ia e Innovaci\'on Productiva) and the USP (Universidade de Sao Paulo). The infrared-optimized 1 meter telescope was delivered to C\'ordoba (Argentina) in late 2015, and the dome was built in   Macon Ridge in 2012. 
 
{\bf Local facilities and resources}: 
The  Astronomical Site and Dark Sky Preserve was established by the Governor of Salta Province in 2011. 
The preserve  is managed by IATE.  Roads, electricity and Internet access  are available.
The local collaboration also involves site characterization, infrastructure, logistics,  maintenance and operation.
Telescopes at the site can be remotely controlled  from  village of Tolar Grande  (9 kms distance along the line of the sight). The provincial government constructed a road from Highway 27, which joins Tolar Grande with San Antonio de los Cobres highway, to the Macon ridge.  

\noindent TOROS  plans to execute their own photometric and spectroscopic follow up of the GW triggers  using  the Bosque Alegre 1.5-m telescope in C\'ordoba and additional resources from the members of the TOROS collaboration in Chile and Argentina (CASLEO observatory in San Juan, Argentina and the Mammalluca observatory from La Serena University in Chile). The spectroscopic followup will be conducted using the ABRAS telescope but also the Gemini, Gran TeCANVLT 8-m telescopes, through target-of-opportunity time to be obtained via the US share of Gemini, and by their Chilean and Mexican collaborators with access to the other facilities.

{\bf Technological advances}: The  considerations sketched above led  the collaboration to designe a system with the largest possible aperture within the budget restrictions (primary mirror diameter of 0.6m) and a large field-of-view (10 square degrees) camera with a very broad bandpass ($0.4-0.9\, \mu m$, equivalent to a combination of the Sloan griz filters). 
There are no commercially available telescope systems (or specific corrector optics) that can meet TOROS requirements. Therefore, a custom corrector optical assembly will replace the secondary structure on the optical tube and will provide the interface between the primary mirror and the CCD focal plane.
 
{\bf Timeline and major milestones}:
The TOROS telescope is ready to start operations. If funding is secured, the ABRAS telescope could start operations soon.
The telescope and its dome have been  installed at the site.  The camera and the manufacture of the prime-focus corrector are already under construction with commissioning expected by December 2020. The focal corrector is currently under construction and the CCD has already been purchased with expected delivery in 2020.

\subsection{{\bf Q}\&{\bf U} {\bf B}olometric {\bf I}nterferometer for {\bf C}osmology (QUBIC)}\label{QUBIC}
  QUBIC is an experiment designed to measure the polarization of the CMB \cite{QUBIC:WP}.   
The main scientific goal  is the measurement of  a primordial  B-modes signal. This is a smoking gun for primordial gravitational waves whose relative strength is given by $r$.   The current  bound  is $r < 0.06$ \cite{Planck2018}, which already disfavours the simplest inflationary models of the early Universe. 
 In the near future,  experiments are expected  to achieve  a sensitivity  to $r$ of order $\sigma(r)=0.01$, with  improvements  up to a few times $0.001$.
QUBIC is one of the international efforts currently  pursuing this  goal, and is using  a novel kind of instrument, which combines the extreme sensitivity of two arrays of bolometric detectors, operating respectively at 150 GHz and 220 GHz,   with the control of the systematics offered by  the interferometric operation of the instrument.  
 
\noindent At the unprecedented sensitivity level that the new generation of  experiments aim to reach,  the systematic effects control and the possibility to remove foregrounds  is a  major concern.     The primordial B-mode signal can be affected by instrumental systematic effects and is contaminated by astrophysical foregrounds, particularly polarization by dust grains aligned by the galactic magnetic field and also synchrotron emission by relativistic electrons. The latter can in principle be accounted for through their angular power spectra and frequency dependence, distinct from that of the CMB.

\noindent The QUBIC Collaboration \cite{Bernardis2018} reports the outcomes of end-to-end simulations of the QUBIC results from 2 years of operation of the full instrument, setting the input for primordial B-modes to $r=0$. There it is shown
that using the spectro-imaging capabilities  it is possible to reconstruct  frequency  sub-bands and obtain  5 maps that are used to estimate the parameter $r$ as well as three-parameters B-modes dust emission model (from latest Planck model).  The width of the likelihood on r shows a  sensitivity  $\sigma(r)=0.013$.  Table \ref{ForecastQUIBIC},  reproduced from \cite{QUBIC:WP},  
 lists the expected sensitivity of QUBIC and other ground-based experiments in the same frequency range, either running or expected to be deployed in the near future. 
It is highlighted in \cite{QUBIC:WP} that not only the reachable sensitivity should be taken into account, but also the systematic effects control and the possibility to remove foregrounds. In this sense, QUBIC has a unique status due to its particular architecture as a bolometric interferometer.

\begin{table}[h]
\caption{Sensitivity of the main B-mode ground experiments operating in a frequency range similar to QUBIC. The label ``fg'' or ``no fg'' corresponds to the assumption on the foregrounds. 
 Taken from   \cite{QUBIC:WP}.}
\begin{tabular}{llllll}
\hline
Project                  & Frequencies (GHz)         & $\ell$ range  & $\sigma(r)$ goal (no fg.) & $\sigma(r)$ goal (with fg.) \\
\hline
QUBIC                    & 150,220                   & 30-200        & $6.0 \times 10^{-3} $        & $1.0 \times 10^{-2}$     \\
Bicep3/Keck              & 95, 150, 220              & 50-250       &$ 2.5 \times 10^{-3}  $       &$ 1.3 \times 10^{-2}$       \\
CLASS *                  & 38, 93, 148, 217          & 2-100         & $1.4 \times 10^{-3}  $       & $3.0 \times 10^{-3}$         \\
SPT-3G $\dagger$                 & 95, 148, 223              & 50-3000       & $1.7 \times 10^{-3} $        & $5.0 \times 10^{-3}$         \\
AdvACT $\ddagger$       & 90, 150, 230              & 60-3000       & $1.3 \times 10^{-3}$         & $4.0 \times 10^{-3}$          \\
Simons Array   & 90, 150, 220              & 30-3000 &      $1.6 \times 10^{-3}$         & $5.0 \times 10^{-3}$          \\
SO (SAT) **              & 27, 39, 93, 145, 225, 280 & 30-300        & $1.3 \times 10^{-3} $        & $3.9 \times 10^{-3}$      \\
\hline
\end{tabular}\\
**CLASS: Cosmology Large Angular Scale Surveyor; $\dagger$ SPT-3G: South Pole Telescope-3rd gener- \\
ation; $\ddagger$ AdvACT: Advanced Atacama Cosmology Telescope; **SO (SAT)=: Simons Observatory\\ Small Aperture Telescopes.\label{ForecastQUIBIC}\
\end{table}

{\bf Current LA involvement}: 
QUBIC International Collaboration is integrated by France, Italy, UK, Ireland, USA and Argentina. In Argentina, 43 members (researchers, engineers, technicians and students) are part of different working groups. The involved institutions are CNEA (CAC, CAB, NOA and Cuyo), ITeDA (CNEA, CONICET, UNSAM) Instituto Balseiro and UNLP.  
As part of the local work 5 PhD thesis are  in progress.
For the first module, Argentina is the only country in Latin America involved. In the future, for operation and installation of all the observatory, it is possible that new agreements with more countries in the region could be signed.

{\bf Local facilities and resources}:  
 Development of  Alto Chorrillo site  (a new scientific/astronomical pole in Salta, Argentina,   where LLAMA will be also located).  This site is  180 km away from the Chajnantor site in Chile where other millimeter-wave experiments and observatories are located (ALMA, APEX, Advanced ACTPol, POLARBEAR and CLASS) and offers similar atmospheric properties.  This location has been characterized during several years within the framework of the site selection process for the CTA project, and for the LLAMA project, which has atmospheric requirements similar to those of QUBIC. 
    The specific contributions from  Argentina involve roads, energy, telecommunications, qualified human resources, contact with the community.

{\bf Scientific advances}:
The contributions of the  argentine QUBIC collaboration  can be grouped in  simulations for the data acquisition, map-making process and component separation to clean the CMB and obtain forecasts on the parameter $r$.
In particular,  the  study of the angular resolution of the reconstructed CMB maps obtained with the simulation pipeline for different sub-frequencies within the 150GHz band, taking into account the spectro-imaging capabilities of QUBIC 
\cite{Gamboa}.
Another important contribution is the improvement of the end-to-end pipeline, and the simulations of the reconstructed maps, that are useful for the study of the instrumental noise. The development of a component separation framework to be able to separate the CMB signal from the foregrounds and studying the ability of QUBIC for distinguishing between different dust models in the data analysis procedure to obtain $r$ and the dust model parameters. 

{\bf Technological advances}:
 Several Argentine research laboratories have started to work on topics related to the   instrument. In order to be able to provide detectors for future instruments, the development of new detector architectures with possibly reduced complexity but enhanced performance, has begun. Also, new readout electronics is being developed. 
 
{\bf Advanced training}:
The  collaboration is planning to develop a series of advanced schools for young scientists and engineers (see for example \url{http://lapis2018.fcaglp.unlp.edu.ar/index.html}).
A program of grants to cover specific topics and allow the incorporation of new students in different stages of their careers is also part of the medium and long term educational purposes of the project.
 
{\bf Connection with industries}:
There is no direct involvement of local industry in the project. Nevertheless, for the fabrication of non-standard equipment and the QUBIC mount, a rotary mechanics with high precision requirements, private companies will be contracted.

{\bf Timeline and major milestones}:
The  instrument is currently in   laboratory calibration phase of its so-called ``Technological Demonstrator''.  The next steps are to test and to install the first module in the site, with first light  expected  for 2021.
The access road to the LLAMA site and from there to the QUBIC site has already been built, and construction for the installation of the first module is under way.

{\bf Computing requirements:} 
For the data analysis, several Tb of storage will be needed, which will be provided at NERSC computing center, where the final analysis will be performed. 
\subsection{{\bf S}outh {\bf A}merican {\bf G}ravitational-Wave {\bf O}bservatory (SAGO)}
 
 SAGO  \cite{SAGO:WP} is a proposed next generation detector for gravitational waves (GW).  The emerging field of gravitational wave astronomy had its first experimental breakthroughs in the last five years, with the detection of mergers of binary black holes \cite{GW150914} and binary neutron stars \cite{GW170817} by a network of three advanced gravitational wave detectors: two in the United States (LIGO--Hanford, and LIGO--Livingston) and one in Europe (Virgo). These two mergers, are the most prominent amongst dozens of such events detected in three observing campaigns to date. Together, LIGO and Virgo observations have enabled various high-impact studies ranging from constraints on the equation of state of neutron stars, to tests of general relativity, and cosmological measurements. In the area of cosmology, in particular, advanced gravitational wave detectors have led to novel independent measurements of the present rate of cosmic expansion (the Hubble constant, $H_0$)  \cite{GW170817:H0,Soares-Santos:2019,Palmese:2020} which are crucial to shed light onto one of the greatest scientific problems of our time: the physical nature of the accelerated expansion of the universe.
While the international GW community continues to upgrade and expand the existing network (e.g. adding a detector in Japan, KAGRA), they have also already started planning the next generation of detectors, known as the 3rd generation (3G) GW observatories. Two such detectors have been proposed: Cosmic Explorer (in the U.S.) and Einstein Telescope (in Europe). These new detectors will represent a giant leap in technology, enabling for the first time detection of mergers at high redshifts ($z \sim 30$). The Latin American community is proposing to build its own 3G GW detector, SAGO.
Such a detector will definitely put Latin America in a leadership position in the exciting new field of gravitational wave astronomy, with particular impact on multi-messenger cosmology.
 
\textbf{Current and planned involvement}:
Latin America has active groups of scientists invested in both the LIGO and Virgo collaborations. The community seeks to grow in size, capability, and impact in the next decade aiming at the ambitious goal of becoming the hosts of a 3rd generation GW detector: SAGO.  

\textbf{Local facilities and resources:}
While the selection of the SAGO site will require an extensive technical study, a  preliminary assessment performed by the USP-IAG seismic group identified three suitable locations: 1) near the Brazil Rio Grande do Sul and Uruguay border, 2) near Brazilian states of S\~ao Paulo and Mato Grosso do Sul, 3) ``Tri\^angulo das Secas" in the Brazilian northeastern region. Two other regions (not seismically calm) are also considered: 1) The Uyuni salt flat in Bolivia, and 2) The AUGER facility.

\textbf{Scientific advances:} The proposed 3G projects are complementary to each other: the larger the network, the better the results (including localization area, distance determination, and physical properties such as mass and spin). As part of the 3G network, SAGO will bring about a new cosmology breakthrough, enabling measurements of the cosmic expansion history since 100 Myrs after the Big Bang. 

\textbf{Technological advances:}
The community plans to focus on detector technology, learning from the experience with the current detectors and performing R\&D work towards the 3G era. Advances in areas such as core optics, light sources, coatings, and cryogenics, are anticipated. 

\textbf{Advanced training:}
The Latin American GW community will grow by training PhDs locally (gaining experience through participation in current state of the art interferometers) as well as by attracting and retaining experts from abroad. 

\textbf{Connection with industry:}
Laser technology and optics are examples of areas of synergies with industry. 

\textbf{Timeline and major milestones:} 
Over the coming decade, the priority is to build a critical mass of researchers with technical expertise. Goals are to perform R\&D and site determination studies. SAGO will then be built in the following decade, and join the 3G GW network in the 2040s.

\textbf{Funding profile:}
SAGO is roughly estimated to be in the range of US\$ 1-2 Billion.

\textbf{Computing requirements:}
Computing support is needed for: data processing, analysis, simulations.
\subsection{Vera Rubin Observatory's {\bf L}egacy {\bf S}urvey of {\bf S}pace and {\bf T}ime (LSST)}

  LSST  is a survey of over 1 billion objects in the sky, using a wide field of view camera on a large aperture telescope in Chile. Among other goals, LSST will explore the nature of DM  and DE. It will do so by analyzing over 20 terabytes of data per night to measure the distribution of galaxies, clusters of galaxies, and supernovae, as well as  the distortions of the shape of galaxies produced by intervening matter. The community plans to implement an independent Data Access Center (iDAC) for LSST in Brazil \cite{LSST:WP}.

\textbf{Current LA involvement}:
The Brazilian participation in  LSST is coordinated by the Laborat\'orio Interinstitucional de e-Astronomia (LineA, \url{linea.gov.br}) and includes 9 PIs and 19 junior members. 

\textbf{Local facilities and resources:}
LIneA's computer facilities presently dedicated to LSST work consist of three clusters yielding a combined total of approximately 25 Tflops of computing power that will operate at 100Gb/s. Moreover, the Laborat\'orio Nacional de Computa\c{c}\~ao Cientifica (LNCC) provides access to a supercomputer called Santos Dumont which has 1 PFLOPS available for shared use.
 It is expected that, by 2021, dedicated computing power will increase to over 100 Tflops with funds already allocated to LIneA by the federal funding agency FINEP. 
 
\begin{figure}[htbp]
\begin{center}
\includegraphics[width=15cm]{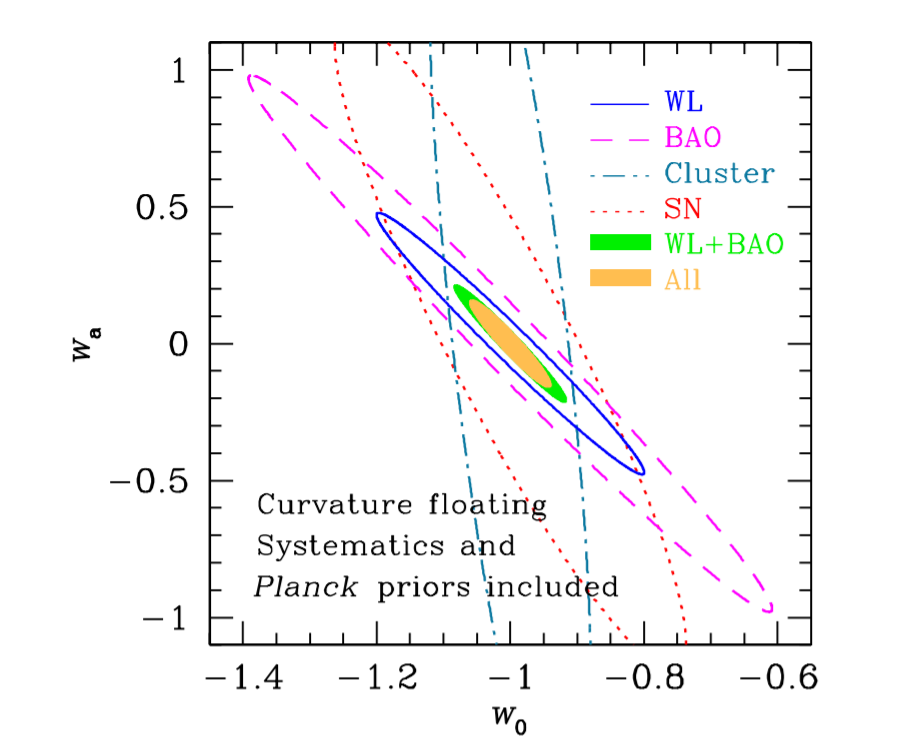}
\caption{Forecast for the accuracy in the estimation of Dark Energy parameters $w_0$ and $w_a$ from multiple probes at the LSST. Figure reproduced from the LSST Dark Energy Science Collaboration (DESC) Science Requirements Document. }
\label{lsst}
\end{center}
\end{figure}

\textbf{Scientific advances:} LSST will use measurements of the galaxy distribution, weak gravitational lensing, supernova and cluster counts to find the best fit cosmological model parameters with unprecedented accuracy and precision. In figure \ref{lsst} it is shown the forecast on the parameters of the dark energy equation of state parametrized as $w(a) = w_0 + w_a (1-a)$. 

\textbf{Technological advances:} Rapid image processing and advanced machine learning techniques are among the core developments of LSST.

\textbf{Advanced training:} LSST research groups in Brazil will produce PhDs with expertise in state of the art cosmic survey analyses. 

\textbf{Connection with industry:} Synergistic opportunities involving computational methods include machine learning and ultra-fast image processing.

\textbf{Timeline:} 
LSST is scheduled to start observations in 2022, lasting until at least  2032.

\textbf{Computing requirements:}
Massive computational resources are necessary for photometric redshift estimation of a billion galaxies, production of galaxy catalogs, weak lensing catalogs, cluster catalogs, supernova catalogs and for running the analysis pipelines.

\subsection{Latin American  PhD  program }

The goal is to create a collaborative LA PhD  program in Astrophysics, Cosmology and Gravitation,   as a network of   institutions from different countries  in the continent.
  Progress in this field  with global impact not only needs  research facilities and possibilities to access observational data, but also qualified human resources, a very important part of research infrastructures.  
  The proposal  described in \cite{LAGS:WP} is based on the  PPGCosmo  experience  (which is a Brazilian PhD program based at UFES) and on the observation  that participation of many LA  countries in  astronomical facilities installed in LA is below what could be expected.
 To create the program, firstly, it is important to form an advisory scientific committee composed of leading scientists from different countries working on those subjects. The institutions and research groups must express explicitly the interest to join the proposed program, with the necessary agreement terms. Hence, the necessary scientific and administrative personnel, as well as the infrastructure, are in principle already available. The important new contributions to this project would be a significant number of PhD fellowships, which should be provided by the different countries, together with a specific financial support for scientific missions within the continent. It would be important that those fellowships have a fixed common value to all students. The main goal is to carry out a common PhD formation, inducing a deep scientific collaboration among all participating institutions and research groups with extensive use of the astronomical facilities existing in the continent. The PhD student will develop their project thesis under the supervision of at least two advisors from at least two institutions in different countries. In order to simplify the bureaucratic work the PhD certificates would be emitted by one host institution and immediately recognized and validated by all institutions joining the project. Since the student internship period in at least two institutes would be a mandatory requirement to obtain the PhD degree this will induce a closer interaction among the institutions and research groups participating in the program.

\section{Areas of Excellence}
 
Latin America is home to a vibrant research community in cosmology. With strong participation in projects ranging from cosmic surveys \cite{LSST:WP,Bernardis2018} to gravitational wave detectors \cite{SAGO:WP}, Latin American groups contribute to cutting-edge results in the field. Remarkable success has been achieved despite their small size relative to groups in U.S. and Europe.  

For example, in the recently concluded Dark Energy Survey (DES, \cite{DES}), the LineA team (in Brazil) has led data processing pipelines and important analysis developments, taking leadership roles within the collaboration and producing high-impact publications. The team has grown to include dozens of researchers many of whom now form the core of the Brazilian participation in the upcoming LSST project.   Similarly, the experience of Latin American researchers in LIGO and Virgo has seeded a community that is now ready to take a bold step towards a leadership role in the next generation of GW experiments. 

Specific areas of excellence include: 1) high performance computing and instrumentation for cosmic surveys and telescope facilities operating in wavelengths ranging from optical to sub-millimeter regimes; 2) detector development, operations, and data analysis for GW observatories. 

Despite challenges regarding sustained funding for long term projects, the community has successfully leveraged its resources and has demonstrated ability to take leadership roles in larger scientific endeavors in the coming decades.

\section{Synergies}
 
In synergy with QUBIC CMB polarization measurements at large-scale angular detection (degree), a bolometer has begun to be designed to be placed in the LLAMA Cassegrain cabin to measure polarization on smaller scales (arcmin or smaller) for the same part of the sky.  The experiment would complement QUBIC search for primordial B-modes, by measuring the secondary B-modes. In addition, it could also have an impact on other scientific goals, such as  testing  the  main pillars of the base cosmological model,  the properties of  the dark sector,  the sum of neutrino masses.

In synergy with neutrino experiments, cosmology can probe the sum of the neutrino masses.   In the based $\Lambda$CDM model with three massive neutrinos, the current limits obtained from   CMB and BAO data, is  $m_\nu = \sum_i m_{\nu_i} < 0.12 $ eV (95\% CL) \cite{Planck2018}. This  bound  is close to the minimum allowed by the inverted hierarchy.  If the mean value of $m_{\nu}$ is  sufficiently less than $0.1$ eV (but not if $m_{\nu} > 0.1$ eV \cite{RoyChoudhury:2019hls}), which can be the case  only   if the neutrino masses follow the normal hierarchy,  
results from future surveys such as EUCLID are expected to be able to  differentiate between the two hierarchies  \cite{Sprenger:2018tdb}. The sensitivity of Planck+Euclid to a neutrino mass sum of 0.06 eV  (close to the minimum  level allowed by oscillations)
 is $\sigma(m_{\nu})=0.02$ eV \cite{Sprenger:2018tdb}.
 
Studies of models of DM, DE, and gravity are extremely complementary, and the understanding of one of them cannot be achieved without the understanding of the others. 
Astrophysics and cosmology offer a complementary approach to study the fundamental properties of DM. They probe DM directly through gravity, the only force to which DM  is known to couple. 
In synergy with direct and indirect DM detection experiments, LSST is expected to provide complementary  tests of well-motivated theoretical models of DM (such as  warm DM, self-interacting DM, DM- baryon interactions,  axion-like particles, ultra-light DM, and primordial BHs) and  help to constrain DM-baryon scattering, DM  self-annihilation, and DM decay. GW experiments can also test some DM models (such as ultra-light DM and primordial BHs  candidates).
 
In addition, CTA \cite{CTA:WP} will also be able to measure the interaction of Very High Enery (VHE) photons with the Extragalactic Background Light (EBL). The EBL is difficult to measure and detect, due to its intrinsic faint surface brightness and strong contamination by foreground sources, but provides an important tool to measure Cosmic Voids and the overall baryonic matter distribution in the Universe, providing important cosmological constraints. The imprint of this structure on the VHE photons, however, will be detectable by CTA, offering a new way to measure the evolution of cosmic voids with cosmological time.

Likewise, GRAND \cite{GRAND:WP} is expected to provide important information on the birth of galaxies in the dark ages, with  profound cosmological impact. By operating at  radio MHz frequencies, GRAND will be able to detect the absorption from the hyperfine transition of the hydrogen atom at very high redshifts ($z\sim 20$), following on the first tentative detection of the signal from the Epoch of Reionization (EoR) by EDGES \cite{Bowman2018}.

Finally, one should emphasize the many synergies of cosmological studies with the evolution of galaxies, as the building blocks of the  LSS  in the universe. These areas are inherently complementary, and as such could benefit strongly from interactions and a joint observational structure -- with the Vera Rubin Observatory being a prime example of complementary goals achievable by the same observations.


\section{Conclusions}

While the LA cosmology community remains rather sparse and scattered about the continent, significant progress has been made towards greater coordination among groups. 
Leveraging local facilities and resources, individual Latin American groups that have traditionally grown in isolation are now pursuing or considering joint ventures. 
The current landscape of  projects is balanced between Latin American led initiatives and global collaborations with Latin American participation. 
Proposed Latin American led experiments include BINGO and SAGO. 
Global   experiments located in  Latin America  include  QUBIC,  TOROS and ABRAS, and LSST. This involves the development of the new astronomical sites at the  Macon Ridge (home to the Toros and ABRAS experiments) and Alto Chorrillos (where QUBIC and LLAMA are located) which are both  in the northwest corner of Argentina, in the Atacama Plateau shared with Chile.  

Apart from these notable examples which involve  part of the Chilean, Brazilian and Argentinean communities,  most of the LA community members are currently dedicated to theoretical aspects not directly connected with any experimental facility. Such theoretical studies continue  to be  essential to exploit observational data and experimental results, deepening  the understanding of their connection with fundamental physics and helping to motivate new ways to explore nature.

However, the LA community faces several challenges when addressing its main science drivers.
One  problem  is the lack of long term stability of funding and variations in the landscape from country to country. This  discourages large scale joint projects despite the fact that  know-how and critical mass have already been achieved. 

The white papers indicate broad community agreement on the benefits of joining forces between different groups and countries despite these challenges.    
Capacity building and collaborative work are    important elements. 
 LA-CoNGA \cite{CONGA:WP} Physics project and the proposal of the Latin American PhD  program (which emphasizes the need of increasing Latin American participation in facilities installed in the region and involves diverse topics), are clear examples highlighting these key elements. The need to participate in international collaborations has also been emphasized.



 

\chapter{Dark Matter}\label{chapt:DM}
\noindent {Marcela Carena (Fermilab, US)\\ 
Diego Restrepo\footnote{\href{mailto:restrepo@udea.edu.co}{restrepo@udea.edu.co}}(Instituto de Física, Universidad de Antioquia, Colombia)
}

\lstset{language=[LaTeX]TeX
,basicstyle=\ttfamily
,showstringspaces=false
,keywordstyle=\color{blue}
,commentstyle=\normalfont\ttfamily\color{red}
,stringstyle=\color{Green}
,inputencoding=utf8
,extendedchars=false
,frame=shadowbox 
}


\section{Introduction}
This chapter focuses on activities in the Latin American region, for a complementary worldwide range of activities see the recent European Strategy update~\cite{Strategy:2019vxc}.

\noindent One of the questions for fundamental science which clearly points to
the existence of Beyond Standard Model physics is Dark Matter (DM),
an unknown,
non-baryonic matter component, whose abundance in the Universe exceeds
the amount of ordinary matter roughly by a factor of five.  
The evidence for DM stems from different times and distance
scales of our Universe. The precise measurements of the cosmic
microwave background (CMB) power spectrum, galaxy rotation curves,
galaxy clusters, gravitational lensing, baryonic acoustic
oscillations and Big Bang Nucleosynthesis (BBN) have shaped our understanding
of the universe and represent a whole panoply of evidence for DM. However, all these data points to
DM only through it gravitational interactions, giving no specific information about its nature.
Despite the fact that DM has been searched for decades in terrestrial and extraterrestrial experiments and probes, those searches 
have given no specific information of what  DM actually is.

\noindent There is lack of information on how  DM was produced in the
early universe, before BBN, and  therefore, the expected DM
signals of the relic density  observed today depend significantly on the specific assumptions of its unknown history.  Supposing  that  the same standard cosmology  holds before BBN, the simplest hypothesis is to assume that the DM  was thermally produced along with the other SM particles.  Then, by considering a weak-like interaction strength of the DM with SM particles, the relic density can be easily accounted for. This is the so call Weakly Interacting Massive Particle (WIMP) miracle.  In the previous decades,
the WIMP mechanism scenario has received by far the biggest attention, both theoretically and experimentally. WIMPs typically carry
electroweak scale mass and couple to the SM with a strength that is
reminiscent to that of the weak interactions. Theories beyond the SM (BSM), such as supersymmetry, naturally provide excellent DM candidates (e.g. electroweakinos) that are among the most studied~\cite{Masiero:2004vk,Profumo:2004at, Bernal:2007uv,Sierra:2009zq,Bernal:2009jc}.%
%

\noindent The non-observation of WIMP-like DM, however, puts this mechanism under scrutiny, both at experimental and theoretical levels.
Moreover, as the WIMP paradigm becomes constrained, there is nowadays no single compelling theoretical framework for DM, but it rather opens a wealth of cosmological possibilities.
Despite our ignorance concerning the DM particle,
we have accumulated important information on the  basic requirements for a DM candidate: (i) it
should yield the correct relic density; (ii) it should be
non-relativistic at the epoch of matter-radiation equality, to form structures in the early Universe in agreement with  observations; (iii) it should be effectively neutral,  otherwise it would have formed unobserved stable charged particles; and (iv) it should be cosmologically stable, with a lifetime much larger than the age of the universe, and consistent with cosmic rays and gamma-rays
observations. Having these requirements in place,  we delineate four classes of DM candidates to be considered:

\begin{itemize}
\item \emph{WIMPs}: They  appear in several, well motivated model building setups. They are based on the thermal decoupling paradigm which is a key input in  observables such as the CMB and BBN.  Under
these assumptions, the correct relic density is obtained for an annihilation cross-section of the order of the electroweak
scale and DM particles with roughly weak scale ${\cal{O}}(100)\ \text{GeV}$ masses. 
Typically the parameters that govern the relic density are deeply intertwined with those in the direct and indirect detection signatures as well as with collider searches \cite{Ajaj:2019imk,Aprile:2019dbj,Aprile:2018dbl,Amole:2017dex,Abdallah:2016ygi,Ahnen:2016qkx,Akerib:2015rjg}.
Many WIMP models can reproduce the correct relic density and yield
signals that are within reach of current and upcoming experiments \cite{Arcadi:2017kky},  with  prospects  of  leading to a very rich experimental program.

\item \emph{Hidden Sector DM}, including Light DM:
As the WIMP paradigm is being challenged, models of Hidden Sector DM, that provide thermal relics  in the mass range between about a few keV to 100 TeV and may interact with laboratory detectors via new exotic, extremely weak interactions  are  under high scrutiny. In particular, the keV-GeV mass range, also referred as Light DM (LDM) demands new technologies, new detection strategies and more sophisticated analyses that typically fail above the GeV region.  In particular, in the LDM region the direct detection searches become powerful through the exploration of  DM-electron interactions. Moreover, in order to reproduce the observed DM relic density, models with LDM typically require the existence of light mediators that themselves can be searched for at laboratory experiments.

\item \emph{Ultralight DM} (ULDM), including Axions and Axion-like Particles (ALPS): It is possible to have  non-thermal ULDM candidates whose masses expand the range  $10^{-22}\ \text{eV}\lesssim m_\Phi\lesssim 1\ \text{eV}$, with extraordinary weak  couplings to SM fields. Such ULDM (lighter than about 0.7 keV) must be bosonic~\cite{Marsh:2015xka}, and the effects of the halo DM in terrestrial experiments are best described  as wavelike disturbances.  
Axions are well-motivated DM candidates originally thought of as an outcome of the Peccei-Quinn (PQ) solution of the strong CP problem~\cite{Peccei:1988ci,Kim:2008hd}. 
The axion is the pseudo-Nambu-Goldstone boson of the  $\operatorname{U}(1)_{\text{PQ}}$ symmetry~\cite{Weinberg:1977ma,Wilczek:1977pj}, 
which is spontaneously broken at an energy scale related to the axion mass and couplings.
The axion DM mass range is
$10^{-12} \text{eV} \lesssim m_a \lesssim 10^{-2}\text{eV}$.
More general ALPS - with a more flexible range of axion mass and couplings - are also possible, allowing for axions and ALPS in the sub-eV region to be stable at cosmological mass scales, and provide suitable DM candidates. Many BSM scenarios, some  inspired by string theory, contain ALPS and other sub-eV dark sector particles such as dark scalars  or dark photons  that can provide non-thermal DM candidates.  
A new set of table top experiments with novel technologies are being explored worldwide~\cite{Battaglieri:2017aum}.
 
\item \emph{Non-thermal or quasi-thermal DM}; There  is a broad class of DM scenarios with masses above the  ULDM range but that are also produced non-thermally. 
The DM masses can span from the keV to 100 TeV, with  interaction strengths varying  quite vastly. Examples of such models include the Strongly Interacting Massive Particle (SIMP), the asymmetric DM (ADM), the Feebly Interacting Massive Particle (FIMP), etc.
It is important to keep an open mind towards  non-thermal histories that can still fulfill the aforementioned requirements and therefore constitute viable DM candidates. 
 \end{itemize}

\section{Astrophysical and cosmological probes of DM}
 Dark Matter - if made of particles and compact objects - is the dominant  form of matter in the universe, and, hence, there are several experiments that try to measure its interactions with  detector materials, as well as  the results of annihilation from regions in which the density is expected to be sufficiently high. 
 There is a large array of experiments exploring the DM paradigm through terrestrial and extraterrestrial probes. In this section we shall only mention the ones with  active Latin American involvement.

\subsection{Direct detection}
Direct detection refers to the observation of nuclear recoils at low energy experiments located at deep underground laboratories. By measuring the energy recoil and the shape of the scattering rate of the events, one may reconstruct the DM-nucleon (or DM-electron) scattering cross-sections and infer the DM mass. Using different targets and readout techniques several experiments have been proposed. 

\noindent DarkSide~\cite{Agnes:2018oej} is one of the leading  experiments currently searching for   WIMP-like DM  and  LDM candidates, having excellent opportunities for
future term projections. The experiment is based on  Liquid Argon (LAr) technology and efficiently searches for DM using scintillation and ionization readout techniques~\cite{Agnes:2018fwg}. After a DM-Argon nucleus scattering, a prompt scintillation light is emitted (S1 signal), and an ionization process occurs at the same time. In the latter event, electrons are drifted towards the anode of the time projection chamber finding a gas phase of argon where they emit electroluminescent light (S2 signal)~\cite{Agnes:2018ves}. Both signal measurements are used in constraining the DM-nucleon scattering cross sections.
It is notable that, just by taking advantage of the high trigger efficiency of the S2 signal events for low energy recoils~\cite{Aalseth:2017fik} experimenters  constrain DM-electron scattering and produce competitive bounds in the 20 MeV-1 GeV DM mass window. DarkSide current results are based on 50 Kg of LAr, and the collaboration is now commissioning a 20 tonnes LAr detector.  Brazil is participating in this experiment through the group of Professor Ivone Albuquerque of University of Sao Paulo (USP), and Mexico  is contributing through Professor Eric Vázquez Jáuregui of Universidad Autonoma de México (UNAM).

\begin{figure}
  \centering
  \includegraphics[scale=0.6]{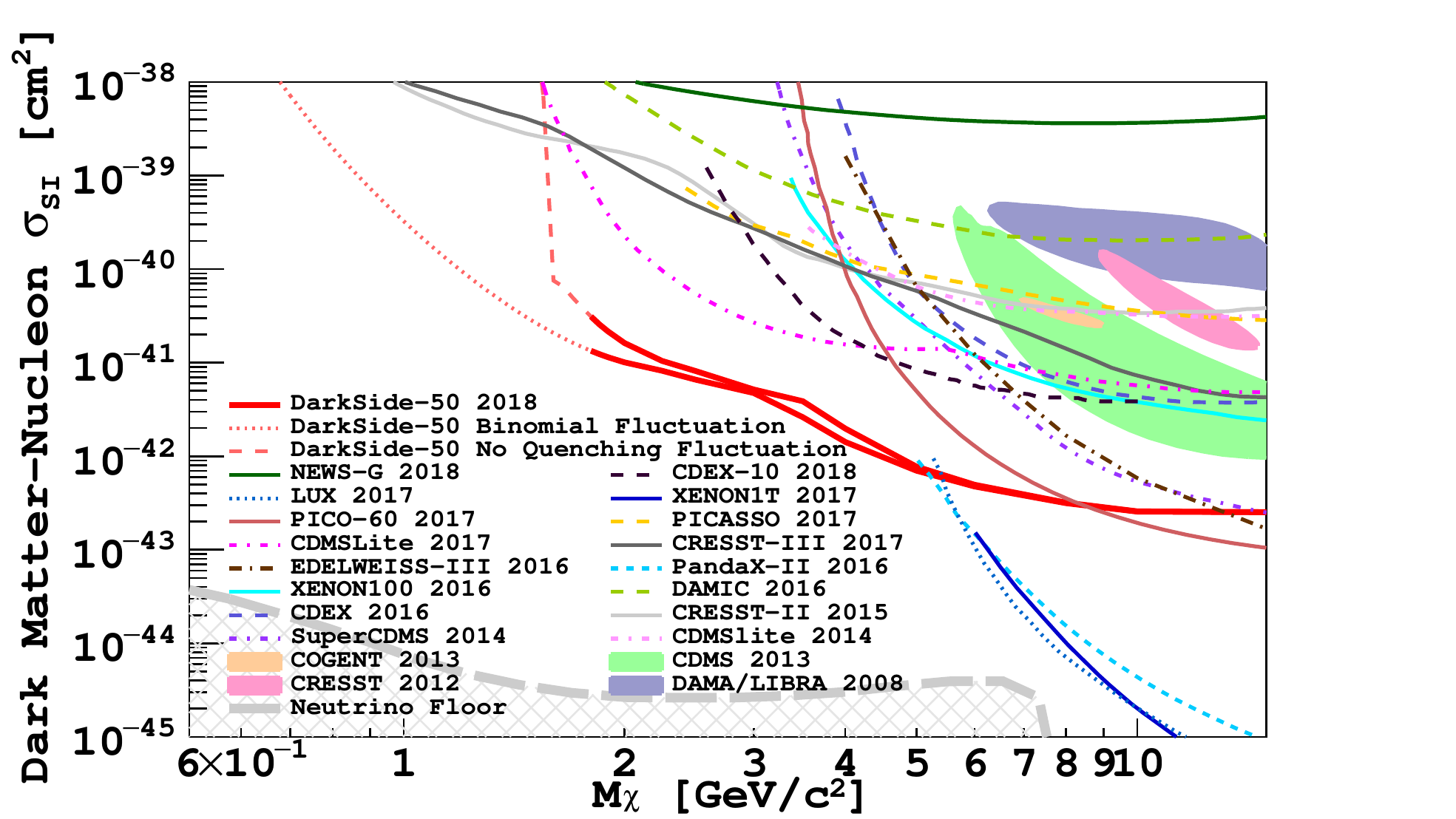}
  \caption{Upper limits on spin independent DMnucleon cross sections from DarkSide-50 in the range
above $1.8\ \text{GeV/c}^2$, from \cite{Agnes:2018ves}}.
  \label{fig:darksidelim}
\end{figure}

\begin{figure}[t]
  \centering
  \includegraphics[scale=0.35]{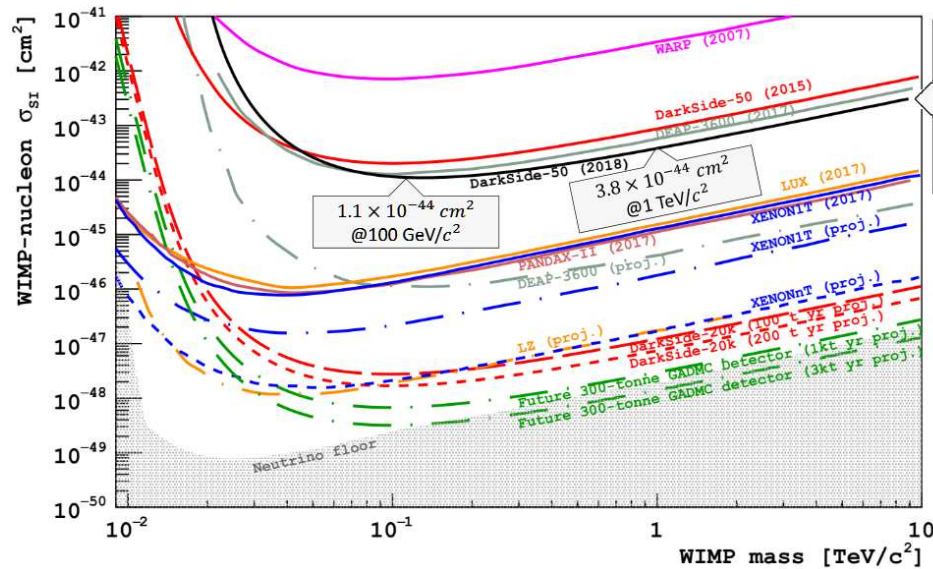}
  \caption{Spin-independent  DM-nucleon  cross  section 90 \% C.L. exclusion limits from the analysis detailed in \cite{1802.07198}   (in black), compared with selected results and projections for XenoNnT (dash-blue) LZ (dash-orange) DarkSide-20k (dash-red) and
  the Global Argon DM Collaboration (GADMC)  (dash-green), reaching the so-called neutrino floor~\cite{Billard:2013qya}. }
  \label{fig:darkside}
\end{figure}

\noindent The DarkSide experiment currently has the best limits for DM-nucleon cross-section in the low mass region,  DM masses in  the 2-5 GeV range, as illustrated in Fig.~\ref{fig:darksidelim}.
New phase DarkSide-20k will feature 20 tonnes of LAr with a future reach very close to the neutrino floor as illustrated in Fig.~\ref{fig:darkside}. While other experiments such as XENONnT and LZ will be highly competitive in  the region of 10 GeV masses and above,  DarkSide seems to have a leading edge in the low energy region of 1-10 GeV. Future experiments like the Global Argon DM Collaboration (GADMC) are designed to reach the neutrino floor over a broad range of DM masses in the 100 GeV to 100 TeV range.~\cite{Billard:2013qya}.

\noindent The LA region also has a very strong presence   in the 
SENSEI~\cite{Barak:2020fql} and future DAMIC-M~\cite{Castello-Mor:2020jhd} experiments, that have a unique opportunity to explore the light DM window through  DM-electron cross-section measurements.
The participation in SENSEI is through the Universidad de Buenos Aires (UBA), the Instituto de Fisica de Buenos Aires (IFIBA) and the Centro Atómico de Bariloche  (CAB) in Argentina, while DAMIC-M is  contributing  through CAB, Argentina. Both experiments use Skipper-CCD technology optimized to detect electronic recoils, 
a technology that has been developed with relevant involvement of Latin American scientists.  
This line of research is done in close collaboration with  Fermi National
Accelerator Laboratory (Fermilab) in the U.S.  Skipper-CCDs are silicon pixelated detectors with a spatial
resolution of $15\ \mu\text{m}$ and the ability to measure the deposited charge per pixel with a readout noise Root Mean Squared (RMS) of $0.1\;e$. This unprecedented low noise level permits to count individual electrons, which translates to the lowest possible threshold detection energy.  
Thanks to its low readout noise level, Skipper-CCD technology is particularly competitive to
test theories that predict LDM in the sub-GeV range. 

\begin{figure}[h]
  \centering
  \includegraphics[scale=0.35]{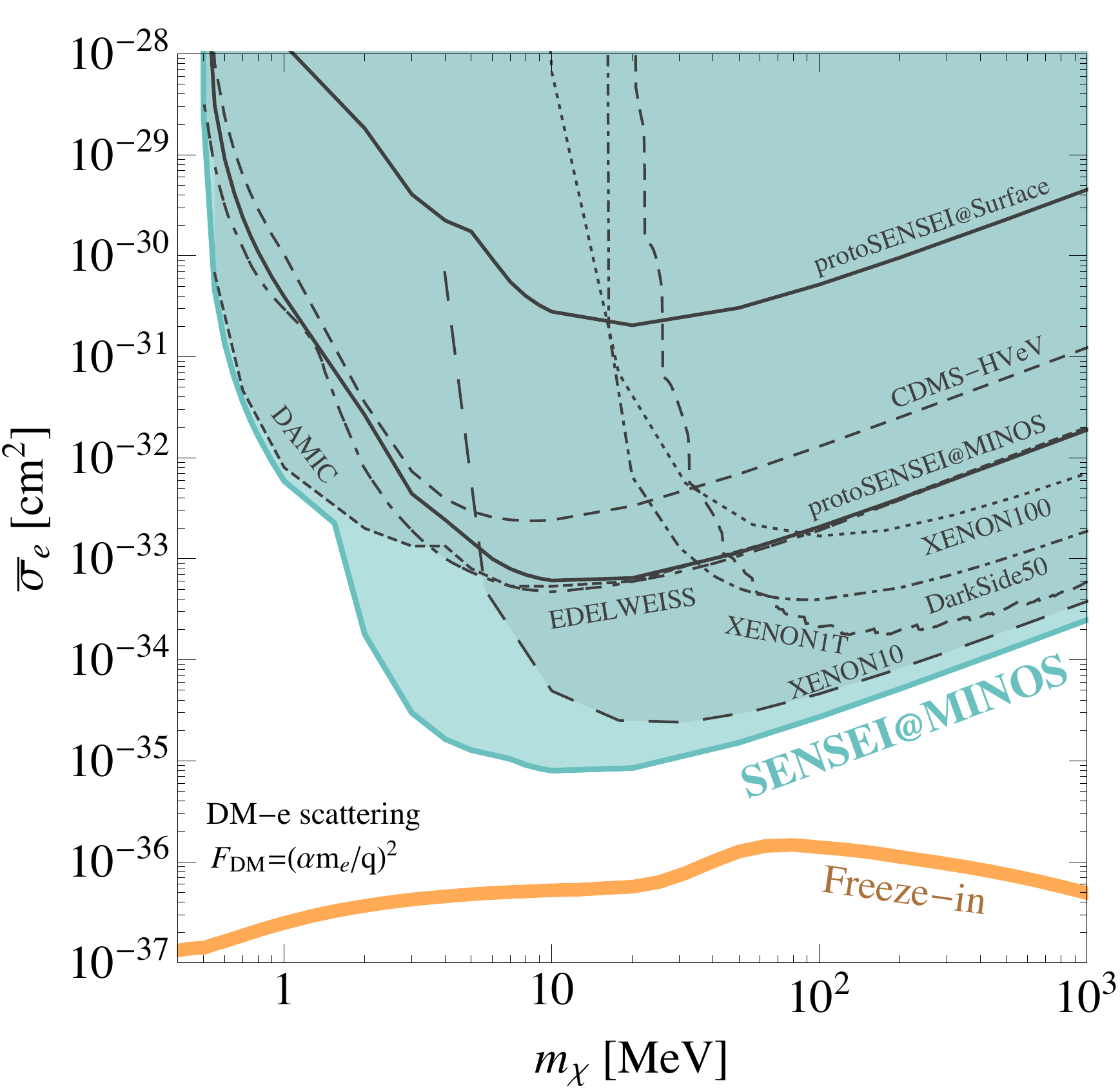} \includegraphics[scale=0.35]{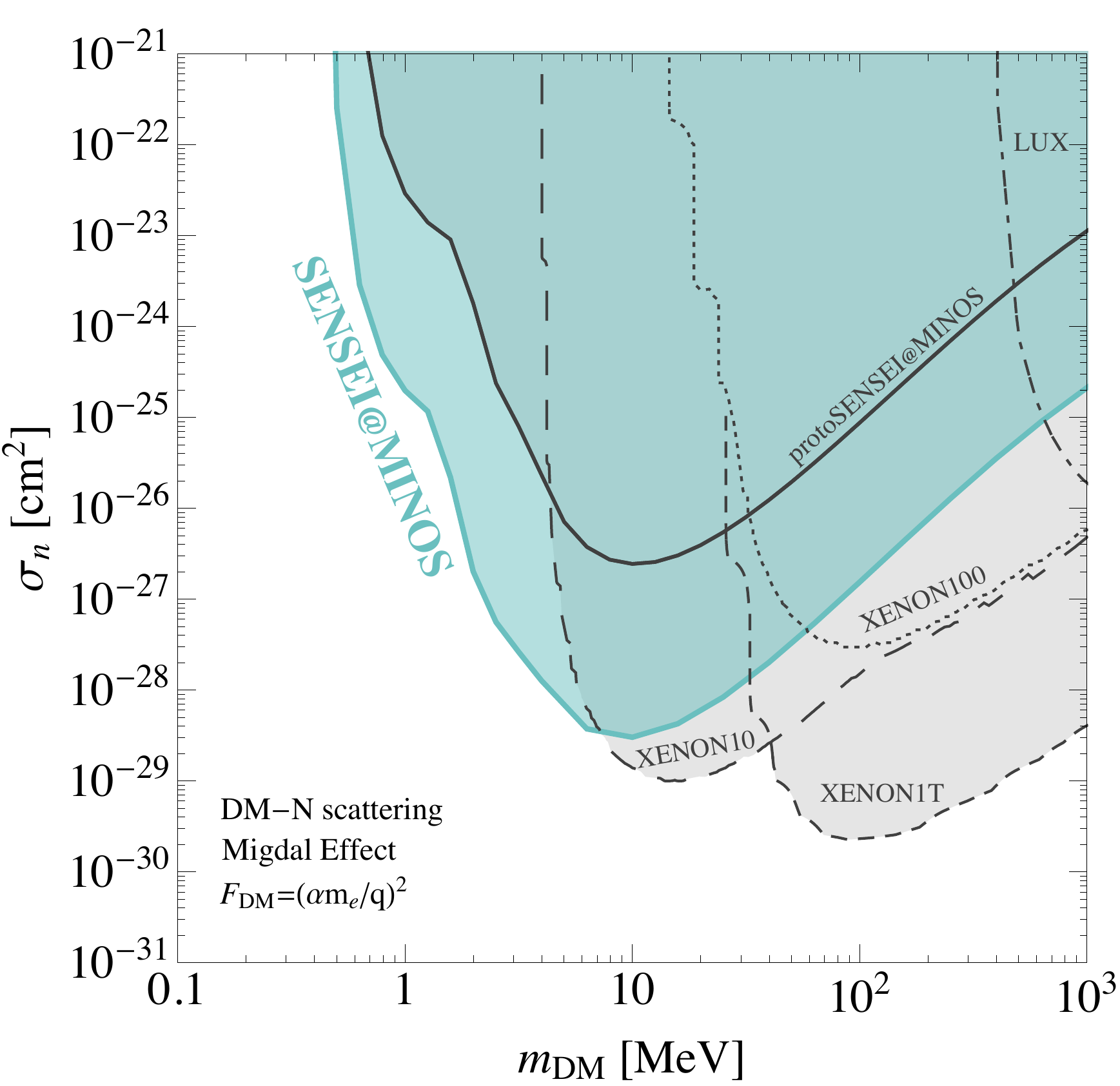}
  \caption{DM limits from SENSEI from~\cite{Barak:2020fql}, for DM-electron scattering cross-section and DM-nucleus scattering cross-section as a function of the DM mass. See text for details.}
  \label{fig:sensei}
\end{figure}

\noindent SENSEI has the current
best limit for LDM-electron scattering cross-section, $\bar{\sigma}_e$, with form factor $F_{\text{DM}}(q)=\left(\alpha m_{e} / q\right)^{2}$, which is illustrated in Fig.~\ref{fig:sensei}-left, and a competitive limit in LDM-nucleus scattering cross-section, $\bar{\sigma}_n$ for a light mediator, which is illustrated in Fig.~\ref{fig:sensei}-right. These are searches for direct
observation of DM in the low mass domain, from the MeV range up to tens of GeV.
In these type of experiments, special care is needed to shield the detector from ambient radiation, be it unstable isotopes or cosmic showers, requiring the
installation of the experiment in underground caverns. In this sense, SENSEI-100-gr-yr (next generation) will upgrade from its current location in the SOUDAN mine by moving to SNOLAB. A similar experiment,  DAMIC-M is planned in Modane in the near future. In addition, already in the plans is the  next generation of Skipper-CCDs detector based DM searches at the OSCURA experiment at SNOLAB, where two orders of magnitude gain in sensitivity with respect to SENSEI-100-gr-yr is expected. It is  of great relevance to note that the prospective, $1700\ \text{m}$ underground
ANDES laboratory,  to be constructed in parallel with the Agua Negra tunnel between  Argentina and Chile, will provide a unique auspicious  location for future generations of Skipper-CCD based DM experiments in the southern hemisphere.

\subsection{Indirect detection}

Indirect DM detection refers to the observation of stable particle fluxes such as electrons, protons, neutrinos and gamma-rays produced by DM annihilation or decay in dense astrophysical environments, such as Dwarf Spheroidal Galaxies and the galactic center.

\begin{figure}
  \centering
  \includegraphics[scale=0.38]{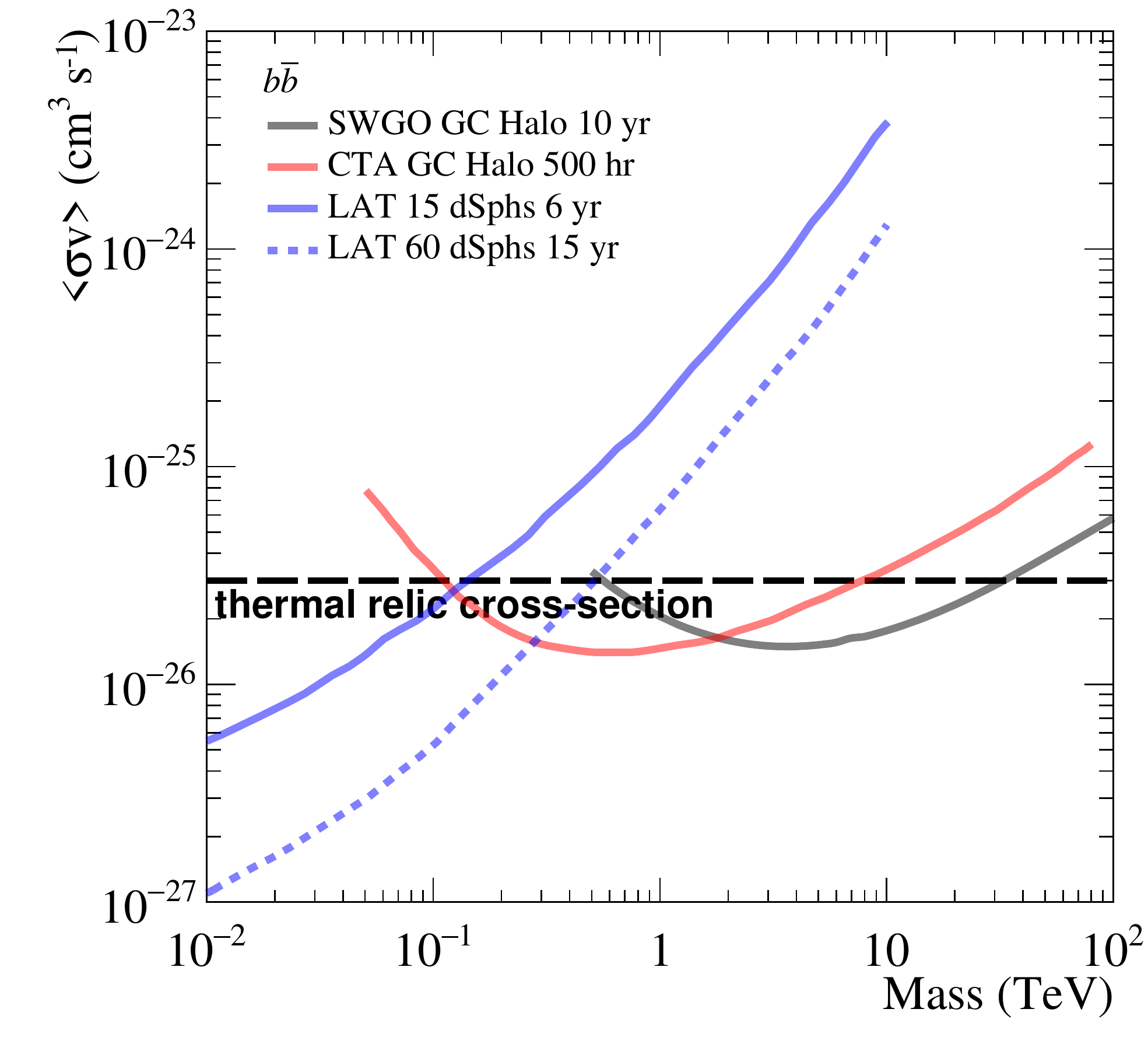}
  \caption{Expected 95\% C.L upper limit on the velocity-weighted cross section for DM self-anniliation into $b\bar{b}$ as a function  of DM for CTA and SWGO observation of the galactic center halo \cite{Viana:2019ucn}}
  \label{fig:ctareach}
\end{figure}

\noindent Chile and Brazil participate in the Cherenkov Telescope Array (CTA), while Argentina, Brazil, Mexico and Peru are part of the Southern Wide-field Gamma-ray Observatory (SWGO). Both experiments are described in the Astrophysics chapter of this report. 
In combination, both experiments will cover a wide  range of heavy DM masses annihilating to bottom quark pairs,  illustrated in Fig.~\ref{fig:ctareach}~\cite{Viana:2019ucn}.

\noindent Hence, both CTA and SWGO will give important information about models of DM above the 100 GeV mass range and in the absences of a signal will interestingly constrain  heavy WIMP scenarios. In particular, for the large mass solution of the Inert Doublet Model (IDM) with DM masses in the range 0.5-3~TeV,  one can conclusively say that CTA will exhaustively probe  the IDM at the TeV scale~\cite{Queiroz:2015utg}.

\section{DM production at colliders}
The nature of DM and its interactions can be probed at accelerator based experiments, complementing experiments and observations from astroparticle physics and cosmology.
The prospect of creating DM in the laboratory, either produced directly in beam-beam collisions with other SM particles or in decays of SM or yet to be discovered new ones is tantalizing. In all cases, the dark matter signal will consist of significant transverse missing energy plus high energetic SM objects, or complex SM final states in the case of cascade decays.
There is a significant number of articles related to DM models, that have been published with strong involvement and leadership of Latin American scientists working on analysis techniques, and strategy searches  at the LHC. For example, novel ideas and models to explore the production of anapole DM~\cite{Florez:2019tqr},   DM in supersymmetric scenarios~\cite{Bernal:2007uv,Das:2017mqw,Avila:2018sja},  models where lepto-quarks couple to dark matter~\cite{Alves:2018krf},  among others have been conducted. Additionally, new analysis techniques using vector boson fusion and initial state radiation jets~\cite{Florez:2016lwi}, have been proposed using simulated data. Several of these searches have lately  been investigated by the  ATLAS, CMS and LHCb  collaborations, all of them with participation from Latin American groups. We next illustrate the case for DM searches with the Compact Muon Solenoid (CMS) experiment.

\noindent The CMS  detector at the LHC has a broad physics program to search for physics beyond the SM, including DM. The effort of several  experimental groups of the Latin American community, in particular scientists at the Universidad de los Andes and Universidad de Antioquia, in Colombia,  are working on DM searches using supersymmetric models and compressed mass spectra scenarios with tau leptons. The searches are carried out using two different techniques,  vector boson fusion~\cite{Khachatryan:2015kxa, Khachatryan:2016mbu,Sirunyan:2019zfq} and jets from initial state radiation~\cite{Sirunyan:2019mlu}.  These efforts have led to many publications that have facilitated the involvement of Latin American students and young scientists in the searches for DM and dark sectors at the LHC.

\section{DM portals}
As the WIMP paradigm becomes experimentally scrutinized  with no evident positive results, extended scenarios with Hidden Sector DM and DM mediators (additional dark sector particles)  become of interest. Dark matter mediators from the dark sector can then interact with SM or even BSM portals, including scalar/pseudoscalar, vector and fermion portals.  In particular, it is promising  to search for DM mediators decaying to SM particles or into the DM particles  in a broad array of possibilities, from accelerator based experiments at colliders and  fixed target probes (see discussion of DM/Dark Sector chapter in Ref~\cite{Strategy:2019vxc} ) as well as through their possible  interactions in reactor neutrinos experiments.  

\begin{figure}
  \centering
  \includegraphics[scale=0.24]{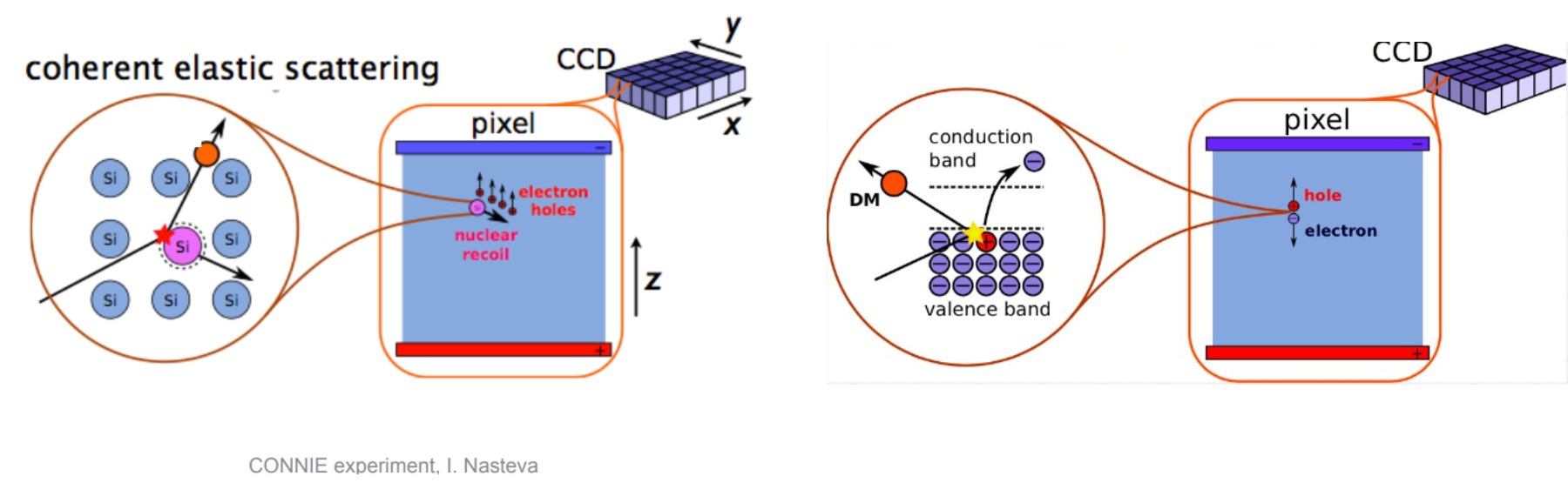}
  \caption{CCD detectors in reactor and direct detection experiments.}
  \label{fig:skipper}
\end{figure}

\noindent The particle detection technology employed in SENSEI is  directly
applicable to the study of neutrino physics exploiting coherent neutrino-nucleus interactions,  recently observed by the COHERENT Collaboration~\cite{Akimov:2017ade}, as illustrated in Fig. \ref{fig:skipper}.
The Skipper-CCD detectors may  prove to be  sensitive to the low energy threshold of the coherent neutrino-nucleus interaction and can be used to measure the high intensity low energy neutrinos escaping from a nuclear reactor. By placing these
detectors at different distances close to the reactor core one could search for new particles, such as $Z'$s  ( that appear in some DM models),  that could contribute to the coherent neutrino-nucleus interactions.

\begin{figure}
    \centering
    \includegraphics[scale=0.22]{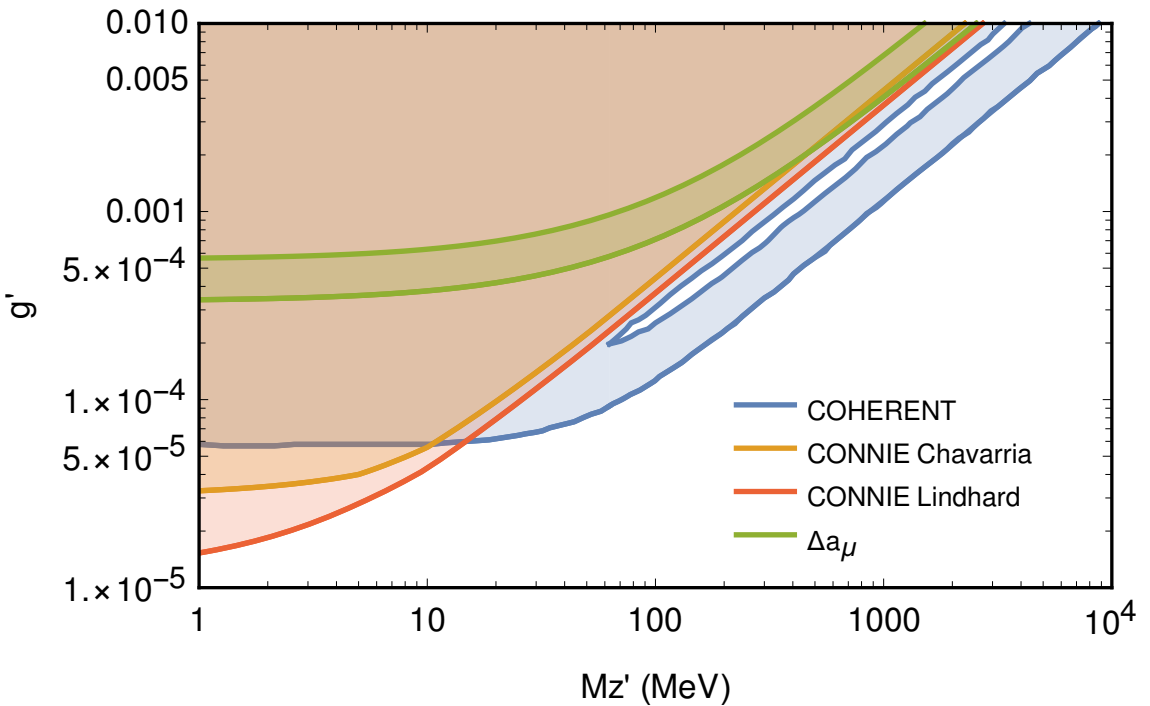}
    \caption{Exclusion region in the $(M_{Z'} , g')$ plane from  CONNIE results in ~\cite{Aguilar-Arevalo:2019zme}.}
    \label{fig:connie}
\end{figure}

\noindent The current Coherent Neutrino-Nucleus Interaction
Experiment (CONNIE), plans to upgrade to Skipper-CCD detectors as a step in the  search for coherent elastic scattering processes mediated by BSM Z' interactions; see first  limits  for such a BSM portal in the current CONNIE experiment
in Fig.~\ref{fig:connie}~\cite{Aguilar-Arevalo:2019zme}. 
Similarly, $\nu$IOLETA\footnote{See \url{https://www.violetaexperiment.com/}} is a
future experiment with very similar goals as those discussed for CONNIE, based on Skipper-CCD technology  and  with the advantage of more flexibility towards its location vis a vis the neutrino reactor core.
Two positions were identified to place $\nu$IOLETA within the Atucha II Nuclear Power Plant in Argentina, at 8 and 12 meters away from
the center of the reactor core.

\noindent The Latin American region is building the Lambda laboratory \footnote{
\url{http://lambda.df.uba.ar/portfolio/}}
 at the Physics Department at UBA, that is set to exploit  Latin America's privileged position in  Skipper-CCD technology. This specialized lab will permit   to  develop experimental collaborations for forefront  DM
searches and neutrino physics experiments.
By the end of 2020 Lambda lab expects to have a first working setup, including
a $50\ \text{m}^2$ laboratory with turbo-molecular vacuum stations, cryo-refrigerators, low-radioactivity vacuum chambers and
 ultra low noise readout electronics for CCDs.
The Lambda lab is working in close collaboration with the Skipper-CCD group at Fermilab.

\noindent There is a large participation of the Latin American region in the neutrino experimental program at Fermilab, including the  future flagship experiment: the Deep Underground Neutrino Experiment (DUNE). This includes leading institutions in Brazil, Colombia, Mexico, Paraguay, and Peru.
The DUNE experiment is based on liquid argon detectors of mega-proportions exploiting the most powerful neutrino beam ever built.  Among its many capabilities it has the  possibility to  produce and then detect relativistic LDM particles that can pass through the shielding material, enter the detector and scatter off the atoms, whose relativistic recoils are detectable~\cite{2008.12769}.
Due to highly intensified beam
sources, large signal statistics is usually expected so
that this sort of search strategy can allow for significant sensitivity to DM-induced signals, despite the feeble interaction of DM- SM particles. Assuming that DM is produced, DUNE will be capable to  search for the relativistic scattering of LDM  at the Near Detector, as it is close enough to the beam source to sample a substantial level of DM flux. The current strong involvement of the Latin American region in DUNE calls for successful participation on the DUNE DM search program.

\section{DM Phenomenology community in LA}
 The Latin American DM community has a successful and long history of involvement in DM model building and DM phenomenology that started about three decades ago.
 The community is ripe to move to the next step of strongly supporting new developments of experimental groups within and beyond the currents efforts.
 
Latin American scientists have played a relevant role in both formulating and studying simplified WIMP-like DM
scenarios~\cite{Masiero:2004vk,Profumo:2004at, Bernal:2007uv,Sierra:2009zq,Bernal:2009jc,Yaguna:2008hd,Goudelis:2009zz,Honorez:2010re,LopezHonorez:2010tb,Alvares:2012qv,Esch:2013rta,Martinez:2014ova,Kelso:2014qka,Martinez:2014rea,Martinez:2015wrp,Arbelaez:2015ila,Horiuchi:2016tqw,Dutta:2017lny,Arbelaez:2017ptu,Filippi:2005mt},
\cite{vonMarttens:2019ixw,Davari:2019tni,Pordeus-da-Silva:2019bak,Bhattacharya:2019ucd,vonMarttens:2018iav,Rodrigues:2018duc,Montero:2017yvy,Hipolito-Ricaldi:2017kar,Marttens:2017njo,Rodrigues:2017vto,deAlmeida:2016chs,Casarini:2016ysv,Pigozzo:2015swa,Piattella:2015nda,Velten:2014xca,Rodrigues:2014xka,Bernal:2014mmt,Esmaili:2013gha,Esmaili:2010wa,Mizukoshi:2010ky,Esmaili:2009ks},
\cite{Penacchioni:2020xhg,Izaurieta:2020xpk,Gomez-Vargas:2019vci,AristizabalSierra:2019ykk,Arias:2019uol,Leite:2020wjl,Hernandez-Arellano:2019qgd,Bonilla:2019ipe,Faber:2019mti,Cabrera:2019gaq}. At the same time, in many theoretical studies across high energy physics,
astrophysics and cosmology  Latin American communities have been dedicated to  explore  DM solutions based on axions, ALPs , dark photon particles~\cite{Ringwald:2016yge,Borsanyi:2016ksw,Alves:2016bib,Graham:2015ouw,Dias:2014osa}, and more general ULDM~\cite{Blas:2016ddr,LopezNacir:2018epg,Blas:2019hxz,Armaleo:2019gil,Castellanos:2019ttq}, including a recent review~\cite{Ferreira:2020fam}.

\noindent Beyond the various non-thermal scenarios related to ULDM, other intriguing  non-thermal and quasi-thermal DM scenarios for DM masses above the keV range have been extensively developed by the LA community.   A challenging scenario, in the hope to find non-gravitational signals
for dark matter, is expected for Feebly Interacting Massive Dark Matter Particles
(FIMPs) in which the relic density is generated out of equilibrium by the
so-called freeze-in mechanism. In this scenario, DM particles
couple to the visible SM sector extremely  weakly, so that they never entered
chemical equilibrium. Instead, the DM particles were produced by decay
or annihilation processes from the visible sector. 
The LA community has been particularly active in the exploration of FIMPs,
pioneering studies on simplified models where the FIMP mechanism is
realized~\cite{Yaguna:2011qn,Molinaro:2014lfa,Bernal:2018kcw} (see Ref.~\cite{Choi:2010jt,Restrepo:2011rj} for
supersymmetric scenarios with gravitinos non-thermally produced).
Additionally, the Latin American theory community has contributed to the  understanding of the impact of  non-sudden reheating on the UV
freeze-in~\cite{Bernal:2020qyu}.  The only existing
review on FIMP DM has a major contribution from Latin American scientsits ~\cite{Bernal:2017kxu}.

\noindent Another non-thermal cosmological possibility is the Strongly Interacting Massive Particles  (SIMP) mechanism.
DM with strong self-interactions provides an appealing  solution to several small-scale structure puzzles.
Under the assumption that the coupling between DM and the SM particles is suppressed, SIMPs allow for a successful dark freeze-out through $N$-to-$N'$ processes, where $N$ DM particles annihilate to $N'$ of them, with $N>N'\geq 2$.  The Latin American community has been actively exploring  this alternative production mechanism, whereas searching  for strategies to probe it, in particular through astrophysical and cosmological observables~\cite{Bernal:2020gzm}.
A review on the different production regimes for SIMPs and general self-interacting DM has been produced by LA scientists~\cite{Bernal:2015ova}.

\noindent Many generations of Latin American scientists has been trained in DM phenomenology and model building and the community is well prepared to continue contributing revolutionary ideas to the field, helping to build a stronger bridge between theory, cosmological observations and experiments.

\subsection*{DM Community Activities}

Many activities, including schools and workshops have helped solidify interactions among  the Latin American DM community and bring international experts to help
train  the younger generations. Notable examples include the ICTP-SAIFR  regular workshops and meetings related to DM with high Latin American involvement  and significant international participation. In addition, 
there is also the yearly Colombian workshop on DM (MOCa). Both initiatives  arose in response  to the community's desire to further strengthen  efforts in DM related topics.  In several Latin American countries like Argentina, Brazil, Chile, Colombia and Mexico there is  a highly motivated critical mass of researchers working on different subjects of particle physics, cosmology, and astrophysics and  future DM endeavours can capitalize on them.

\section{Synergies}
The Dark Matter conundrum is at the boundary of  particle physics, astro-particle physics and cosmology, and as such, interdisciplinary collaboration and cross-fertilization is needed.

\noindent The understanding of DM will demand signals from multiple experiments, together with compatibility with indirect measurements from DM annihilation and decay products. The general purpose Gamma Ray Observatories which aim to improve the
existing catalogues of Gamma-ray sources in the sky have also the potential to discover DM and help  understand its properties.

\noindent There is an important overlap both in technology and search opportunities between DM and neutrinos. From the experimental side the liquid argon detectors and related technologies are playing a crucial role both in DM searches, e.g. DarkSide, and the neutrino detectors, including the mega-project DUNE. Furthermore, the Skipper CDD technology is powerful in detecting low thresholds signals that are relevant for DM direct detection as well as for exploring neutrino properties.  

\noindent The recent experimental success in observing gravitational waves has opened up a new avenue for observational cosmology, which will be further enhanced by the future LISA mission. Gravitational waves produced in  phase transitions, possibly involving dark sectors,  could be detected with the LISA  observatory, regardless of the feeble strength  of non-gravitational interactions between the dark and visible sector particles~\cite{Almeida:2018oid,Bernal:2020qyu}.

\noindent Baryon acoustic oscillations (BAO) are fluctuations in the density of the visible baryonic matter of the universe, caused by acoustic density waves in the primordial plasma of the early universe. BAO is one of the main evidences for DM and therefore its exploration has direct implications for dark sectors~\cite{1603.08299}. 
\noindent A possible experiment along this line - based on a Radio Telescope and leaded by Brazil - is BINGO  (BAO from Integrated Neutral Gas Observations). It is  designed to measure BAO at radio frequencies and its description and scientific scope can be found in the Cosmology chapter of this report. 
 
\noindent The future  Legacy Survey of Space and Time (LSST) will complement direct DM detection experiments by improving measurements of the local phase-space density of DM using precision astrometry of the Milky Way stars. For dark matter-baryon scattering, small-scale structure measurements with LSST can probe
DM masses and cross sections outside the range accessible to direct detection experiments. For details see the Dark Matter white paper from LSST at \url{https://lsstdarkmatter.github.io/}

\section{Conclusions}
Astrophysical probes of DM require large underground facilities to have competitive limits in the exploration of  various types of DM candidates. At present, the experimental activities in which the Latin American community is involved are  based at  European, US and Canadian underground laboratories. Researchers in Latin America are active in the DarkSide experiment, that has provided the best limits in the GeV region for  DM-nucleon cross-sections, and  has future plans to test DM candidates with annihilation cross sections close to the neutrino floor.
The expertise and capabilities contributing to large direct DM search experiments could be further exploited by encouraging Latin American scientists to actively engage in the upcoming Global Argon DM Collaboration (GADMC), that is the long term future for liquid argon based WIMP-like DM searches.
In the future there is the unique opportunity for an underground laboratory in the region: The  prospective ANDES Agua Negra tunnel between Chile and Argentina. Such underground laboratory would be the only one of its kind in the southern hemisphere, with extraordinary  capabilities  for direct detection DM searches, as well as neutrino physics exploration, and will make the region a central international hub for underground experiments.

\noindent Other type of searches for DM candidates in the sub-GeV regime require very low energy threshold detectors. A new technology based on  Skipper-CCD detectors can identify electron recoils at the level of one to  few electrons from elastic scattering of MeV to  GeV DM candidates. The SENSEI experiment at MINOS has currently set the stringent limits for light DM-electron scattering cross-sections in this mass range and it can also probe  nuclear recoils successfully. In the near future, Latin American participation in DAMIC-M at Modane and  SENSEI-100-gr-yr at  SNOLAB will reach DM-electron scattering cross sections  in the $10^{-40}$ to $10^{-41}$  $\text{cm}^{2}$ sensitivity. It will be of interest to the LA community to capitalize on this experience and actively participate in the next generation of Skipper-CCDs detector based DM searches  at the OSCURA experiment at SNOLAB, where two orders of magnitude gain in sensitivity for the DM-electron scattering cross sections are expected.

\noindent The Latin American community has acquired a strong expertise in  Skipper-CCDs technology, and already plays a relevant role in a group of experiments based on it.  Skipper-CCDs  detectors are relevant not only for light DM searches, but also for exploring dark portals at detectors in the vicinity of nuclear reactor cores,  such as the current, terrain-level  experiment  CONNIE in Brazil and the near future $\nu$IOLETA  experiment Argentina. A new laboratory at UBA, Argentina is taking on the excellent opportunity to lead and train the next generation of Latin American experimenters in Skipper-CCDs technologies that are of use well beyond the DM efforts.

\noindent Latin America is playing a unique role in the next generation of gamma ray observatories, such as CTA South and SWGO, which offer excellent opportunities to perform DM experiments. This opens a remarkable possibility  to the region of playing a leading role in indirect DM detection experiments.

\noindent The  international particle physics  community has been dedicated to search for DM candidates created in collisions at high energy collider experiments since the LEP era. At the same time,  Latin American scientists have played an important role  in   Standard Model and beyond the standard model physics at colliders, including LEP, the Tevatron and now the LHC, and its upcoming High Luminosity (HL) upgrade. This opens an opportunity for the Latin American community to now engage in DM/dark sector collider searches. DM particles  and long lived and/or feeble interacting particles, that may be related to extended DM sectors, may open the path to discoveries. An example of Latin American efforts  is the CMS group in Colombia that has lead DM searches in various channels. The expertise and long term involvement  in collider experiments in the region makes it possible  for a broad group of young researchers  to  participate/lead DM searches and new analyses strategies  at the HL-LHC run.

\noindent In addition to participating in large international projects such as the LHC, the Latin American community is also getting involved in the next neutrino mega-science project hosted by Fermilab: the Deep Underground Neutrino Experiment (DUNE). Although DUNE is an experiment primarily designed to  explore neutrino properties, and  its possible new interactions, it also offers a unique window into dark matter  exploration. 

\noindent Latin America has a large, mature theoretical community with vast expertise in DM model building and phenomenology, with active collaborations across the region. The theoretical community also leads the organization of schools and workshops that facilitate the training of the next generation. Such a strong theoretical effort will naturally evolve into an even  deeper involvement in experimental activities ranging from direct and indirect detection experiments to specific accelerator based  searches and new analysis techniques. The Latin American community is ripe to move to the next step in the exploration of DM and dark sector signals, and the theoretical community can bridge efforts,  helping to prepare the next  generation of DM/Dark sector experimentalists.

\input{./DM/DM.bbl}

\chapter{Neutrinos}\label{chapt:neutrinos}
\vspace{1cm}
\noindent Alfredo Aranda (U. de Colima, Mexico)\\ Jorge Molina (U. Nacional de Asunci\'on, Paraguay)\\ Harold Yepes Ramírez (YTU, Ecuador)






\section{Introduction
}\label{sec01}

\noindent Neutrinos are key to solving open questions in our understanding of matter and the subatomic world. Latin America's participation in international neutrino experiments  is relatively new. Nevertheless, it is at the core of  the efforts in Latin America to participate in future local and non-local research infrastructures for global neutrino data analysis networks, through upgrades and scaling of already existing infrastructures or the development of new ones.\\
The neutrino puzzle is still not fully solved. 
Current main concerns regarding neutrinos may fall into two categories: neutrino properties and the role of neutrino as  probes of astrophysical processes. In this respect,  active topics address fundamental questions, such as precision measurements of neutrino oscillation parameters (mixing angles), the leptonic CP-violation phase ($\delta_{CP}$) of the mixing matrix, the ordering of the neutrino mass states (Neutrino Mass Hierarchy - NMH),  their Dirac or Majorana  nature, the absolute neutrino mass scale, identification of possible astrophysical sources, and new neutrino states (such as a sterile neutrino). In particular, the main unknown parameters being pursued are: to resolve the $\theta_{23}$ octant degenracy, the NMH and $\delta_{CP}$. Notwithstanding the concerns still to be solved about neutrinos, there are potential applications of neutrino related-technology with some prototypes already running.

\subsection{Neutrino oscillations, mass hierarchy and leptonic phase}\label{sec:1a}

\noindent Neutrino oscillations are described by the Pontecorvo–Maki–Nakagawa–Sakata (PMNS) mixing matrix, which relates the flavour eigenstates ($\nu_{e}$, $\nu_{\mu}$, $\nu_{\tau}$) to mass eigenstates ($\nu_{1}$, $\nu_{2}$, $\nu_{3}$)  \cite{ref27}, as shown in Equation (\ref{eq01}) for the Dirac neutrino case:

\begin{eqnarray}\label{eq01}
\begin{bmatrix*}[c]
    \nu_{e}  \\
    \nu_{\mu} \\
    \nu_{\tau}  
\end{bmatrix*}
= 
\begin{bmatrix*}[c]
    U_{e1} & U_{e2} & U_{e3} \\
    U_{\mu1} & U_{\mu2} & U_{\mu3} \\
    U_{\tau1} & U_{\tau2} & U_{\tau3}    
\end{bmatrix*}
\begin{bmatrix*}[c]
    \nu_{1}  \\
    \nu_{2} \\
    \nu_{3}  
\end{bmatrix*}.
\end{eqnarray}
\\
The simplest approximation to the oscillation probability between two neutrino states (2$\nu$), having a neutrino state $\nu_{\alpha}$ being detected as $\nu_{\beta}$ is:

\begin{equation}\label{eq02}
P_{2\nu}(\nu_{\alpha} \rightarrow \nu_{\beta}) = \sin^{2}\left(2\theta\right)\sin^{2}\left(\frac{\Delta m_{ij}^{2}L}{4E_{\nu}}\right),
\end{equation}
\\
with $\Delta m_{ij}^{2} = m_{i}^{2}-m_{j}^{2}$ ($m_{j}$ is the mass of the \textit{j}-th neutrino mass eigenstate). Equation (\ref{eq02}) is able to explain almost all the data available on neutrino oscillation parameters (except Short Baseline (SBL) anomalies) even in a 3-flavour neutrino (3$\nu$) scenario \cite{ref28}. The PMNS matrix can be parametrized  as the product of three independent rotations, through three mixing angles ($\theta_{23}$, $\theta_{13}$, $\theta_{12}$), where the second (unitary) rotation depends on a phase ($\delta_{CP}$), and of a diagonal matrix of phases (\textit{P}):

\begin{eqnarray}\label{eq03}
\newcommand\tab[1][1cm]{\hspace*{#1}}
\begin{aligned}
 U_{PMNS} & = 
\begin{bmatrix}
    1 & 0 & 0 \\
    0 & \cos\theta_{23} & \sin\theta_{23} \\
    0 & -\sin\theta_{23} & \cos\theta_{23}    
\end{bmatrix}   
\begin{bmatrix}
    \cos\theta_{13} & 0 & e^{i\delta_{CP}}\sin\theta_{13} \\
    0 & 1 & 0 \\
    -e^{i\delta_{CP}}\sin\theta_{13} & 0 & \cos\theta_{13}
\end{bmatrix} 
\begin{bmatrix}
    \cos\theta_{12} & \sin\theta_{12} & 0 \\
    -\sin\theta_{12} & \cos\theta_{12} & 0 \\
    0 & 0 & 1
\end{bmatrix}P  \\ & =
\tab[1.4cm]
\begin{bmatrix}
 c_{12}c_{13} & s_{12}c_{13} & e^{i\delta}s_{13} \\
 -s_{12}c_{23}-e^{i\delta}s_{13}c_{12}s_{23} & c_{12}c_{23}-e^{i\delta}s_{13}s_{12}s_{23} & s_{23}c_{13} \\
  s_{12}s_{23}-e^{i\delta}s_{13}c_{12}c_{23} & -c_{12}s_{23}-e^{i\delta}s_{13}s_{12}c_{23} & c_{23}c_{13}
\end{bmatrix}P 
\end{aligned}
\end{eqnarray}
\\
where the first, second and third matrices represent the accessible regions to atmospheric, reactor and solar neutrino fluxes, respectively, and \textit{P} is a unitary matrix for the Dirac neutrino case or a diagonal matrix (depending on two phases) for the Majorana neutrino case \cite{ref29}. The three mixing angles have all been measured and current efforts are focused on precision measurements. In contrast $\delta_{CP}$ is still unknown. In addition to these parameters, there are two independent mass-squared differences $\Delta m^{2}$ scales present in the puzzle ($\Delta m^{2}_{21}$, $\Delta m^{2}_{32}$). The remaining mass difference is given by: 
\begin{equation}\label{eq04}
\Delta m^{2}_{31} = \Delta m^{2}_{32} + \Delta m^{2}_{21} 
\end{equation}
\\
\noindent 
One usually defines two possibilities for the ordering of the neutrino masses: a ``Normal Hierarchy" (NH) with $m_3>m_2>m_1$ and an "Inverted Hierarchy" with 
$m_2>m_1>m_3$.
In the latter case,  $\Delta m^{2}_{31}$ and $\Delta m^{2}_{32}$ are both negative, and can be precisely measured in reactor experiments \cite{ref30}, and represent the mass difference between the heaviest and lightest neutrino or the opposite. The 3$\nu$ oscillation parameters from fits based on most recent data available are summarized in Table \ref{tab0a} \cite{refosc1}\cite{refosc2}.

\noindent

\begin{table*}[htp]
\caption{The 3$\nu$ scenario fit to global data (July 2020) obtained with the inclusion of the tabulated $\chi^{2}$ data on atmospheric neutrinos provided by the Super-K Collaboration. The best fit point (bfp) is shown as the central value.}\label{tab0a}
\centering
\begin{adjustbox}{width=\textwidth}
\begin{tabular}{ccccc}
\hline\noalign{\smallskip}
Parameter & bfp $\pm$ 1$\sigma$ (NH) & 3$\sigma$ range (NH) & bfp $\pm$ 1$\sigma$ (IH) & 3$\sigma$ range (IH) \\
\noalign{\smallskip}\hline\noalign{\smallskip}
$\sin^{2}\theta_{12}$ & $0.304^{+0.012}_{-0.012}$ & 0.269 $\rightarrow$ 0.343 & $0.304^{+0.013}_{-0.012}$ & 0.269 $\rightarrow$ 0.343 \\
$\theta_{12}/^{\circ}$ & $33.44^{+0.77}_{-0.74}$ & 31.27 $\rightarrow$ 35.86 & $33.45^{+0.78}_{-0.75}$ & 31.27 $\rightarrow$ 35.87 \\
$\sin^{2}\theta_{23}$ & $0.573^{+0.016}_{-0.020}$ & 0.415 $\rightarrow$ 0.616 & $0.575^{+0.016}_{-0.019}$ & 0.419 $\rightarrow$ 0.617 \\
$\theta_{23}/^{\circ}$ & $49.2^{+0.9}_{-1.2}$ & 40.1 $\rightarrow$ 51.7 & $49.3^{+0.9}_{-1.1}$ & 40.3 $\rightarrow$ 51.8 \\
$\sin^{2}\theta_{13}$ & $0.02219^{+0.00062}_{-0.00063}$ & 0.02032 $\rightarrow$ 0.02410 & $0.02238^{+0.00063}_{-0.00062}$ & 0.02052 $\rightarrow$ 0.02428 \\
$\theta_{13}/^{\circ}$ & $8.57^{+0.12}_{-0.12}$ & 8.20 $\rightarrow$ 8.93 & $8.60^{+0.12}_{-0.12}$ & 8.24 $\rightarrow$ 8.96 \\
$\delta_{CP}/^{\circ}$ & $197^{+27}_{-24}$ & 120 $\rightarrow$ 369 & $282^{+26}_{-30}$ & 193 $\rightarrow$ 352 \\
$\frac{\Delta m_{21}^{2}}{10^{-5}eV^{2}}$ & $7.42^{+0.21}_{-0.20}$ & 6.82 $\rightarrow$ 8.04 & $7.42^{+0.21}_{-0.20}$ & 6.82 $\rightarrow$ 8.04 \\
$\frac{\Delta m_{3l}^{2}}{10^{-3}eV^{2}}$ & $+2.517^{+0.026}_{-0.028}$ & +2.435 $\rightarrow$ +2.598 & $-2.498^{+0.028}_{-0.028}$ & -2.581 $\rightarrow$ -2.414 \\
\noalign{\smallskip}\hline
\end{tabular}
\end{adjustbox}
\end{table*}

\noindent As shown, the best fit is obtained for NH data ($\Delta m_{31}^{2} > 0$), while IH data is disfavoured with a difference of $\Delta\chi^{2} = 7.1$ between both hypothesis, relative to the corresponding local minimum. Additionally, it is possible to observe $\Delta m_{32}^{2} , \Delta m_{31}^{2} > 0$ for NH and $\Delta m_{31}^{2} , \Delta m_{32}^{2} < 0$ for IH. \\
Determination of $\delta_{CP}$ is being searched through enormous efforts of future upgrades to artificial neutrino sources in the main accelerator facilities worldwide at Fermilab, JPARC and Protvino. Outstanding synergies are taking place with leading expectations from the analysis of data collected by experiments such as MINOS, NO$\nu$A and T2K \cite{refosc1}. Nonetheless, very competitive sensitivity is also estimated by the next generation of experiments such as DUNE \cite{ref55} and KM3NeT-ORCA (the P2O side project \cite{refdelkm}), with planned upgrades on the beam at different power and exposures, in turn being suited for precision on oscillation parameters with NH. P2O, with a sufficiently long beam exposure ($\sim$ 4 year at 450 kW), can  reach a 2$\sigma$ sensitivity to $\delta_{CP}$, comparable with the projected sensitivity of NO$\nu$A and T2K. Recently, T2K presented results that exclude the complete absence of CP violation at 95$\%$ CL \cite{T2KCP20}. An advanced phase of P2O would give a 6$\sigma$ sensitivity to $\delta_{CP}$ after 10 years of operation at 450 kW, competitive with the projected sensitivity of also next generation experiments DUNE and T2K/T2HK. \\
Table~\ref{tab0b} presents a list of earth-based neutrino factories and detectors with current and possible future Latin American participation,  where (-) stands for no detector planned yet.

\begin{table*}[htp]
\caption{Neutrino sources and detectors with Latin American participation. The tags stands for: $^{\S}$unused (but of interest for local small/large scale experiment), $^{\dagger}$already in use (by local/international experiment), $^{\ddagger}$planned (for international experiment). For reactors ($\bar{\nu}$ flux) \cite{ref51} thermal power is assumed. Baselines stand for running (or approved) experiments where computed/expected fluxes are also indicated. Space-based neutrino sources availability are assumed by default. POT stands for Protons On Target.}\label{tab0b}
\centering
\begin{adjustbox}{width=\textwidth}
\begin{tabular}{lllllll}
\hline\noalign{\smallskip}
Source & Detector & L [m] & P [GW] & $\phi_{\nu}$ [s$^{-1}$] & Operator & Type \\
\noalign{\smallskip}\hline\noalign{\smallskip}
$^{\S}$Embalse (AR) & - & - & 2.1 & - & NASA & PHWR \\
$^{\S}$Atucha I (AR) & - & - & 1.2 & - & NASA & PHWR \\
$^{\dagger}$Atucha II (AR) & $\nu$IOLETA & 8(12) & 2.2 & 9.0(4.0)$\times$10$^{20}$ & NASA & PHWR \\
$^{\S}$Angra I (BR) & - & - & 1.9 & - & Eletronuclear & PWR \\
$^{\dagger}$Angra II (BR) & $\nu$-Angra/CONNIE & 30 & 3.8 & 1.21$\times$10$^{20}$ & Eletronuclear & PWR \\
$^{\S}$Laguna Verde I (MX) & - & - & 2.3 & - & CFE & BWR \\
$^{\S}$Laguna Verde II (MX) & - & - & 2.3 & - & CFE & BWR \\
$^{\S}$Yangjiang/Taishan (CN) & JUNO & 5.3$\times$10$^{4}$ & 26.61 & $\sim 10^{20}$ & YJNPC/TNPJVC & PWR \\
$^{\dagger}$NuMI (US) & MINER$\nu$A/NO$\nu$A & 1.0$\times$10$^{3}$ & 0.7$\times$10$^{-3}$ & $\sim 10^{20}$ (POT) & Fermilab & LINAC \\
$^{\ddagger}$LBNF (US) & DUNE & 1.3$\times$10$^{6}$ & 2.4$\times$10$^{-3}$ & $\sim 10^{21}$ (POT) & Fermilab & LINAC \\
$^{\ddagger}$J-PARC (JP) & Hyper-K & 2.95$\times$10$^{5}$ & 1.3$\times$10$^{-3}$ & $\sim 10^{22}$ (POT) & J-PARC & LINAC \\
$^{\ddagger}$Protvino (RU) & KM3NeT-ORCA & 2.6$\times$10$^{6}$ & 4.5$\times$10$^{-4}$ & $\sim 10^{20}$ (POT) & Protvino & LINAC \\
\noalign{\smallskip}\hline
\end{tabular}
\end{adjustbox}
\end{table*}

\noindent The above equations refer to the standard 3-flavour neutrino scenario. However, for a (3+1)-flavour scenario (e.g., addition of a hypothetical sterile neutrino), Equation~(\ref{eq01}) can be  modified accordingly by adding new flavour ($\nu_{s}$) and mass ($\nu_{4}$) eigenstates connected through the $[U_{s1}, U_{s2}, U_{s3}]$ row vector in the PMNS matrix, in order to give room to include a fourth flavour, hence, obtaining a 4$\times$4 PMNS matrix and three independent squared-mass differences $\Delta m^{2}$ where $U_{e4}^{2}+U_{\mu4}^{2}+U_{\tau4}^{2}+U_{s4}^{2} = 1$, is a PMNS unitarity condition. 

\noindent Additional details about the main challenges in neutrino oscillations, related to the determination of the mass hierarchy and leptonic CP phase for experiments with Latin American participation, are reviewed in Section~\ref{sec02}.

\subsection{Neutrino masses and nature}\label{sec:1b}

\begin{figure*}[htp]
  \includegraphics[width=1.0\textwidth]{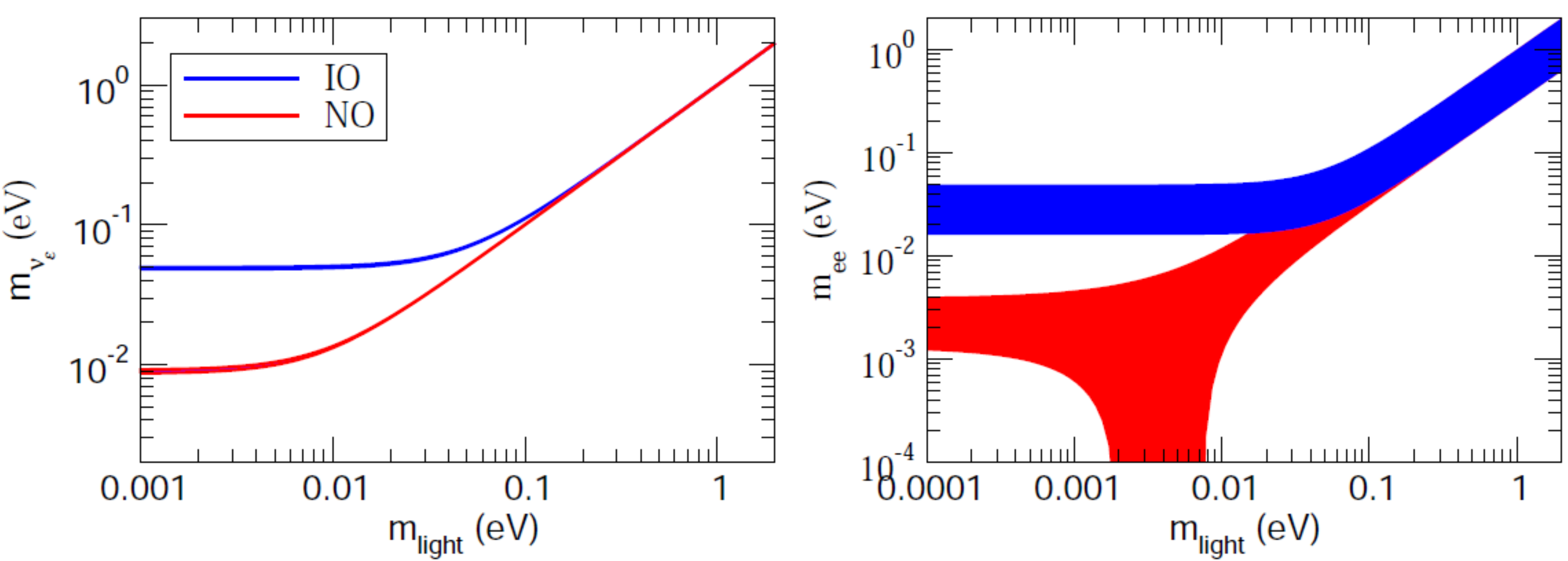}
\caption{Allowed regions at 95\% of Confidence Level ranges, one degree of freedom, in the 3$\nu$ mixing scenario as a function of the lightest neutrino. Left: neutrino mass observable in $^{3}$He $\beta$-decay. Right: neutrino mass expected in $\beta\beta0\nu$.} \label{fig00}
\end{figure*}

\noindent Neutrinoless Double Beta Decay ($\beta\beta0\nu$) searches have the potential to solve the questions related to the neutrino nature and absolute mass scale. This process requires the neutrino to be its own antiparticle, thus neutrinos are Majorana particles, \textit{L} ( \textit{R} subindexes denoting the Left- or Right-handed nature of neutrinos).  Experimental searches are looking for the so-called effective (Majorana) mass \cite{ref31}:

\begin{equation}\label{eq06}
\langle m_{\beta\beta} \rangle = \left\lvert \sum_{i=1}^{3}U^{2}_{ei}m_{i} \right\rvert .
\end{equation}
\\
\noindent In the  case that $\beta\beta0\nu$ is mediated by light neutrino exchange, Equation (\ref{eq06}) can be linked to the corresponding isotope half-life ($T_{1/2}^{0\nu}$) as a function of either $\langle m_{\beta\beta} \rangle$ and the nuclear matrix element ($M^{0\nu}$) (describing all the nuclear structure effects):

\begin{equation}\label{eq07}
T_{1/2}^{0\nu} = G^{0\nu}(Q,Z) \left\lvert M^{0\nu} \right\rvert ^{2} \left(\frac{\langle m_{\beta\beta} \rangle}{m_{e}} \right) ^{2},
\end{equation}
\\
\noindent where $G^{0\nu}(Q,Z)$ corresponds to the lepton phase-space (kinematic) factor (function of the Q-value of the decay, charge and mass of the final state nucleus).  Figure~\ref{fig00}, extracted from \cite{refmass}, represents the allowed regions plots in terms of the allowed range $m_{\nu_{e}}$ (from $^{3}$He $\beta$-decay) and $m_{ee}$ ($m_{\beta\beta}$) (from $\beta\beta0\nu$ decay) as a function of $m_{light}\equiv m_{0}$.

\noindent The ranges are obtained by projecting the results of the global analysis of oscillation data, with the exclusion of the tabulated $\chi^{2}$ data on atmospheric neutrinos provided by Super-K. The region for each ordering is defined with respect to its local minimum. To date, the best neutrino mass limit is set by the KamLand
Zen experiment, with $m_{\beta\beta} = 50 - 160$ meV, the best half-life sensitivity is set by the GERDA experiment, with $T_{1/2} > 11 \times 10^{25}$ years (90\% Confidence Level). Scalability of this kind of experiments projects sensitivities below 10 meV, indeed covering the allowed parameter space for IH as in Figure~\ref{fig00}.

\subsection{Astrophysical probes}\label{sec:1c}

\begin{figure*}[htp]
\centering
  \includegraphics[width=0.5\textwidth]{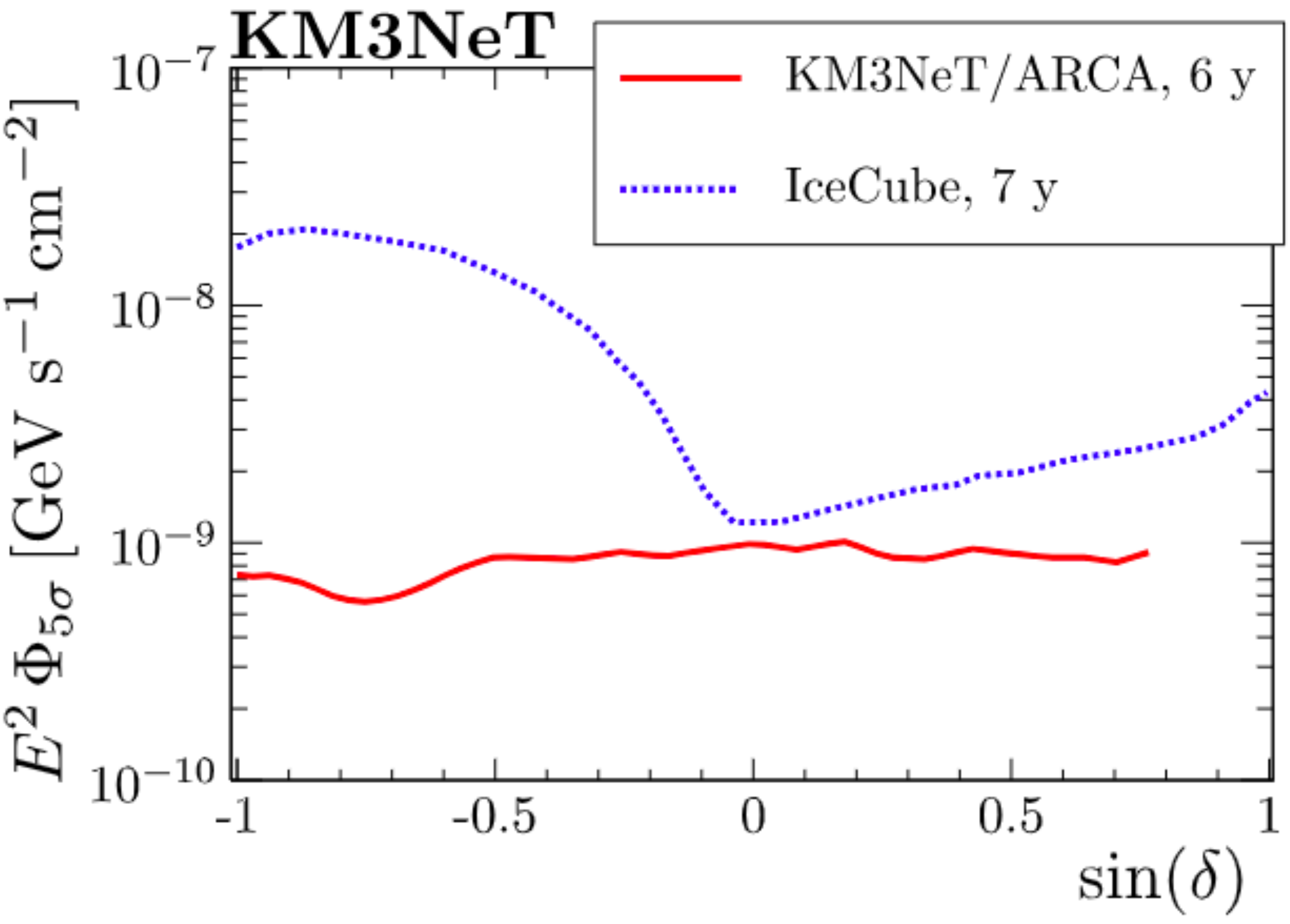}
\caption{Discovery flux ($\Phi_{5\sigma}$) as a function of the source declination angle ($\delta$) for the Mediterranean (two-sided gaussian probability) and South Pole (one-sided gaussian probability) very-large volume HE Neutrino Observatories. Convention in gaussian probabilities for KM3NeT impacts weakly on the results ($<$ 4\%).}\label{fig000b}
\end{figure*}

\noindent A lot of progress has been made in the last few years in the design of experiments, techniques and sensitivities aimed at unveiling neutrino properties, their nature, and usefulness as a probe, with huge remarkable achievements. This is due to  the strategic cooperation between experimentalists, theorists and funding capabilities. The discovery of the TXS 0506+056 Blazar as the first extragalactic neutrino source by the IceCube Collaboration \cite{ref32} boosted the potential of Neutrino Observatories to complete  the picture of astrophysical neutrinos. The complementary observation of whole the sky, by the Mediterranean and South Pole Neutrino Observatories, KM3NeT and IceCube, is strategic and crucial in the current understanding of the High-Energy (HE) neutrino astrophysical phenomenon. Figure~\ref{fig000b}, extracted from \cite{refsource}, represents the expected sensitivity to point-like neutrino sources by assuming a generic, unbroken power law energy spectrum proportional to E$^{-2}$ flux (compatible with IceCube findings) as a function of the declination angle at 5$\sigma$ significance, for both of the aforementioned observatories.

\noindent As can be seen, an excellent sensitivity in the Southern Hemisphere can be reached with KM3NeT for an observation time of six years, comparable to results reported by IceCube in similar conditions with seven years of data of all sky search. The detection media (sea water and polar ice) play a fundamental role in the scientific goals of such experiments. The better the understanding of light propagation in the medium the better detectors performance is achieved (e.g., angular resolution, effective area). In ice there is less light absorption so there is better energy reconstruction. In water there is less scattering thus better pointing accuracy (point source identification). The location of KM3NeT in the Northern hemisphere provides a full sky coverage, visibility of Galactic Plane, Galactic Center, and a 1.5 sr common view per day. The contribution to strategic activities like this plays an important role and sets challenges regarding the \textit{know-how}. Contribution from Latin America to KM3NeT, where an Optical Calibration System (OCS) is of special interest, a device that might be extended to other Water Cherenkov Detectors (WCD). The arrival of the KM3NeT Collaboration to Latin America guarantees at least the participation in one of these strategic infrastructures optimized for HE astrophysical neutrino studies, joining to the workforce on this strategic area in the world. Additionally, further results of KM3NeT regarding sensitivity to diffuse neutrino flux (full sky) report discoveries at 5$\sigma$ significance in less than one year. The biggest scientific challenge in the newborn HE neutrino astronomy, is of having a combined single distributed planetary instrument, known as HE PLanEtary neutrino Monitoring system (PLE$\nu$M), an United Nations Open Universe initiative \cite{ref69}. The Latin American community is ready to join the global efforts, either on motivations/experience on HE astrophysical data analysis, as well as in construction and commissioning of observatories, and flux predictions (Table~\ref{tab03}). \\
Additionally, as the multi-purpose capability of KM3NeT for neutrino parameters studies, ORCA, with atmospheric neutrinos as beam, will provide a 3$\sigma$ sensitivity on NH after three years of exposure and competitive precision in $\Delta m_{32}^{2}$ (2-3\%) and $\sin^{2}\theta_{23}$ (4-10\%). See section \ref{sec02a} for a detailed description of a new proposal for a facility in Latina America focused on $\nu_{\tau}$ astronomy.

\subsection{Search for new neutrinos states: light sterile neutrinos and heavy neutral leptons}\label{sec:1d}

\noindent The study of neutrino oscillations is based on the 3$\nu$ paradigm suggested by the structure of the Standard Model (SM), namely the existence of three complete generations. This means that whether right-handed neutrinos exist or not, there are only three light species that oscillate among themselves. This is of course the minimal possibility and additional sterile neutrino states could exist that would in principle oscillate with the three usual states, leading to a richer oscillation parameter space. In fact, the anomalies that have been observed in experiments at short-baseline accelerators, reactors, and source (Gallium)  (LSND, MiniBooNE, Gallex, SAGE) can not be accounted for by the 3$\nu$ paradigm and sterile neutrinos have been proposed as a possible remedy. It turns out, however, that even with several additional sterile neutrinos, it is not possible to explain all available data \cite{Dentler:2018sju}\cite{Gariazzo:2017fdh}. The existence of these anomalies has motivated several reactor- and accelerator-based experiments such as the NEutrino Oscillation at Short baseline (NEOS) in South Korea, the Search for Sterile Reactor Neutrino Oscillations (STEREO) in France, Neutrino-4 in Russia and the Detector of Anti-Neutrino based on Solid Scintillator (DANSS) in Russia, Precision Reactor Oscillation and Spectrum Experiment (PROSPECT) in the USA, the Search for short baseline antineutrino disappearance with composite Lithium-6 scintillator detectors (SoLiD) in Belgium, experiments ID137 at Fermilab, the MicroBooNE experiment (BooNE -  Booster Neutrino Experiment) that uses the same neutrino beam as MiniBooNE and that involve a future phase with the implementation of the Short-Baseline Near Detector (SBND) and Imaging Cosmic And Rare Underground Signals (ICARUS) in the near future. Also DUNE contains a program for the study of sterile-active oscillation parameters. \\
The potential existence of sterile states brings about the possibility for the presence of heavy neutral leptons that could be produced in colliders. In fact, from the theoretical and model-building point of view, these states are common in most extensions of the SM, and in some cases predicted/expected to exist with a mass scale close to the electroweak scale, thus available for experimental searches. Such searches have been performed at existing colliders. Sensitivity studies have been made for proposed experiments such as FASER, MATHUSLA and CODEX-b, and for future colliders (e.g., the FCC) as well. For some mass scales of  the heavy neutral leptons it could be also detected in neutrino experiments such DUNE. \\
It is clear from this that there is a community-wide interest in the study and potential discovery of these states, namely sterile neutrinos and heavy neutral leptons, whose motivation comes from the study of Beyond Standard Model (BSM) physics and Non-Standard Interactions (NSI).

\section{Research infrastructures}\label{sec02}

\noindent The diversity of neutrino experiments are providing valuable results to fit the global neutrino picture. Furthermore, underground/ice/water observatories offer excellent conditions for multidisciplinary research in Earth and Life Sciences as geophysics and biology (such as the proposed ANDES laboratory, see below), as well as in subatomic sciences as searches for rare events (e.g., $\beta\beta0\nu$, proton decay), exotic physics (e.g. DM), tests of general relativity, etc. They also offer low background conditions for atmospheric, reactor, solar (space-based accelerators in general), and earth-based accelerator neutrino experiments \cite{ref33}\cite{ref34}. Additionally, nuclear astrophysics processes, involving neutrinos as a probe are also favored. 

\begin{figure*}[htp]
  \includegraphics[width=1.0\textwidth]{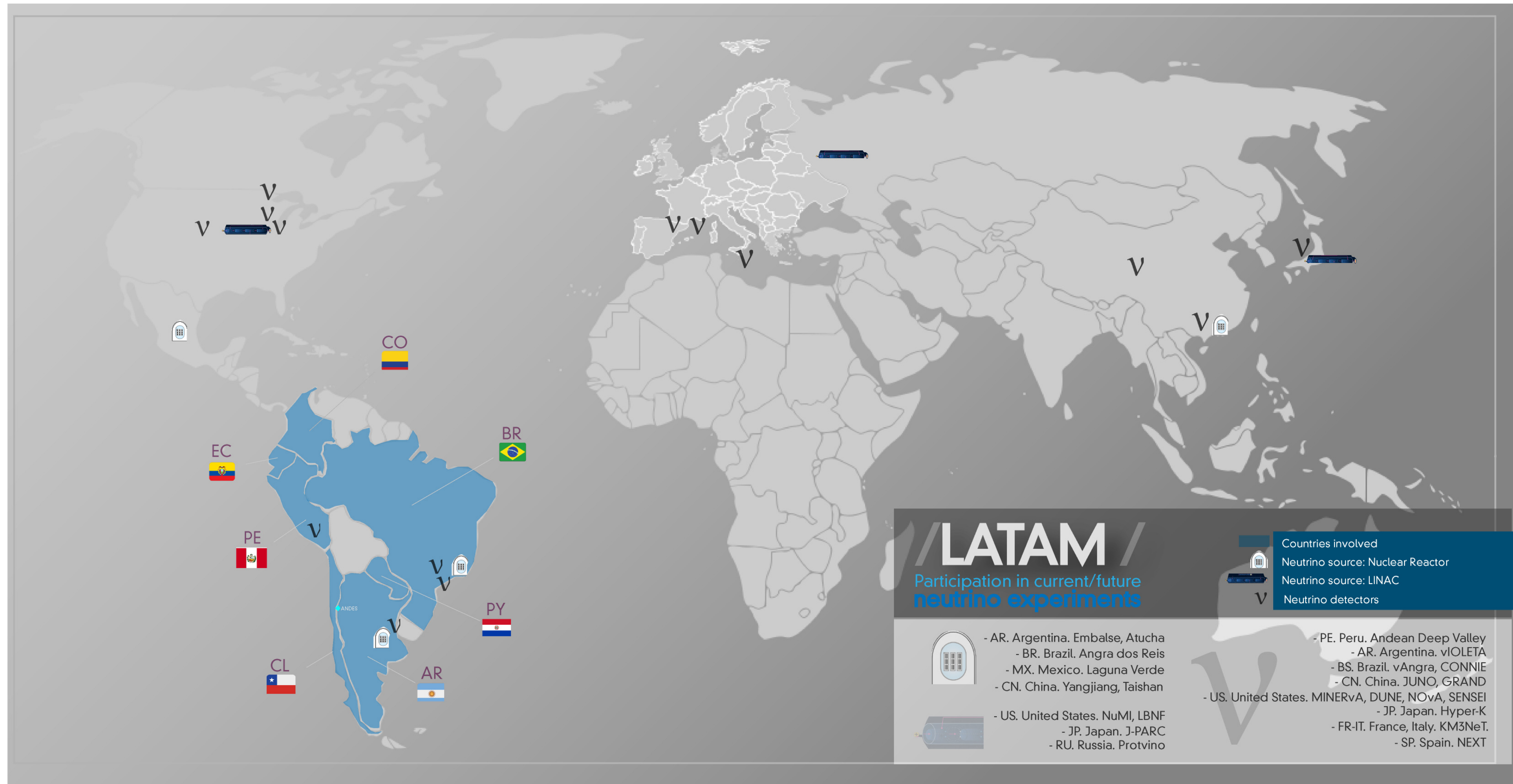}
\caption{Latin American  participation in current/future neutrino experiments: earth-based neutrino sources and detectors are indicated, space-based neutrino sources are assumed by default. MINER$\nu$A, whose data period has ended, is still providing world’s best high-precision neutrino cross-section measurements, is also represented.} \label{fig01}
\end{figure*}

\noindent Neutrinos appear in the Latin America context in sync with global efforts to address the open questions of Neutrino S\&T. Latin American countries currently involved in neutrino experiments either international and local collaborations are shown in Figure~\ref{fig01}.

\noindent Currently,  Latin America has a strong involvement in the  US in neutrino experiments such as DUNE, mainly featured by the design and construction of the Photon Detection System (PDS) for the single phase far detector modules. The PDS consists of a light-trap concept known as ARAPUCA or its variant X-ARAPUCA  and its corresponding readout systems\cite{ref48}. Other important participation of Latin American groups has been in the Miner$\nu$a, No$\nu$a and SBND projects. 
Latin America is also involved in specific programs in EU with CERN and other large-scale multidisciplinary research infrastructures as KM3NeT, setting a promising versatile future for Latin America in Neutrino S\&T global effort.

\noindent The feasibility of the ANDES laboratory in South America \cite{ref03} becomes very strategic towards the future of the neutrino community in Latin America as a flagship local large-scale infrastructure, run by an international consortium. ANDES is aimed to be installed in an international tunnel ($\sim$45 km long) to be constructed (2021-2029) along the Agua Negra Pass between Chile (Coquimbo) and Argentina (San Juan). ANDES will consist of a Main Hall (21 m x 23 m x 50 m), Secondary Hall (16 m x 14 m x 40 m), the Large Pit (30 m diameter, 30 m height), a geoscience area, offices and small labs. The vertical depth of ANDES is $\sim$ 1.8 km, omnidirectional $\sim$ 1.7 km, with horizontal road access, in total the facility reaches 70k m$^{3}$ of laboratory volume plus 35k m$^{3}$ access tunnels. The Large Pit, can be used to place a  next generation neutrino detector, the whole volume of the pit would be filled with ultra-pure water as shielding, waterproof against any leak coming from water of the rocks forming the cavern. The Main Hall could be available for hosting another emblematic experiment, e.g., a $\beta\beta0\nu$ detector. ANDES may host  neutrino experiments with a complementary global science programs, e.g., neutrino nature, cosmic neutrinos, DM neutrinos, Geoneutrinos, NSI, and applications in homeland security as Far Field (FF) site \cite{ref03}, see also [I-32]. A Supernovae neutrino program in ANDES would allow coupling to Global Neutrino Network (GNN) \cite{ref49} and SuperNova Early Warning System (SNEWS) \cite{ref50}, and follow-up under a joint sub-threshold analysis of offline time/space correlation multi-messenger searches (e.g., KM3NeT, CTA, LSST, SKA, HAWC, SVOM, LIGO/VIRGO, TAROT, ELT) some, with already relevant Latin America participation. 

\noindent Globally, the widely explored technologies in neutrino detection can be summarized as: a) Scintillation (noble elements gas, liquid, solid or dual phase) and Hydrid Detectors (e.g., Scintillation + Cherenkov), b) WCDs (liquid, solid/ice), c) Radio Cherenkov Detectors (ice, atmosphere) and Pressure-Waves Acoustic Detectors (water, ice), and d) Pixelated Detectors (semiconductors). The light/charge readout as a result of the light/charge response of the detection medium (LAr, GXe, C$_{x}$H$_{y}$ compounds, sea water, purified water, ice, salt, air, etc.), is usually performed by a synchronized performance of SiPMs, PMTs, WaveLength Shifter (WLS), Front End Boards (FEBs) and Data AcQuisitions systems (DAQ), etc. 

\noindent The NEXT experiment at the Canfranc Underground Laboratory (LSC), with Colombian participation, is a High-Pressure Gaseous Xenon (HPGXe) TPC with independent planes for energy (PMTs) and tracking (SiPMs), a good example of hybrid detection system \cite{ref58}. 

The next subsections present details of neutrino experiments with participation of Latin American groups that submitted a white paper during the LASF4RI-HECAP process. 

\subsection{Latin America-based large-scale infrastructures}\label{sec02a}

\begin{figure}
    \centering
    \includegraphics[width=0.7\textwidth]{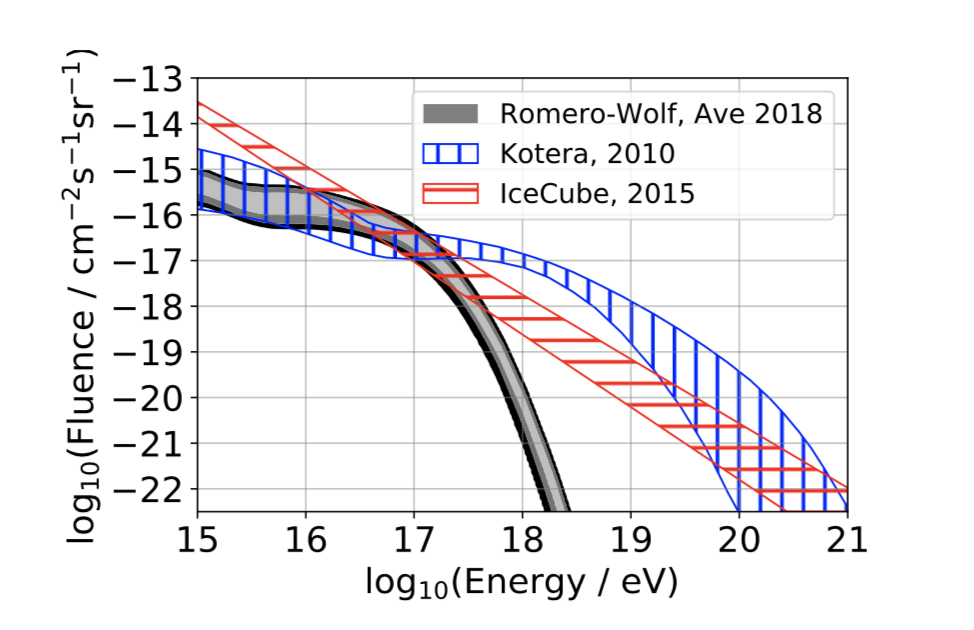}
    \caption{Neutrino fluxes probed by TAMBO, from \cite{ref62}.}
    \label{fig:deep valley}
\end{figure}

\paragraph{TAMBO: Tau Air Shower Mountain-Based Observatory} This is a proposal to be implemented in the Peruvian Colca Valley at the Tau Air shower Mountain-Based Observatory (TAMBO), as an array of small WCDs ($\sim$ 22k, 1 m$^{3}$ each, spaced by 100 m on the mountain slope), for HE $\nu_{\tau}$-astronomy with high background suppression \cite{ref62}. Andean Deep-Valley science objectives span: study of HE neutrino sources above 10 PeV, characterization of sources between 1-10 PeV range by measuring the tau component, and to constrain the particle acceleration potential of point source transients observed via multi-messenger astronomy. The $\nu_{\tau}$ propagates in rock with an interaction length of thousands of kilometers for the energy range of interest, it releases $\tau$s by Charged-Current (CC) interactions (range $\sim$ 50 m - 5 km before decay) transferring $\sim$80\% of its energy to the medium. If the decay happens within this range, it will be knocked out from rock to air and may indeed decay in air and about 50\% of its energy goes into particles that produce Extensive Air Showers (EAS). The longitudinal profile of the EM component of the EAS has a characteristic length of $\sim$10 km (mountains separations) and a diameter of $\sim$200 m near the EAS maximum. The topography where the array would be placed favours the geometric acceptance compared to a flat ground array. It is expected that Andean Deep-Valley will also shed light on discriminating photons from hadrons and contribute to $\gamma$-ray observations at the PeV range and beyond, and also to characterize the Cosmic Rays (CRs) anisotropy at PeV energies with new measurements at other latitudes. The TAMBO is close to producing a Conceptual Design Report, however, very interesting initiatives and a large collaborative effort are currently  shaping up between Brazilian, Peruvian, US and EU counterparts. Andean Deep-Valley would be able to link to GNN and SNEWS mentioned above. Figure~\ref{fig:deep valley} shows the range of neutrino fluxes that would be probed by TAMBO.\\

\subsection{Latin America-based small-scale - high impact- infrastructures}\label{sec02b}

\paragraph{Neutrinos Angra Experiment ($\nu$-Angra)} The $\nu$-Angra collaboration has been running smoothly since 2018 a WCD located at Angra Dos Reis nuclear complex (core II), recording $\bar{\nu}_{e}$ events from Inverse Beta Decay (IBD) during fuel burnout looking to monitor the reactor dynamics \cite{ref52}. About 354 gallons GdCl$_{3}$ doped water (0.2\%) constitutes the radiator and detection medium, contained into 0.90 m $\times$ 1.46 m $\times$ 1.02 m (height-length-width) plastic tank surrounded by 32 eight-inches PMTs (16 top-placed, 16 bottom-placed), 6 internal faces covered by diffuse reflectors (99\% reflectance at 400 nm) and top veto with 4 PMTs strategically located for tagging pulses triggered by cosmic muon events. The readout electronics is made of conventional elements, after amplification and shaping by the FEB, digitization is performed by a VME-based crate system which gathers processes and outputs samples to a FPGA controlling the data stream. The dependence of the neutrino counting rate with the reactor's activity make neutrino detectors a remote monitoring tool of the reactor's activity, composition of burned fuel, thermal power production and targeting of plutonium (main threat-related issue). The $\nu$-Angra detector is located on the surface, 30 m away from the core, the optimal signal-to-background ratio is the biggest issue to overcome, in this sense the goodness on proximity to the core is as important as an optimized shielding (e.g., by overburden). Brazil, leading the $\nu$-Angra collaboration has successfully turned the experimental neutrino expertise into practical roles for the Latin America nuclear sector. This is a R\&D proposal looking for, in a midterm, be considered as an innovative tool in homeland security, synchronized to recognized US and EU efforts as WATCHMAN \cite{ref41} and NUCIFER \cite{ref63}. Such efforts must follow IAEA's protocols and motivations for FF reactor monitoring \cite{ref64}, as potential and novel technologies that can be applied for non-proliferation safeguards. Reliability and cost-effective technology (significantly lower if implemented in the operational budget of the plant) is the big challenge. Remote monitoring reduces dose in the operation, maintenance, verification and inspection staff, by skipping access to the containment building and other restricted areas. \\

\paragraph{Skipper-CCD experiments (CONNIE, SENSEI, $\nu$IOLETA)} These experiments with different physics motivations but same core technology (ultra-low threshold Skipper-Charge Coupled Device (CCD)) are Silicon pixelated detectors systems with a spatial resolution of 15 $\mu$m capable of charge recording per pixel with a background noise RMS $\sim$ 0.1 $e^{-}$ (electrons individual counting). Skipper-CCDs were pioneered at Fermilab for the Sub-Electron-Noise Skipper Experimental Instrument (SENSEI) \cite{ref66} searching for direct DM observation in the MeV-GeV range. Due to the ultra-low background conditions requirements for DM searches, SENSEI would be a good candidate to be host at ANDES, see chapter on dark matter for more details. Low-energy recoils from CE$\nu$NS (cross-section more than two orders of magnitude higher than in the neutrino-nucleon case) of reactor antineutrinos with Silicon nuclei and tests of NSI (including sterile neutrinos searches from anomalies in nuclear reactor SBL experiments) are the main physics goals inspired by this kind of technology. Coherent Elastic neutrino-Nucleus Scattering (CE$\nu$NS) can also be studied from solar, atmospheric and diffuse Supernovae neutrinos, relevant processes have been identified as background components for future DM searches and next generation experiments. CE$\nu$NS directly measured in controlled reactor experiments are useful for the parameterization and suppression of such components.  The Skipper-CCD technology enables exploration of CE$\nu$NS at a SBL nuclear reactor experiment with CONNIE \cite{ref53}.CONNIE is currently a detection array of 12 CCDs ($\sim$ 73 g) running since 2016 at Angra-II in the same hut of $\nu$-Angra. Future upgrades consider Skipper-CCD technology ($\sim$ 10 kg) for significant improvements in the lower energy threshold and sensitivity. Additionally, due to the ultra-low threshold feature of Skipper-CCDs, it is also a  viable option for the quantification of $\bar{\nu}_{e}$ fluxes in experiments as $\nu$IOLETA \cite{ref54}. The latest  proposal is for a multi-kilogram array placed at Atucha nuclear complex (core II), 8-12 m away from the core and inside the dome (concrete shielding against CR muons background and enhancement of $\bar{\nu_{e}}$ flux).  Potential and flexibility of Skipper-CCD technology are reasonable criteria for boosting local cooperation between Argentina and Brazil and their impact on the nuclear industry, with the strategic partnership of Fermilab. \\

\subsection{International large-scale infrastructures}\label{sec02c}

\paragraph{Jiangmen Underground Neutrino Observatory (JUNO)} is an underground liquid scintillator (Linear Alkyl Benzen) experiment of $\sim$ 20 kton of total active mass under construction \cite{ref30}, located 53 km away from the Yangjiang (6 cores) and Taishan (2 cores) nuclear complexes in Southern China (Kaiping, Jiangmen). JUNO will work in antineutrino detection mode through IBD. The main focus of its neutrino physics programs spans: precision measurement of  of $\theta_{12}$ and $\Delta$m$_{21}^{2}$, NHM, Supernova and solar neutrinos, geo-neutrinos and proton decay. Institutions from Brazil and Chile are progressively getting involved in hardware through the sPMT ($\sim$ 25k 3-inches small PMTs) subsystem (Under Water Box and HV-Splitter), software (simulation of ABC FEB) and physics (solar oscillation parameters, physics potential). JUNO is the most representative reactor-neutrino experiment nowadays, continuing the remarkable legacy of experiments as Daya Bay, Double Chooz and RENO. \\

\paragraph{Deep Underground Neutrino Experiment (DUNE)} This experiment is under construction  formed by a Near Detector (ND) (Fermilab) and a Far Detector (FD)  located at Sanford Underground Research Facility (SURF), spaced by 1300 km and exposed along the LBNF beam line \cite{ref48}. DUNE is designed to address precision measurements of $\theta_{23}$ octant, NMH and potential discovery of CP violation in leptons, $\delta_{CP}$. The FD (charge and light detection, calorimetry-positioning, light-yield field) is a LAr TPC of $\sim$ 40k tons of total active mass equally segmented into four modules. Scintillation light from interactions in LAr ($\sim$ 40 photons/keV when excited by a minimum ionizing particle, 127 $\pm$ 10 nm, WLS used) plays a major role in the timing, triggering and calorimetry of the detector. Brazil is strongly connected to DUNE efforts with major involvement in all the strategic areas: hardware (PDS of FD, with the ARAPUCA detector conceived, designed and developed in Brazil, see chapter on Instrumentation and Computing for more details), software (Monte Carlo simulations for light production and propagation) and physics (BSM and Supernova Neutrinos). Colombian, Peruvian and Paraguayan groups are mainly involved in the hardware of the PDS of DUNE, specifically through the digitization boards for SiPMs (required to run at room temperature) for the main readout electronics system so-called ``DAPHNE''. Colombian groups' interest in physics are connected to NSI and DM. The Peruvian groups are also contributing releasing phenomenological analyses of DUNE sensitivity to Beyond Standard Model phenomena\cite{gago}.\\

\paragraph{NuMI Off-axis $\nu_{e}$ Appearance experiment (NO$\nu$A)} An inspiring experiment with valuable operational experience for DUNE. NO$\nu$A has a 810 km baseline (Fermilab to Minnesota), formed by identical two neutrino detectors (ND of $\sim$ 300 metric-ton 1 km from NuMI beam and FD of $\sim$ 14 metric-kiloton), both are made up of 344k cells of plastic PVC filled with liquid scintillator (C$_{9}$H$_{12}$ pseudocumene + special additives, with mineral oil as solvent) \cite{ref59}. Neutrino-scintillator interactions release a burst of charged particles whose energy is collected using WLS connected to Avalanche PhotoDiode (APD) arrays for readout. NO$\nu$A physics programs are focused on the precise determination of $\theta_{23}$ and $\Delta$m$_{32}^{2}$, NMH and $\delta_{CP}$. Colombian contributions are in physics analysis: cross-section data analysis in neutrino or antineutrino-mode (inclusive and exclusive channels) and reduction of uncertainties: axial masses, final state interactions, random phase approximation and the 2p2h. \\

\paragraph{Short Baseline Neutrino Experiment (SBND)} In addition to the experimental observation of the three neutrino picture described above, several experimental “anomalies” have been reported which, if experimentally confirmed, could be hinting at the presence of additional neutrino states with larger mass- squared differences participating in the mixing. SBDN is one of the three detectors part of the Short Baseline neutrino program at Fermilab, where neutrino or anti-neutrino beams can be produced in the same experiment and,  both charged current and neutral-current channels can be explored to clarify these anomalies. The Brazilian group involved in DUNE is also participating in SBND with an R\&D program for novel photon detection systems.

\paragraph{KM3 Neutrino Telescope (KM3NeT)} It is an international collaboration constructing the future km$^{3}$-scale deep-sea neutrino observatory in two abyssal sites at the Mediterranean: ARCA, near Capo Passero in Sicily (Italy), optimized for cosmic and astrophysical neutrinos, and ORCA, near Toulon (France), optimized for atmospheric neutrinos \cite{ref56}. ARCA and ORCA, as multi-purpose neutrino detectors, share the same technology but their layouts are sparser and denser respectively. KM3NeT scaling is in progress aiming to complete 3 building blocks (2-ARCA, 1-ORCA) formed by 115 Detection Units (DUs) each comprising of 18 Digital Optical Modules (DOMs) (31 three-inches PMTs, calibration systems, central logic board and power supply) vertically arranged and equally spaced. The Cherenkov light induced by relativistic charged leptons produced by CC neutrino interactions when crossing the detection media is collected by the DUs array. KM3NeT owns unique features as a multi-disciplinary infrastructure for Life and Earth Sciences: permanent connection, real time, and large band-width for undersea communications with shore stations on the coast or in remote operation. User ports available in the submarine network of the seabed allows to connect specialized instrumentation and modules operating in challenging environments. Main physics programs search for with ARCA for the discovery and subsequent study of extraterrestrial neutrino sources, while with ORCA, the aim is to measure $\theta_{23}$ and $\Delta$m$_{32}^{2}$, and NMH with atmospheric neutrino sources.  In addition searches for to NSI, exotic physics (monopoles, nuclearities), test of fundamental physics and $\nu$-tomography (attenuation and oscillation) are also part of the physics program. The P2O proposal, a Long Base Line (LBL) experiment of $\sim$ 2600 km, aiming to shoot a neutrino beam from Protvino tovthe  ORCA site is the bet towards a measurement of $\delta_{CP}$. The dense core of ORCA offers in turn lots of possibilities for indirect DM searches, and would be open for a strong and strategic collaboration with Latin America. Current activities in Ecuador are focused on hardware (calibration) both for the photonics (DOM angular acceptance, HV tuning of light sources) and acoustic positioning subsystems (digital compasses), as well in the design of the OCS. Regarding physics contribution to Blazars or Supernovae neutrinos is expected to start soon. \\

\paragraph{Hyper-Kamiokande (Hyper-K)} The next generation of the Super-K detector is known as Hyper-K \cite{ref60}, it will continue the legacy of the successful operative experience and challenging physics programs. In this version, two new WCDs are considered: a 0.5 kt ND and 258(188) kton FD of total(fiducial) mass, at $\sim$ 1 km and 295 km from the Japan Proton Accelerator Research Complex (J-PARC) neutrino beam, respectively. The design of Hyper-K, consists of drum-shaped tanks of 10 m in diameter $\times$ 8 m in height for the ND, and 60 m in diameter $\times$ 74 m in height for the FD. The photosensor arrays are $\sim$ 500 multi-PMT(19) DOMs in the ND in charge of measuring flux and cross-section in order to reduce systematics at FD, and 40k twenty-inches inner detector PMTs (first option) or 50\% twenty-inches + multi-PMT DOM, in addition $\sim$ 7k outer veto detector PMTs at the FD. The tanks will be filled with ultra-pure water ($\sim$ 260k tons in the FD) as radiator media for collecting signals coming from CC neutrino interactions with the electrons of nuclei of water, relativistic electrons or positrons creating the distinguished Cherenkov rings. Hyper-K will be located under the Nijugo mountain 295 km away from the J-PARC neutrino beam facility in Tokai, Japan. The main physics goals of Hyper-K span from determination of $\theta_{23}$, $\Delta$m$_{32}^{2}$ and precision measurements of oscillation parameters, and NMH supported by atmospheric and J-PARC neutrino sources, $\delta_{CP}$ observation with J-PARC neutrinos, Supernova and relic neutrinos, and $^{8}$B neutrinos from Sun's fusion reactions, proton decay and DM searches, and other exotic phenomena as sterile neutrinos and tests of Lorentz invariance. Brazil is currently involved in Hyper-K in physics studies of  NSI, as well as possible contributions to GRID computing and data storage. \\
\underline{Major physics drivers:} $\theta_{23}$, $\Delta$m$_{32}^{2}$, precisions, NMH, 

\paragraph{Giant Radio Array for Neutrino Detection (GRAND)} The GRAND's concept is based on a large array of antennas (20 subarrays with 10k antennas each over $\sim$ 10k km$^{2}$) for detection of radio emission from EAS triggered by cosmic-ray interactions with atmosphere \cite{ref65}. The detection array consists of 5m-high antennas optimized for horizontal showers bow-tie design, 3 perpendicular arms, frequency range 50-200 MHz, with a inter-antenna spacing of 1 km. GRANDProto300 prototype would be suited with 300 detection units (radio antennas + surface detectors) over $\sim$200 km$^{2}$ with denser infill array in a prospective site in QingHai province, China. GRAND's primary goals are focused on Ultra-High Energy (UHE) neutrinos (cosmogenic and astrophysical), with secondary goals in cosmology and radioastronomy. GRANDProto300 will be able to study the transition between galactic and extra-galactic CRs. GRAND10k constitutes an intermediate step in  GRANDS's timeline. Once its final layout, GRAND200k, the observatory would be able to detect cosmogenic neutrinos (produced by UHE CRs interactions while intergalactic propagation) with energies larger than 100 PeV by radio signal detection produced by the interaction of UHE CRs, $\gamma$-rays and neutrinos with Earth's atmosphere. Brazil and Argentina are currently involved in GRAND collaboration (planned), and their roles are still an open issue. GRAND will be able to link to GNN and SNEWS mentioned above, and couple to a multi-messenger strategy with Latin America participation. \\

\section{Areas of excellence in Latin America}\label{sec03}

\noindent There are seven countries actively engaged in neutrino projects in the Latin American region: Colombia (CO), Peru (PE), Paraguay (PY), Ecuador (EC), Argentina (AR), Brazil (BR) and Chile (CH) \cite{ref67}. The identified capacities in theory and phenomenology, and from participation in neutrino experiments as presented in Section \ref{sec02} is summarized in Table~\ref{tab03}.

\begin{table*}
\caption{Potential capabilities and involvement in neutrino science by the Latin American community: sources, phenomenology and theory SI/NSI. Based on historical collaborative records and current LASF4RI-HECAP process. \cite{ref67}.}\label{tab03}
\centering
\begin{adjustbox}{width=\textwidth}
\begin{tabular}{cccccccccc}
\hline\noalign{\smallskip}
Country & Geo LE & Cosmic ME & Cosmic HE & SI (XS, CE$\nu$NS) & Masses & $\theta_{23}$ & NMH & $\delta_{CP}$ & BSM, NSI (e.g., DM) \\
\noalign{\smallskip}\hline\noalign{\smallskip}
CO & - & - & - & $\surd$ & $\surd$ & $\surd$ & $\surd$ & $\surd$ & $\surd$ \\
PE & - & - & $\surd$ & $\surd$ & - & - & - & - & $\surd$ \\
PY & - & - & - & $\surd$ & - & - & - & - & $\surd$ \\
EC & - & $\surd$ & $\surd$ & - & - & $\surd$ & $\surd$ & $\surd$ & $\surd$ \\
AR & - & $\surd$ & $\surd$ & $\surd$ & - & - & - & - & $\surd$ \\
BR & $\surd$ & $\surd$ & $\surd$ & $\surd$ & - & $\surd$ & $\surd$ & $\surd$ & $\surd$ \\
CL & $\surd$ & $\surd$ & - & - & - & - & $\surd$ & - & $\surd$ \\
\noalign{\smallskip}\hline
\end{tabular}
\end{adjustbox}
\end{table*}

\noindent Columns 2-4 refers to potential capabilities and experience in neutrino observatories for geo and cosmic neutrinos studies, 5-6 for Standard Interactions (SI) and absolute neutrinos masses, 7-9 for still unknown parameters, and last column for open opportunities in NSI and advanced physics programs. 
On the other hand, Table~\ref{tab04} shows identified capacity building in detectors instrumentation and related technology.

\begin{table*}
\caption{Potential capacities and involvement in neutrino technology by the Latin American community: detector instrumentation and related technologies. Based on historical collaborative records and current  LASF4RI-HECAP process \cite{ref67}.}\label{tab04}
\centering
\begin{adjustbox}{width=\textwidth}
\begin{tabular}{cccccccc}
\hline\noalign{\smallskip}
Country & SiPMs & PMTs & CCDs & Acoustic/Radio & FEB/DAQ & Protocols (EO/OE, WR) & AI, IT, GRIDs \\
\noalign{\smallskip}\hline\noalign{\smallskip}
CO & $\surd$ & $\surd$ & - & - & $\surd$ & $\surd$ & $\surd$ \\
PE & $\surd$ & $\surd$ & - & - & $\surd$ & - & - \\
PY & $\surd$ & - & $\surd$ & - & $\surd$ & - & - \\
EC & - & $\surd$ & - & - & - & - & $\surd$ \\
AR & - & - & $\surd$ & - & - & - & - \\
BR & $\surd$ & $\surd$ & $\surd$ & $\surd$ & $\surd$ & $\surd$ & $\surd$ \\
CL & - & $\surd$ & - & - & - & - & - \\
\noalign{\smallskip}\hline
\end{tabular}
\end{adjustbox}
\end{table*}

\noindent In this case, columns 2-5 stand for photonics and acoustics experience in neutrino detection systems and calibration, 6-7 for readout, DAQ and communications, last column for declared Data Science experience, Information Technology, and computing management and availability in neutrino projects. With respect to the last item, several highly-qualified scientists and engineers from Venezuela, Puerto Rico and Mexico  with a lot of experience running HEP projects \cite{ref68} locally and internationally could contribute to these efforts.  

\section{Synergies }\label{sec05}

\subsection{Local large-scale infrastructures}\label{sec:4a}

\noindent ANDES and TAMBO are two of the future initiatives for a large-scale implementation in the region. The construction of ANDES, supported by the international community, would hopefully be starting in a 5-10 years timeline. Once civil constructions is finalized it could  host two neutrino experiments in the Main Hall and the Large Pit. A Supernovae neutrinos experiment might be an option for the Main Hall. The Large Pit could host a a $\beta\beta0\nu$ neutrino experiment, as this can be a valuable opportunity for Latin America (Table~\ref{tab03}). On the other hand, TAMBO is also as a very good option for $\nu_{\tau}$-astronomy, being  a groundbreaking field, globally attractive and being already supported by US and EU counterparts. TAMBO does not use ``new technologies'', the detection array concept is based on well known WCDs incorporating PMTs with nanosecond time resolution is commercially available, with electronics relatively simple to use (e.g., AUGER, LAGO). WCDs overlap with other HECAP experiments such as HAWC and SWGO also supported by Latin America collaborators with complementary roles. Very-large volume neutrino observatories as KM3NeT are suited for making their science programs looking $\nu_{\mu}$ as the golden channel, multi0messenger programs are therefore another strategic factor for Latin American. Thanks to IceCube's breakthrough discoveries and plans for upgrades, the scaling of KM3NeT and the arrival of new neutrino observatories (e.g., GVD, P-ONE) will make possible to survey the whole sky with unprecedented potential for discovery and increased sensitivity. \\

\subsection{Local small-scale infrastructures}\label{sec:4b}

\noindent The pair of Skipper-CCD experiments $\nu$IOLETA and CONNIE are intimately connected not only by the potential offered by the technology, but by the practical roles of applied antineutrino  technology. 
These outstanding short-scale Near Field (NF)-SBL projects, may garner support and for  the upgrade of the experiments and development of a FF-LBL option. Some proposals have been discussed in the past in national meetings with funding agencies and stakeholders, where the construction of the FF site is the big challenge. 
The bet for FF sites for $\nu$IOLETA, CONNIE, $\nu$-Angra, and detector upgrades may sound reasonable. Several prototypes during R\&D might indeed be early tested in the several research reactor facilities in Brazil and Argentina. Moreover, links between governments, funding agencies and international organizations such as IAEA are supporting the progress of this kind of technology through detailed technical cooperation programs and more.

\subsection{International large-scale infrastructures}\label{sec:4c}

\noindent The participation in many international  collaborations  covering almost all cutting-edge neutrino physics programs have allowed important advances in  training, networking and qualified human resources. The Latin American strategy must also recognize and value  current collaborations mainly started by huge individual efforts. In this sense, initiatives as discussed in Subsections \ref{sec:4a}-\ref{sec:4b}  complete the picture of prioritized  international multidisciplinary large-scale infrastructures, multi-purpose neutrino facilities and common core technologies. \\

In order to facilitate make sustainable and stronger the participation  in these international efforts support is needed for local labs and computing centers, funding for mobility, graduate programs, contributions to large-scale experiment common funds for ensuring construction and operative overheads. Concretely, in terms of hardware, Latin America commitments are in: a) commissioning of the PDS of DUNE, b) commissioning of the OCS of KM3NeT, c) commissioning of the sPMT subsystem of JUNO. Physics, data analysis and high-performance computing tools  are naturally open to all collaborative efforts, implying budget significantly lower, but equally important contributions, than in hardware. Strong theory groups are spread throughout Latin America, participating e.g. in NO$\nu$A and  Hyper-K, with vast experience on neutrino physics.

\section{Conclusions}\label{sec06}

\noindent Beyond physics, multi-disciplinary and multi-purpose  drivers are strategic in the synchronized and visionary efforts of HECAP on a global scale.  KM3NeT is an example with permanent connection, real time, and large band-width for undersea communications with shore stations, and unprecedented experimental conditions for Earth and Life Sciences. In addition, it is  making possible neutrino physics with atmospheric, cosmic and astrophysical sources. TAMBO and ANDES are two proposed local large-scale initiatives. The future ANDES laboratory would also offer  multi-disciplinary and/or multi-purpose science opportunities. The Angra Dos Reis neutrino lab fits this criteria as well providing options for applied neutrino physics prototypes as $\nu$-Angra (nuclear safeguards) and basic science with CONNIE (coherent elastic neutrino-nucleus scattering), with remarkable testing of novel technologies.  This also applies to $\nu$IOLETA, a strategic facility in the Latin America roadmap. The pair of Skipper-CCD experiments $\nu$IOLETA and CONNIE are in turn intimately connected and  important in the strategy for local short-scale infrastructures.

\noindent Regarding international large-scale infrastructures we note that the PDS system of DUNE, the OCS of KM3NeT, the sPMT of JUNO, have been identified as important initiatives in the region. 




\chapter{Electroweak \& Strong Interactions, Higgs Physics, CP \& Flavour Physics and BSM}\label{chapt:colliders}
\vspace{1cm}
\noindent M. Cambiaso (Universidad Andrés Bello, Chile)\\ E. Carrera (U. San Francisco de Quito, Ecuador)\\ A. Gago (Pontificia Universidad Cat\'otica del Per\'u, Peru )\\ M. Mulders (CERN, Switzerland)\\ R. Rosenfeld (IFT-UNESP, Brazil)\\ A. Zerwekh (U. T\'ecnica Federico Santa Mar\'{i}a, Chile)












\section{Introduction}

These are exciting times for Particle Physics. In recent years a multitude of experimental results and advances in theoretical calculations and ideas have yielded unprecedented clarity and detailed understanding of the fundamental forces and building blocks of Nature. 

\noindent In particular, the current generation of flagship collider facilities and experiments has allowed exploration at the high-energy and luminosity frontier. A vibrant participation from the scientific community in Latin America has played an important and growing role in these endeavours. 

\begin{figure}[h]
    \centering
   \includegraphics[width=0.8\textwidth]{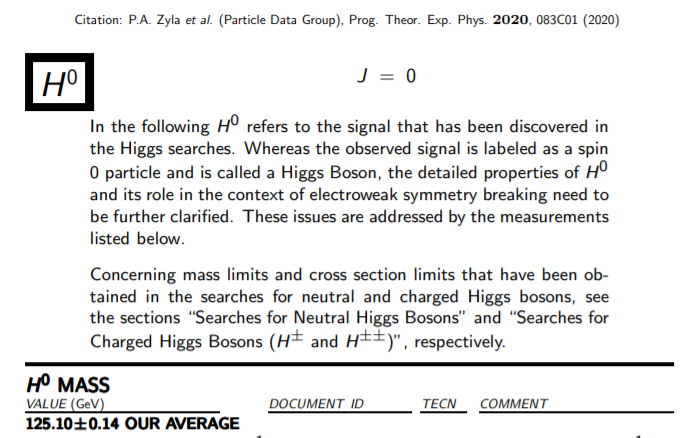}
    \caption{First page of the Higgs boson entry in the 2020 PDG.}
    \label{fig:HiggsPDG}
\end{figure}

\begin{figure}[h]
    \centering
   \includegraphics[width=0.8\textwidth]{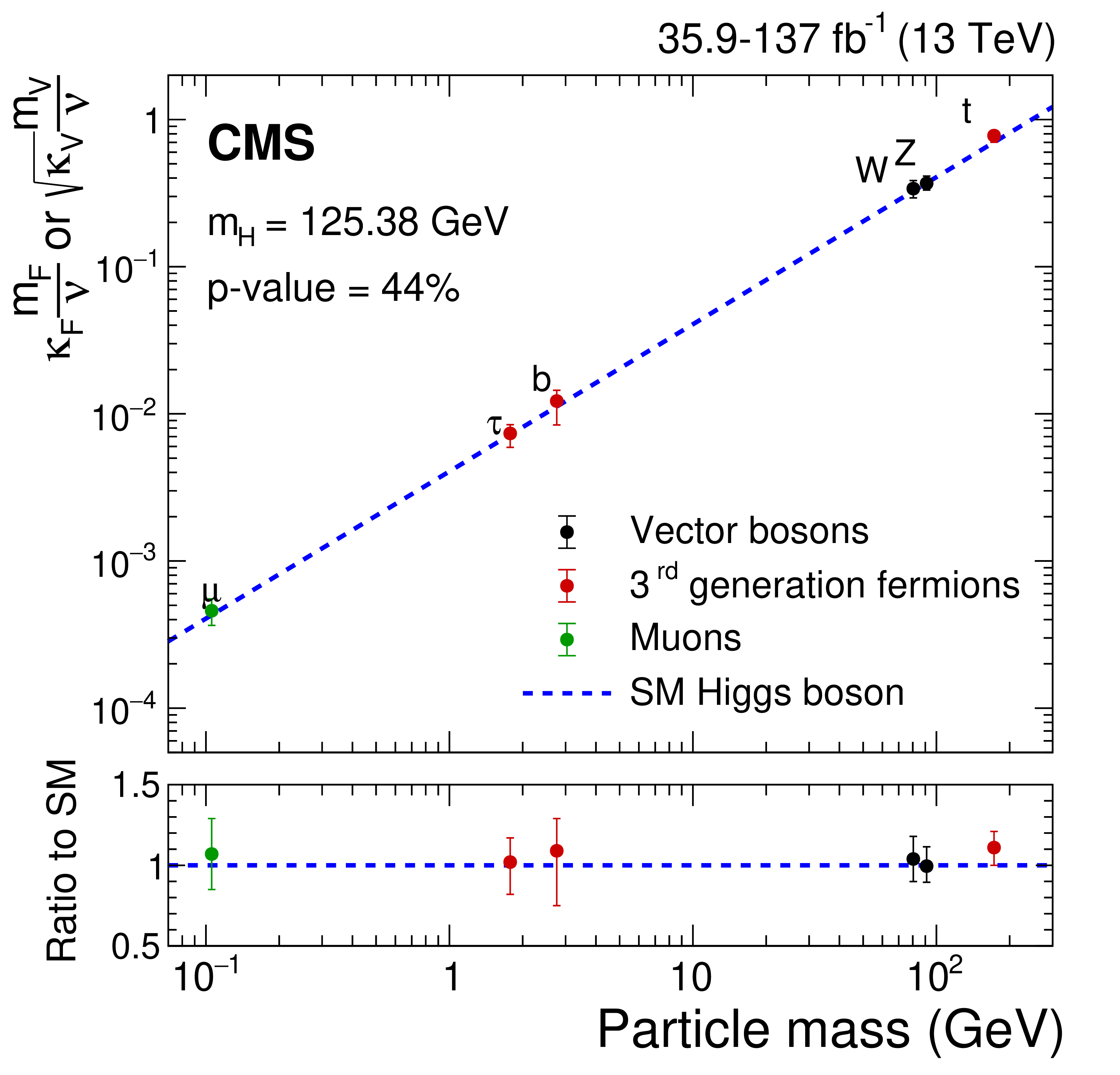}
    \caption{Reduced coupling-strength modifiers $\kappa_F m_F/v$ for fermions (F=t,b,$\tau$,$\mu$) and$\sqrt{\kappa_V}m_V/v$ for weak gauge bosons (V = W, Z) as a function of their masses $m_F$ and $m_V$, respectively. The SM prediction for both cases is also shown (dotted line). The error bars represent 68\%CL intervals for the measured parameters. The coupling modifiers are measured assuming no BSM contributions to the Higgs boson decays, and the SM structure of loop processes such as ggF and $H \rightarrow \gamma \gamma$. The lower inset shows the ratios of the values to their SM predictions. From CMS-HIG-19-006.}
    \label{fig:HiggsCouplings}
\end{figure}

\noindent After several decades the last missing particle of the Standard Model (SM) of particle physics was finally detected in 2012: the Higgs boson. 
It is amazing to see the rapid development in the Higgs measurements.  Figure \ref{fig:HiggsPDG}  shows the first lines of the Higgs boson entry in the 2020 PDG. Notice the mass average of $m_H = 125.10 \pm 0.14$ - a $0.1 \%$ measurement!
The couplings of the Higgs boson are also in very good agreement with the SM. Figure  \ref{fig:HiggsCouplings} presents  a recent CMS  result that includes the first $3\sigma$ measurement of the Higgs couplings to muons, the first time that there is direct evidence that the Higgs interacts with a 2nd generation particle. 

\begin{figure}[h]
    \centering
   \includegraphics[width=0.4\textwidth, scale=0.5]{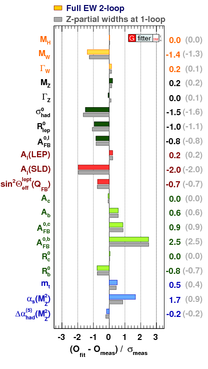}
    \caption{Comparing fit results with direct measurements: pull values for the SM fit, i.e. deviations between experimental measurements and theoretical calculations in units of the experimental uncertainty. [Figure from the GFitter group.}
    \label{fig:EWPrecision}
\end{figure}

\noindent
Furthermore, one should also emphasize the very good agreement among precision measurements of electroweak observables with theoretical calculations, as exemplified in Figure \ref{fig:EWPrecision}.

\noindent Despite its tremendous success in explaining a huge amount of processes studied in different HEP facilities, the SM has shortcomings that point to its incompleteness as the ultimate model of Nature. 
Some of the shortcomings of the SM are:
\begin{enumerate}
    \item origin of its 19 free parameters (related to flavour and CP problems);
    \item origin of neutrino masses;
    \item origin of dark matter;
    \item origin of dark energy;
    \item hierarchy or naturalness problem;
    \item does not contain gravity.
\end{enumerate}

\noindent In particular, the SM describes only roughly 5\% of the components of the Universe. It is known that we do not know what 95\% of the universe is made of; and that is a big gap. Dark matter and dark energy, contributing with 25\% and 70\% to the energy density of the Universe today respectively, are not described by the SM. In addition, we know that neutrinos have masses and this is also not contained in the SM.

\noindent The hierarchy or naturalness problem deserves more explanation. The Higgs boson mass in the SM receives large contributions from radiative corrections due to some new physics appearing at a higher energy scale $\Lambda$. Therefore an exquisite fine-tuning of bare parameters in the Lagrangian is necessary to explain why $m_H \ll \Lambda$. This problem, although a theoretical one, is the main guide to build models beyond the SM (BSM).
We can classify the BSM proposals as:\\
$\bullet$ Supersymmetric (SUSY) extensions of the SM: Higgs mass protected by SUSY;\\
$\bullet$ Composite Higgs: Higgs mass protected by a shift symmetry;\\
$\bullet$ Extra dimension: Higgs mass protected by the geometry of space-time.

\begin{figure}[h]
    \centering
   \includegraphics[width=\textwidth]{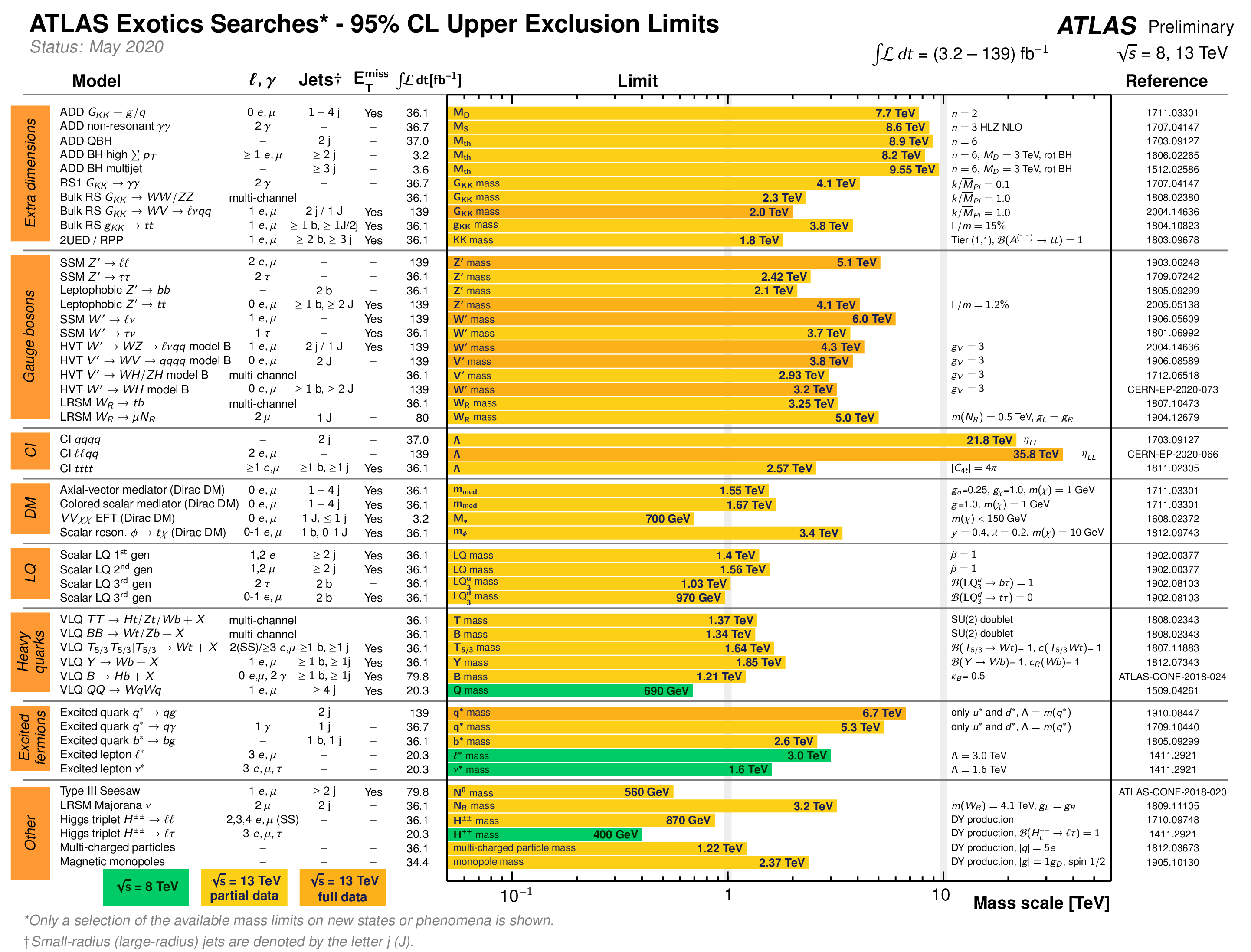}
    \caption{Bar chart with the search of particles predicted by several BSM proposals. Notice that energy scales of the order of 10 TeV have already been explored at the LHC without signs of new physics. From ATLAS Exotics Group (2020). }
    \label{fig:exotica}
\end{figure}

\noindent However, the naturalness problem as a guide to BSM has been challenged by the absence of new physics at the LHC. 
As shown in Figure \ref{fig:exotica}, energy scales of the order of 10 TeV have already been explored.
We may be entering the post-naturalness era \cite{Giudice:2017pzm}. 

\noindent There are also challenges related to the strong interactions. The SM contains the basic ingredients for the description of hadrons in terms of their fundamental constituents, quarks and
gluons. The interaction between quarks occurs through gluons, which also interact with each other. QCD has been very successful in describing a large variety of physical phenomena, from the most basic properties of nuclei found in nature to extreme conditions, such as the dynamics inside neutron stars or collisions between elementary particles, either in the atmosphere (cosmic rays) or in large particle accelerators.
Two important features of QCD are asymptotic freedom and confinement. The first principles
understanding of confinement is still an outstanding open question. An
important way to study confinement is to create a deconfined medium in high energy
heavy ion collisions. In extreme situations of temperature and/or baryon density,
confinement in hadrons is no longer a necessary feature of the medium. In this sense, a phase transition to more complex states of matter occurs. The study of these states is extremely important for the complete understanding of the equation of state of QCD and its properties.

\noindent In particular, when the baryonic density is low and the temperature is high, a phase transition is expected to form a deconfined state of quarks and gluons called Quark-Gluon Plasma (QGP). The observation and characterization of this state has important consequences for the detailed understanding of the strong interactions and it is also
crucial for a better comprehension of the evolution of our Universe since it is expected that it was in a similar state after a few moments from the Big Bang. The observation and
investigation of this state of matter allows us to understand in more detail the evolution
of the primordial Universe, its expansion and hadronization. The experimental
investigation of the existence and characterization of this state of matter is the focus of
large accelerators of heavy ions at relativistic energies, such as RHIC (Relativistic Heavy
Ion Collider) in the USA and LHC (Large Hadron Collider) in Europe.

\noindent This chapter focuses on the Latin American contribution to the advancements in the study of the electroweak and strong interactions, Higgs physics, the origin of CP violation, the origin of flavour, and new physics beyond the Standard Model.
The main physics drivers are:
\begin{enumerate}
    \item precision tests and detailed exploration of the Standard Model;
    \item study properties of the quark-gluon plasma;
    \item better understanding of the proton structure through improved parton distribution functions;
   \item testing the properties of the Higgs boson;
    \item exploring new sources of CP violation that could explain the matter-antimatter asymmetry in the universe;
    \item search for dark matter candidates; 
  \item search for signals of the existence of compact extra dimensions;
  \item search for SUSY;
    \item testing models for the origin of  neutrino masses;
    \item search for other new physics beyond the Standard Model.
\end{enumerate}

\begin{figure}[h]
    \centering
   \includegraphics[width=\textwidth]{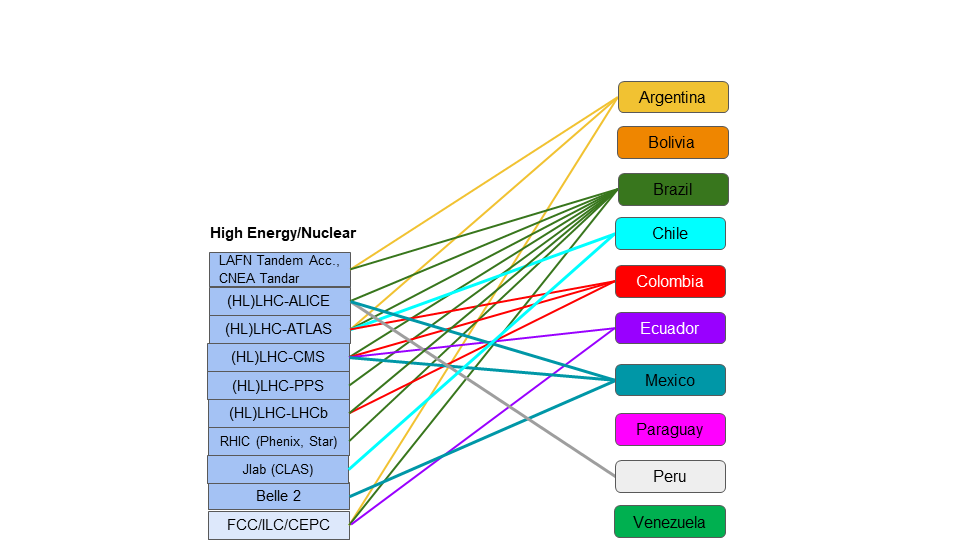}
    \caption{Landscape of LA participation in HEP/Nuclear experiments. }
    \label{fig:landscape}
\end{figure}

\noindent In the following we will describe the Latin American participation in the different experimental facilities in the region and in the world where the physical drivers mentioned above are studied.
The main participation of Latin American countries in HEP/Nuclear experiments is illustrated in Figure \ref{fig:landscape}.

\noindent The majority of the HEP community in Latin America is participating in CERN experiments.  More than 500 scientists, graduate students and engineers associated to institutes in the Latin American region are officially associated with CERN, predominantly as members of the LHC collaborations but also in some cases in other, smaller size experiments. Out of those, about 150 participants are considered Authors, and appear as authors on scientific publications. The exact criteria for authorship depends on the experimental collaboration, but generally reflect a level of scientific, financial or technical contributions. In Table \ref{tab:GreyBook} the participation is listed, separated according to country and experiment. 

  \begin{table}[]
      \centering
      \begin{tabular}{l|lcc}
Country	&	Involvement	&	Participants	&	Authors \\
\hline
Argentina	&	ATLAS	&	23	&	5 \\
Brazil	&	ALICE	&	50	&	7 \\
	&	ATLAS	&	67	&	8 \\
	&	CMS	&	70	&	32 \\
	&	LHCb	&	39	&	19 \\
	&	ALPHA	&	5	&	3 \\
	&	ISOLDE	&	12	&	4 \\
	&	AMS	&	2	&	0 \\
	&	ProtoDune	&	6	&	0 \\
	&	RD51	&	5	&	1 \\
Chile	&	ATLAS	&	70	&	9 \\
	&	CLIC	&	2	&	0 \\
	&	ISOLDE	&	5	&	1 \\
	&	NA64	&	7	&	2 \\
	&	SHiP	&	6	&	0 \\
Colombia	&	 ATLAS	&	7	&	2 \\
	&	CMS	&	21	&	6 \\
	&	ISOLDE	&	2	&	2 \\
	&	LHCB	&	3	&	3 \\
	&	RD51	&	4	&	3 \\
Costa Rica	&	ISOLDE	&	1	&	1 \\
Cuba	&	ALICE	&	5	&	1 \\
Ecuador	&	CMS	&	9	&	0 \\
Mexico	&	ALICE	&	82	&	14 \\
	&	CMS	&	41	&	18 \\
	&	NA62	&	5	&	2 \\
Peru	&	ALICE	&	12	&	3 \\
\hline
\hline
LA Region	&	ALICE	&	108	&	29	\\
	&	ATLAS	&	167	&	24	\\
	&	CMS	&	141	&	56	\\
	&	LHCB	&	42	&	22	\\
	&	OTHER	&	103	&	15	\\
	&	\textbf{TOTAL}	&	\textbf{561}	&	\textbf{146}	\\
	\hline
      \end{tabular}
     \caption{Participation of scientists and engineers from institutes in the Latin American region in CERN experiments [Source: CERN Greybook database 
   (as of September, 2020)]
   }
      \label{tab:GreyBook}
  \end{table}


\section{Participation of LA groups in HEP Activities}
\label{ew-bsm:participation}

\subsection{Nuclear Physics}

There are two particle accelerators dedicated to basic nuclear physics in South America: the one at the Open Laboratory for Nuclear Physics (LAFN) at
the Institute of Physics of the University of Sao Paulo (IFUSP) in  Brazil
and another Pelletron tandem, called Tandar, at Commission Nacional de Energia Atomica (CNEA) in Buenos Aires,  Argentina.
 LAFN is equipped with a Tandem Accelerator, Pelletron 8UD, which produces stable beams of energies up to 5 MeVA.
 
\noindent The experimental hall of LAFN has several beamlines with different equipment: large multipurpose scattering chamber, special beam line for applications, Enge Split-Pole magnetic spectrometer, HPGe array with mini-ball for particle detection.
In 2004, the RIBRAS system, acronym standing for Radioactive Ion Beams in Brasil, became operational at LAFN, connected to the 8 MV Pelletron accelerator. It delivers light radioactive ion beams, such as He, Li, Be, Be, B,  produced by transfer reactions, which are separated, and focused by two superconducting solenoids. It is the first such equipment in the Southern Hemisphere and the only one in Latin America.

\noindent LAFN has about 60-70 users including staff members, post-docs, graduate students and external users. The main interest is in nuclear reactions with weakly bound projectiles, stable or radioactive.

\noindent The nuclear physics community in Brazil has 35 institutions, 140 researcher members and  roughly 60 students and post-doctoral fellows working in three large areas:\\
$\bullet$ 62 researchers in High Energy Nuclear Physics (HENP)\\
$\bullet$ 31 researchers in Nuclear Structure and Reactions (NSR)\\
$\bullet$ 47 researchers in Nuclear Physics Applications (NPA)

\noindent In HENP, the activities are\\
{\bf Theory}\\
$\bullet$ Hadron theory, effective models, QCD sum rules, among others. (17 researchers)\\
$\bullet$ Compact stars, EOS with quarks and hadrons, Strong Magnetic field, among others. (14 researchers)\\
$\bullet$ Relativistic heavy ions, hydrodynamics, Quark- Gluon Plasma, among others. (10 researchers)\\
$\bullet$ QCD phenomenology, low x and Color Glass Condensate. Diffractive pp scattering, among others. (9 researchers)\\
$\bullet$ QCD theory, Confinement and Critical Point, Lattice QCD, eq. Dyson-Schwinger, among others. (6 researchers) \\
{\bf Experiment} \\
$\bullet$ Ultrarelativistic heavy ions, hydrodynamics, QGP, etc. (3 researchers)\\
$\bullet$ 2 Researchers at UFRJ of the  Auger Collaboration and CONNIE neutrino experiment 

\noindent There are very few experimentalists working in Hadron Physics in Brazil, and
most LHC and RHIC experimentalists are not part of nuclear physics projects.

\subsection{Jefferson Laboratory}

A  Chilean group participates in the CLAS collaboration in Jefferson Lab. It is composed of two faculty professors, four postdocs and six graduate students. Its contribution has two primary thrusts: analysis of the EG2 dataset for deep inelastic scattering on nuclei, and development of hardware projects for future CLAS experiments.

\noindent The CLAS EG2 dataset is the world's largest sample of deep inelastic scattering on nuclei with identified hadrons. The Chilean analysis activity includes studies of multiplicity ratios, transverse momentum broadening, two-pion Bose-Einstein correlations, and short-range correlations. Recently, for instance, it finalized an analysis of the eta meson multiplicity ratio in two decay modes ($\eta \rightarrow \gamma \gamma$ and $\eta \rightarrow \pi^+ \pi^- \pi^0$). Members of the group have strongly advanced the measurement of the multiplicity ratio for the omega meson. The results are clearly promising for this world’s-first measurement.

\noindent There are four activities for hardware development.  A dual-target has been developed for the future deep inelastic scattering experiment on nuclei with CLAS (E12-06-117) in an experiment that has four co-spokespersons who are members of the Chilean group. This system allows rapid changes of solid targets for fixed-target experiments with a closely proximate cryogenic target. This activity  has included the construction and testing of the final system based. 
The second activity has been the installation and upgrade of the BAND (Backward Angle Neutron Detector), which will be used in the same experiment as well as in experiment E12-11-003A for two-nucleon correlations studies that include detection of backward-angle neutrons. BAND has been collecting data during much of the first operations of CLAS12.
The third activity has been the development of lightweight mirrors for future Ring- Imaging Cherenkov detectors for CLAS, which will benefit the above experiments as well as many others, such as E12-09-007.  This effort has been strengthened by close collaboration with a group developing mirror technology for astronomy. The fourth activity is to develop polarized target technologies to be used in experiment E12-14-001.  

\subsection{LHC-ATLAS}

Nine institutes from Latin America, approximately 50 scientists and engineers from Argentina, Brazil, Chile and Colombia, had  worked  on ATLAS since well before data-taking started, making important technological and software contributions to the instrument. Latin American universities have so far awarded close to 90 PhD and  MSc-level degrees for work on ATLAS, and work is in progress for many more.
Latin American researchers have made significant, and in many cases highly visible, contributions to ATLAS across the spectrum of activities: detector and trigger operations, the data processing chain, software development, reconstruction of important particle signatures, physics analysis, and detector upgrade developments.

\noindent In terms of detector development activities, the ATLAS hadronic calorimeter system has significant contributions from Brazilian institutes. Chile is one of five countries producing muon detector chambers using novel technologies for the upgrade of the ATLAS muon system. Institutes from Argentina and Colombia have made important contributions to the ATLAS trigger and data acquisition system.

\noindent In Argentina, HEP groups have been members of ATLAS since 2006.
Argentina is committed to the ATLAS upgrade for the HL-LHC with contributions
to the trigger electronics, Data Acquisition, High-Level trigger software and hardware, and photon/electron/tau/jet reconstruction.
The responsibilities in ATLAS have included so far
participating in the online testing of data taking, the development of the programs for the
online selection of events, the calibration of detectors, the measurement and improvement of
their performance, as well as the pursuit of multiple lines of physics analysis, several of them
as leaders of the respective analysis teams.

\noindent The main interests have concentrated on event triggering, reconstruction and physics analyses based on photons and/or jets. These have included SM precision measurements, searches for supersymmetry and other BSM models. Several PhD thesis in the group have addressed these analyses.


\noindent The Argentinian institutions are involved in the design, construction and tests of the Global Trigger component of the first part of the trigger, the hardware-based Level-0 with the commitments: \\
$\bullet$  Design, production and testing of Common Modules 
(a common hardware module with
different functionality implemented in firmware, composed of ATCA blades and FPGAs
with many multi-gigabit transceivers); \\
$\bullet$  Development of algorithmic firmware and software; \\
$\bullet$  Design, production, testing; installation and commissioning of Rear Transition Modules (RTM);\\
$\bullet$  Design, production, testing, installation and commissioning of Fibre Management; \\
$\bullet$  Procurement of general infrastructure.

\noindent A new HEP laboratory is being set up at Instituto de Fisica La Plata (IFLP, UNLP-CONICET). The first stage consists mainly of an electronics laboratory to develop and deploy hardware for high speed, real-time signal processing.  The laboratory is being
equipped and is expected to be fully operational by the end of 2020. 
Some of the
equipment that this lab will have when fully operational includes: \\
$\bullet$ FPGA evaluation kits \\
$\bullet$  High-performance computers for firmware synthesis \\
$\bullet$ Soldering station and tools for handling fine pitch PCBs SMD components\\
$\bullet$  High-speed oscilloscopes and high-speed function generators \\
$\bullet$  ATCA shelf \\
$\bullet$  High-speed optic fibers management.

\noindent The
Argentine ATLAS groups are also involved in applying Machine Learning techniques in current and near-future
analyses (for instance, in jet tagging, fast simulation, calorimeter calibration, and more), and
foresee a much heavier involvement and usage in Artificial Intelligence
tools. This requires preparation, both in terms of peripheral computing and software power but also in terms of education and supporting human resources.
However, one difficulty mentioned in the white paper is the current lack of computer resources.

\noindent The  ATLAS/Brazil Cluster is composed of physicists, engineers and computer technicians from UERJ, UFBA,
UFJF, UFRJ, and USP and started research activities at CERN in 1988. The Cluster is involved in both electronic instrumentation and
algorithm development for calorimetry. 
The activities aim at developing the new back-end electronics
(FPGA-based) and new energy estimation algorithms based on deconvolution strategies to be
used both online and offline.  For the
electromagnetic calorimeter (Lar), activities are mainly related to the implementation of the Super Cells,
which refer to a new trigger piece of hardware. Additionally, signal processing and machine
learning techniques are being applied for cross-talk mitigation.

\noindent Additional (level one) trigger from the Tilecal last segmentation layer is currently being provided for the extended barrel part of ATLAS. It may also cover the barrel so that the muon trigger may be assisted from Tilecal information. Still, within Tilecal, an effort is being made to provide finer granularity through a multianode PMT readout. The aim is to achieve two or four times finer detector cell granularity without changing the
mechanical structure of the calorimeter. 
Artificial intelligence and blind source separation
techniques are being used envisioning operation in the second upgrade phase.  For the triggering system, they participate in developing new computational intelligence-based techniques for the high-level calorimeter trigger (HLT) system, which are also to be extended to the offline filtering. Besides, efforts are being made to improve the trigger calibration for electrons.
Possible extensions of the machine learning techniques developed for HLT to level-one triggering (embedded electronics) are also being pursued (FPGA-based designs). Concerning the physics analysis, activities include the search for ALP (Axion Like Particle) and Standard Model analysis.

\noindent The ATLAS management has mainly been structured upon the  Glance/FENCE
technology, which was developed by the ATLAS/Brazil Cluster.
It is worth mentioning that the Brazilian industry has been participating in the
construction of ATLAS, with two electronic circuit boards designed by the Cluster and produced in Brazil, which were both operating in the ATLAS trigger system until the end of the
Run 2 data acquisition in 2018. Finally, the Atlas/Brazil cluster has a strong outreach component.

\noindent In  Chile, there is a group working in three primary analysis.
The first analysis is of the production and suppression of $J/\psi$ mesons and $\psi'$ mesons in collisions between lead ions, with the normalization taken from pp collisions at the equivalent energy.  The second is based on the study of the flow coefficient $v2$ for these mesons.
Both of these groundbreaking studies are of high international interest in the worldwide heavy-ion community. Finally, the third analysis is the search for double Higgs production in the ATLAS data.

\noindent A large, ongoing technical project with ATLAS continues to be the Chilean contribution to the muon system upgrade. This work involves producing more than 140 small-strip Thin Gap Chambers (sTGC) that will be used for triggering and precise measurement of muons. This project, a collaboration between Chile, Israel, China and Canada (with contributions and coordination from Russia and CERN) consists of fabrication and testing activities.

\noindent In  Colombia, the experimental High Energy group of Universidad Antonio Nariño has been part of the ATLAS Collaboration since 2007. In these years, the group has been part of different activities, including
detector performance studies and its calibration, physics analysis as well as activities related to the
ATLAS detector maintenance and operation. The group is interested in searches for BSM physics such as charged Higgs, invisible Higgs, SUSY signatures, heavy neutral leptons and other possible long-lived particles. ATLAS work has focused on the e/gamma trigger, assessment and design of ATLAS Control network and upgrade of TDAQ with network engineers. Also, for the upgrade, there has been work done on the temperature
sensors calibration and readout for the ITk. A new detector laboratory has been built at this institute over the last few years focused on R\&D for SiPMs and MPGDs.

\subsection{LHC-CMS}

In total, 12 groups from Latin America (Colombia, Ecuador, Brazil, Mexico) contribute to CMS with almost 50 scientists, representing roughly 3.5\% of the collaboration. Latin American countries have so far awarded around 60+ PhDs and  MSC-level degrees for working in CMS, and work is in progress for many more. The collaboration with CMS was instrumental in establishing the Master- and the PhD programme in some institutes. There are also many engineering and computing students/specialists that are working in CMS.

\noindent Latin American researchers have made significant, and in many cases highly visible, contributions to CMS across the spectrum of activities: Detector hardware, data taking at CERN, operation, maintenance and upgrade, physics analysis, data processing including validation and certification, computing with the operation of Tier-1, Tier-2 and Tier-3 sites, and central software developments.

\noindent The Latin American countries and institutes contribute crucially to various detector systems in CMS, for example, in the area of the Muon detectors in two different technologies and their upgrades in hardware and especially in electronics. They also contribute to the HCAL (hadronic calorimeter), the beam radiation monitoring system BRIL, as well as to the very forward Precision Proton Spectrometer (PPS).

\noindent In  Brazil, there is a group of roughly 30 researchers from CBPF, UERJ, UFABC and UNESP working in CMS. A Tier 2 of the WorldWide LHC Computing Grid (WLCG)is hosted at UNESP. The group has had a key role in the CMS-TOTEM Precision Proton Spectrometer (CT-PPS). They have also participated in the analysis of monojet events, heavy-ion physics and vector-boson fusion.

\noindent In Colombia, the High Energy Physics group of Universidad de Los Andes (UNIANDES), has been a full member of the CMS collaboration since March 2006, contributing both in the detector operation and physics analysis. 
The group has been involved in the operation and upgrades of the
resistive plate chambers (RPC) and gas electron multiplier detectors (GEMD) used for muon identification and triggering. About 30 engineers from UNIANDES have been part of the international team in charge of
the computing and data operations, through the GRID system, of the CMS experiment. Some of the engineers have been stationed for about two years at CERN, others at FERMILAB. 
The group has also been heavily involved and leading searches of supersymmetric signatures and physics beyond the Standard Model. Few of them focused on the experimentally challenging compressed mass
spectra scenarios, which require special physics tools such as jets from initial-state radiation or vector-boson fusion topology. 
The UNIANDES HEP group has graduated seven PhD and more than a dozen MSc students. As a bridge for technology transfer to Colombia, the group has setup an  instrumentation laboratory in the Bogota university campus, where interdisciplinary applications of semiconductor
and gaseous radiation detectors are being investigated.

\noindent As an example from the white papers, in  Colombia, the group of Phenomenology of Fundamental Interactions of Universidad de Antioquia (UDEA) has
been working for several years in phenomenology. Since 2019 the group is a member of the CMS
Collaboration. They have been working on searches for physics beyond the Standard Model. Currently, three professors, one post-doc, two Phd students, one master student and two undergraduate students, and members of the group are associated to CMS projects.

\noindent In  Ecuador there are groups at Escuela Polit\'ecnica Nacional (EPN) in Quito and Universidad San Francisco de Quito (USFQ) participating in CMS.
Ecuadorian scientists and engineers have contributed to the High Luminosity Trigger, the Beam Radiation Instrumentation and
Monitoring (BRIL) subsystem and are also deeply involved in the data preservation and
open access activities. Several analysis addressing top physics, Higgs physics and exotic
particles searches have had Ecuadorian contributions. The community has of the order of 30 PhDs in HECAP and expressed interest in joining one future collider experiment in the energy frontier, such as the ILC, FCC.

\noindent  In Mexico there is a group of 15 PhDs contributing with M\&O to the CMS experiment. CMS-Mexico is heavily involved in the Resistive Plate Chamber System part of the CMS Muon system and in the Beam Radiation, Instrumentation, and Luminosity (BRIL) system  since (2019).
Universidad Iberoamericana is one of the Institutes responsible for the improved Resistive Plate Chamber (iRPCs) Production for the Phase II Upgrade. Benemérita Universidad Autónoma de Puebla (BUAP) is responsible for the  Detector Control System of the current Resistive Plate Chambers system and the corresponding for the validation of the iRPCs. BUAP, CINVESTAV and Universidad Iberoamericana participated in the build and test of the first prototypes of the iRPCs. At Universidad Iberoamericana in Mexico city a laboratory  for R\&D and for the construction and validation for iRPC’s will be completed by the end of 2020. CINVESTAV  holds a TIER-3 cluster for CMS-GRID, while BUAP is in the process of certification to hold a TIER-3 cluster too. Universidad Iberoamericana is responsible to coordinate the Simulations \& Performance Estimations subgroup of the BRIL Phase-2 Upgrade Group. Universidad de Sonora is involved in the BRIL project taking leading roles in tasks related to the estimation of the radiation environment, luminosity measurements in pp and Pb-Pb collisions, and coordinating the Online Luminometry group for Phase-2 upgrade. Physics analyses where Mexicans are leading and participating include Heavy Flavor Physics, Quarkonia, Higgs, Top, Heavy Ions and Beyond the Standard Model Physics.

\subsection{LHC-LHCb}

\noindent In Brazil, researchers from CBPF, UFRJ and PUC-Rio are members of the LHCb collaboration since 1998. The group today has around 20 faculty.
During its 20-years long participation in LHCb, the Brazilian team gave several crucial contributions, with key roles in the development, installation and commissioning of the Muon System, notably in the readout electronics, multi-wire proportional chamber design and characterization, and experiment control system. The Brazilian groups developed the first muon identification algorithms that played a major role in LHCb flagship measurements. Brazil gave important contributions also in trigger (L0 and HLT), and
computing. Brazilian physicists are contributing substantially to several physics analysis projects. There was a recent review on direct CP violation in beauty and charm hadron decays authored by 2 Brazilian members of the collaboration \cite{Bediaga:2020qxg}.

\noindent The Brazilian team is involved in the development of two new detectors for the LHCb upgrade which
will be installed in the Long Shutdown 2 of LHC in 2019-2021:\\
$\bullet$ the upgrade of the VELO. This new detector features silicon pixel sensors and a state-of-the-art
readout chip with very high data throughput. Sensors and readout ASICS are placed very close
to the LHC beam (~5 mm) and operated in vacuum, in a very high radiation environment. The
group contributes to the test of the pixel sensors, to the design, optimization and tests of the
detector system and to the highly sophisticated motion system; \\
$\bullet$ Scintillating Fibre Tracker (SciFi). This is the largest scintillating fibre tracker ever built and
operated. Brazil is contributing to the development and test of the so-called front-end boards, hosting the ASICS that read out the Silicon Photomultiplier arrays connected to the scintillating fibres.

\noindent Brazil is providing valuable computing resources through a TIER2 for the GRID. The main challenges mentioned are the increasing costs for M\&O and mobility.

\noindent In  Colombia, the  Universidad Nacional de Colombia
(UNAL)  was accepted to participate as an associated group of LPNHE
(Laboratoire de physique nucléaire et de haute énergies) in the LHCb Collaboration in 2014.
The group is composed of two professors, three undergraduate students, four master students and one PhD student. Their members are collaborating in the physics of charm quarks, quarkonia and exotic states analyses.

\subsection{LHC-ALICE}

Scientists from Latin America (from 13 institutions from Brazil, Chile, Colombia, Mexico and Peru) make up 5\% of the collaboration. A major contribution to the ALICE upgrade is the design and production of the SAMPA chip, a dedicated analog/digital ASIC for the readout of the 500k channels of the TPC and 1M channels of the tracking chambers of the Forward Muon Spectrometer, that was developed and funded in Brazil. Another of the key contributions  to the upgrades is a scintillator detector with a high time resolution called V0+, which is part of the Fast Integration Trigger and is developed and funded in Mexico. 

\noindent In Brazil, there is a group of approximately ten physicists from the University of São Paulo (USP), the University of Campinas (UNICAMP), the Federal University of ABC (UFABC), and the Federal University of Rio Grande do Sul (UFRGS) working in ALICE.

\noindent The ALICE groups in Brazil have been providing their fair share contribution for computation to the
experiment since 2007 through the SAMPA cluster (acronym from the Portuguese,
Sistema de Análise e Multi-Processamento Avançado), hosted at USP.
Currently, the cluster has 2408 CPU cores, totalizing 18.7
kHS06 of processing power and 0.85 PB of storage. However, it is important to highlight that part of this processing capability, around 6 kHS06, is used by local users, outside the GRID. And an increase in computing power and data storage will be needed for Run-3.

\noindent For the Run-3 upgrade, the Brazilian groups have provided major contribution through the full development of a new front-end readout SAMPA chip for two ALICE detectors, the Time
Projection Chamber (TPC) and the Muon Chamber (MCH), and the construction of the
mechanical support of the new Muon Forward Tracker (MFT). 

\noindent Recently, the ALICE collaboration proposed an upgrade for the LHC Run-4 
to build a Forward
Calorimeter (FoCal). 
The Brazilian group plans to contribute to the  R\&D and construction of the FoCal Silicon Pad layers readout system. 
Given the large experience of the group with the development of the SAMPA chip, which will instrument the ALICE TPC and MCH
detectors beyond Run-3, it is natural to expect that a significant contribution to this part of the project can be achieved.
They also plan the study of aging
due to the surface chemical reaction
that occurs at the electrodes inside the Gas Electron Multiplier (GEM).

\noindent  In Mexico, there is a group of 15 PhDs contributing with M\&O to the ALICE experiment.
CINVESTAV and Universidad Autónoma de Sinaloa are responsible for the  design, construction and operation of the Forward Diffractive Detector (FFD).
This detector is now in the construction phase.
ICN-UNAM is responsible for the design, construction and operation of a pico-amperemeter for the TPC, in collaboration with BUAP. This employs new technology, and there is already a prototype tested in Mexico and CERN, with the final instrument expected for the end of 2020. In addition, it holds a TIER-2 cluster for ALICE-GRID.
IF-UNAM is responsible for the V0-plus detector. Ten prototypes have been built and tested at IF-UNAM and CERN. The construction of all parts of the detector, which contains 50k optical fibers, was finalized in October 2019 and shipped to CERN for final assembly.
BUAP has also been responsible for the operation of the ACORDE detector and the design and construction of the CTP (Central Trigger Processor) in collaboration with the University of Birmingham. The physics analyses where Mexicans have active visibility are in cosmic rays, diffractive physics, multiplicity measurements and underlying events.

\noindent  In Peru the PUCP (Pontificia Universidad Cat\'olica del Per\'u) started working at ALICE in 2004. Since then, the group has contributed to the development of an analysis tool for the software, detector design, performance studies, and many more. In particular, the group has had a major role in the  design, performance studies, and the test-beam experiments of the ALICE diffractive (AD) detector system. The AD detector increases the ALICE detector's forward physics coverage and its precision on the measurements of the inclusive diffractive cross-sections. Currently, they have a leading role in the analysis aimed to report the ALICE's updated (taking into account the AD detector system) measurements of the diffractive cross-sections.  

\noindent They have had contributed to the characterization tests of the CMOS Monolithic Active Pixel Sensors that are the cornerstone of the Muon-Forward-Tracker (MFT), a Si-tracking detector part of the ALICE upgrade programme planned to be working after the Second Long Shutdown. More recently, and still in the context of the MFT, the PUCP team is contributing in the implementation of the geometry of the MFT within the new ALICE computing model.

\subsection{SuperKEKB}
Mexico requested to join the Belle II experiment in SuperKEKB in 2013. 
There are currently 8 faculty researchers and 6 PhD students from
three institutions: CINVESTAV IPN, UAS and UNAM (IF).
A tier 2 center for Belle II has been online since May 2015. The group is developing electronics for the Large-angle beamstrahlung monitor (LABM).
The group participated in the $\tau \rightarrow 3\pi \nu$ analysis and will  concentrate on the
search for Lepton Flavor Violation in $\tau$ decays.

\subsection{Future Colliders}

Despite the LHC success, there are still many open scientific questions regarding the fine details of the electroweak and strong sectors of the SM and its possible projections into a new, more complete framework for the understanding of the Universe. Therefore, the majority of groups have special interest in continuing their efforts during the period of operation of the upgraded, high-luminosity version of the accelerator, the HL-LHC~\cite{ApollinariG.:2017ojx}. 
The LHC has a clear schedule until 2035.
There is a clear understanding, among Latin American researchers, that the medium and long-term future of particle physics lies mostly on the next generation of colliders, whose construction is currently being discussed at the international, global scale.

\begin{figure}[ht]
\begin{center}
  \includegraphics[width=0.95\textwidth]{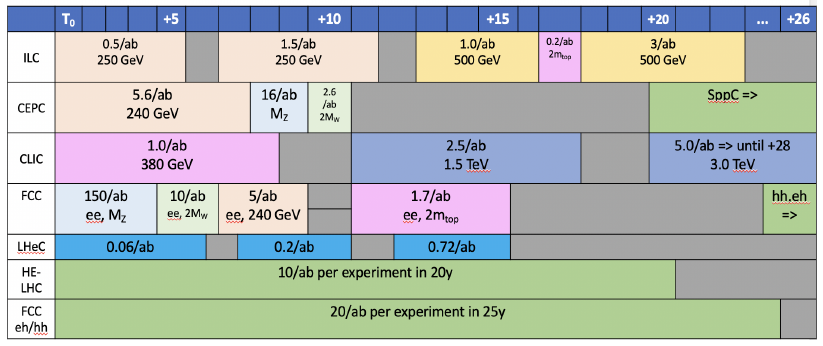}
  \end{center}
  \caption{Timeline (starting at a future time $T_{0}$) of various collider projects.  For details, see reference~\cite{deBlas:2019rxi}.}
  \label{fig:futurecoll}
\end{figure}

\noindent While the HL-LHC is already approved~\cite{hl-lhc-approval}, the next generation of colliders are still under scrutiny.  The reference point for the development of such large scale devices will be, precisely, the HL-LHC.  Latin American groups have expressed their interest in the high-energy version of the LHC (HE-LHC)~\cite{Dainese:2019rgk}, all the possible versions of the Future Circular Collider~\cite{Abada:2019lih} (i.e., ee, eh and hh), the Circular Electron-Positron Collider (CEPC)~\cite{CEPCStudyGroup:2018ghi}, the International Linear Collider~\cite{Fujii:2019zll}, the Compact Linear Collider (CLIC)~\cite{Charles:2018vfv}, and the Large Hadron electron Collider~\cite{AbelleiraFernandez:2012cc} (LHeC or HE-LHeC). A recent reference~\cite{deBlas:2019rxi} presents a summary of the timeline and parameters of these future accelerators (see Figure~\ref{fig:futurecoll}.)

A white paper presented by a group of 16 young experimentalists working in Brazil details their interest in the participation in the next generation of collider experiments.
In particular, they aim at organizing the community of experimental HEP in Brazil to become a member of  only
one large experimental collaboration in the next high-energy collider, joining forces to contribute to the development, construction, commissioning and operations of the detector as a coordinated effort. In particular they would like to avoid very specific short-aim and small-visibility projects and favour longer-term contributions that can make a greater impact within the chosen collaboration and in our research field. They estimate the computing requirements for each possibility of future experiments and foresee difficulties with the funding of such endeavours that should start to be addressed. 

\noindent Argentina ATLAS group has shown interest in FCC$_{hh}$, and Ecuador
also considers joining one next-generation experiment. There is clearly room for a joint LA effort in this direction.

\noindent Here we highlight the opportunities that these large scale experiments would bring about in the context of the current SM and BSM physics interest in the region and their extensions into the future.

\noindent Many groups in Latin America are deeply involved in the pursuit of a deeper understanding in Higgs physics.  Among the Higgs properties that generate greater interest in the region are the couplings to other SM particles (and posibly BSM ones) and to itself. The latter characteristic plays a key role in determining the Higgs potential experimentally, which is of high priority for all future colliders. In the framework of the Effective Field Theory, the most direct way to explore the Higgs cubic self-interaction is through the study of double Higgs production, which has notable interest within the Latin American community.  While these studies will advance greatly during the HL-LHC rule, it will take the next generation of colliders to establish unequivocally the existence of such self-interactions and, even more remarkable, to start probing the realm of quantum corrections to the Higgs potential~\cite{deBlas:2019rxi}.

\noindent Reaching evidence level for two-photon Higgs production in ultraperipheral proton and nuclear collisions could be a possibility at the future HE-LHC.  Observation level of this production mechanism could be reached at the FCC.  This would open and independent area of study (with leading roles in the region) for measurements of the $H-\gamma$ coupling, not based on decays but in production modes~\cite{dEnterria:2019jty},  

\noindent All future colliders will enhance electroweak precision measurements~\cite{Mangano:2016jyj,Contino:2016spe}.  Discrepancies between the electroweak precision observables and its measurements could indicate the presence of new physics.  This modifications can be gauged by the so-called oblique parameters~\cite{Peskin:1991sw}, which is one of the possible topics that can spark interest in the region with great potential for success.

\noindent Opportunities for dedicated flavour physics studies will be dramatically enhanced in the HL-LHC.  With the expected LHCb upgrades, for instance, it will be possible to test, with unprecedented precision, CP-violating phases, transitions like $b\to s\ell^{+}\ell^{-}$ and $b\to d\ell^{+}\ell^{-}$ (in both muon and electron decay) and flavor violation.  The rather new area of charm sector studies will also benefit from the high luminosity version of the LHC and may lead to significant discoveries~\cite{Bediaga:2018lhg}.  In the future, with the enhanced statistics and cleaner environments at the next generation of colliders, there will be great potential to study flavor physics.

\noindent In the same context of flavor physics, groups in Latin America, which are involved in the Belle II  experiment at SuperKEKB, will most likely continue their participation until completion of its research program, whose highlight is the search for lepton flavour violation.  The target integrated luminosity (50 ab$^{-1}$) for Belle II will be accumulated by 2029~\cite{Uno:2020jrc}.

\noindent The study of QCD in very dense and high temperature conditions will also benefit from this new generation of colliders~\cite{Citron:2018lsq}.  The increase in luminosity and energy will allow for stronger collective signatures in QGP, the characterization of dense QCD matter through hard-scattering processes (even Higgs boson production), and the exploration of saturated parton densities in never-seen-before conditions.  

\noindent FCC will give ample opportunities to exploit the regional expertise with jets in order to study parton radiation and fragmentation, jet properties and substructure, heavy quark jets and non-perturbative QCD phenomena~\cite{Anderle:2017qwx}. 
  
\noindent From supersymmetry to dark matter to the search for additional dimensions, the search for physics beyond the standard model has generated great interest in the region.  The increase in energy and luminosity for the next generation of colliders will enable potential discoveries that have been so far denied to us by Nature.   A recent reference~\cite{Golling:2016gvc} presents a thorough study of the wide range of possibilities.  
  
\noindent Lastly, revolutionary technologies, like those used in potential muon-muon colliders to alleviate the problem of energy loss by radiation~\cite{Alexahin:2013vla,Delahaye:2019omf} while keeping the clean conditions of lepton colliders, or the plasma-wakefield methods for particle acceleration~\cite{Shiltsev:2019rfl} could take leading roles in the future of the field.

\subsection{Theory}
We identify a large phenomenology community in Latin America doing excellent work in precision calculations in the SM (for example QCD) and in BSM signatures (especially SUSY, Composite Higgs, Extra Dimensions, Effective Field Theories and 3-3-1 extensions of the SM).

\section{Training, outreach, exchange programmes}
 
 The membership of international experimental collaborations gives access to global networks of universities, laboratories, and eminent scientists and engineers. The structures inherent in these scientific collaborations provide natural opportunities to strengthen regional networks connecting research teams in LA countries, as well as worldwide contacts, mobility and visibility.
 
\noindent Exchange programs for students, scientists and engineers between LA countries and Europe in the past have had very beneficial effects for capacity building. Also, direct access to the experiments where Scientific discoveries are made, and connections with world-class scientists offer a great resource for training and capacity building of higher education in the LA region. Here we list a few examples of educational programs directly linked to the involvement in the HEP community: 
 
 \begin{itemize}
     \item realizations of several MasterClasses in different countries to familiarize high school students with particle physics in the context of CERN's International Particle Physics Outreach Group (IPPOG);
     \item CERN Latin-American Schools in High Energy Physics (CLASHEP) have been occurring regularly every two years in different countries in the region, since 2001;
     \item several activities in HEP held at the ICTP South American Institute for Fundamental Research;
    \item contribution to capacity building through education programs such as LA-CoNGA. 
 \end{itemize}

\section{Areas of excellence and leadership}

\noindent We identified a few examples of excellence and leadership in the region. An incomplete list is:

\begin{itemize}
    \item  Diffractive physics where UERJ CMS group in Brazil plays a leading role
in CT-PPS detector and physics;
\item Analysis of three-body decays of hadrons and CP violation in LHCb;
\item Precision calculation of Higgs boson cross-section 
\item Software development for triggers 
\item HPC with several Tier 2 grids in the region
\end{itemize}
\noindent In addition, the participation of LA groups in these cutting-edge experiments led to the development of important capabilities in the region. Among these capabilities, we highlight a few:
\begin{itemize}
    \item design of a chip called SAMPA in Brazil for readout of ALICE detectors to be installed for run 3;
    \item design and test of Global Trigger and Rear Transition Modules for ATLAS in Argentina;
    \item Electronics for the Large Angle Beamsstrahlung Monitor for Belle II developed in Mexico;
    \item testing of Low Gain Avalanche Detectors for ATLAS in Brazil;
    \item expertise and leading roles in several experimental analysis in all collaborations; 
    \item several computer clusters dedicated to these experiments are installed in the region;
    \item expertise in theoretical calculations in most aspects of SM and BSM processes.
\end{itemize}

\section{Synergies}


The SM provides a framework that has close connections to understanding the Universe at large scales, as studied in astronomy and cosmology. The search for Dark Matter is an example of such a synergy. Neutrino physics is deeply intertwined with the electroweak force, flavour physics and arguably may provide a window on BSM processes and a possible origin for the CP violation necessary to explain baryogenesis.

On the experimental side, many developments in the region, such as electronic readout capabilities, the construction of RPCs and other detector components, and software development for triggering, are common to many experiments.
In addition, several analysis topics have overlap among the different groups in different experiments. An example that stands out is diffractive physics, with strong groups in CMS-Brazil and ALICE-Mexico. Certainly, greater coordination among the different groups can be achieved.
 
\section{Conclusions}

Latin America has a large community of approximately 500 physicists working in High Energy Physics. Most of the community is involved in the LHC experiments, and several impactful contributions to detector design and construction have been performed. Participation in the LHC experiments provides opportunities
such as:

\begin{itemize}
   \item Sophisticated radiation-hard and high-speed electronics development;
\item Firmware development for state-of-the-art FPGAs with advanced tools;
\item Detector development including high-granularity silicon detectors, and ultra-fast precision timing detector technologies;
\item High data volume, and high data rate distributed computing infrastructures;
\item Advanced computational techniques, such as machine learning;
\item And, very importantly, training of young scientists and engineers in the high-
technology sector.
\end{itemize}

\noindent In 10 years of LHC operation we have had a rich harvest of physics results, documented in over 3000 LHC publications \cite{Mangano:2020icy}. So far only 5\% has been collected of the total data expected by the end of the HL-LHC for ATLAS and CMS. Additionally, ambitious upgrades of ATLAS and CMS will significantly improve the detectors beyond their current performance, and even more exciting prospects exist for LHCb and ALICE. A rich physics program until 2035 is therefore guaranteed, and the ongoing detector upgrade projects offer outstanding opportunities for the LHC groups in LA countries to participate in a more coordinated manner in beyond-the-state-of-the-art detector and electronics R\&D and production, and for new teams to get involved. In addition, it is the right time to start thinking  strategically about participation in the next generation of colliders.

\noindent Latin America has a leading role in some aspects of HEP and synergies with other activities in HECAP were identified. Moreover, HEP has provided important capacity building in the region through training and outreach programs.




\chapter{Instrumentation and Computing}\label{chapt:instru}
\vspace{1cm}
\noindent 	Hiroaki Aihara (U. of Tokyo, Japan)\\
Reina Camacho Toro (LPNHE/CNRS, France)\\
		Arturo S\'{a}nchez Pineda (ICTP/INFN/ U. of Udine, Italy)\\
		Hernan Wahlberg (U de la Plata, Argentina)\\




		

		







\section{Introduction}\label{chapt:intro-adsc}
The scientific goals outlined in the previous chapters pose unprecedented experimental challenges. The unique nature and magnitude of these challenges drives technological innovation and development of new instrumentation and efficient high-performance computing solutions to meet unprecedented demands. Participation in global international collaborative efforts to develop such new technologies and exposure to state-of-the-art computing infrastructure and expertise offers great opportunities, as well as certain challenges, in some cases unique to the Latin American region.

This chapter presents a summary of instrumentation and computing related submissions to LASF4RI-HECAP. Both the state-of-the-art and challenges for the main technologies and projects are highlighted. Supporting instrumentation and computing research and development (R\&D) ecosystems is essential to attain the physics goals of the high energy physics, cosmology and astroparticle communities, as well as to build a relationship with industry, creating practical applications and transferring them to society.

\noindent The structure of this chapter is as follows. First, the main scientific key questions and non-scientific associated drivers are outlined. Second, common developments in the region are identified  and described, emphasizing  existing collaborations, their current status and technical challenges. Third, synergies and opportunities for are presented. Fourth, the topics of capacity building and capacity keeping are discussed. Finally, conclusions are presented.\bigskip

\subsection{Main key scientific questions and highlights}

\noindent In the context of \textbf{accelerator science and technology}, the key scientific questions identified are the following:

\begin{itemize}
  \item How to achieve proper complementarity for the high-intensity frontier vs the high-energy frontier?
  \item What are the options and challenges for accelerator technology?
\end{itemize}



\noindent For groups involved in collider experiments, successful completion of High-Luminosity Large Hadron Collider (HL-LHC) upgrade remains a priority (see also Sections~\ref{ew-bsm:participation}). Institutions already participating in LHC experiments in Argentina, Brazil, Chile, Colombia, Cuba, Ecuador, Mexico and Peru expressed interest in continuing doing so in the HL-LHC concerning instrumentation and physics prospects studies. The situation is similar for groups in Brazil, Colombia, Paraguay and Peru participating in accelerator-based neutrino experiments. The groups closely follow the global/international discussions regarding future colliders: CLIC, ILC, HE-LHC, FCC-ee, FCC-hh and CEPC [I-2,I-3]. A detailed summary of the expected performance of the future colliders (in answer to the scientific drivers presented above) can be found in the Physics Briefing Book of the European Strategy for Particle Physics Update 2020~\cite{Ellis:2691414}.\bigskip


\noindent In the last few decades, the Latin American experimental community has grown considerably. There is a large diversity of activities ongoing in the region covering collider experiments, neutrino physics, nuclear physics, cosmology, astrophysics, astroparticles, gravitational waves and concrete capacity-building collaborations. The community's contributions to ambitious programs -- like LHC experiments, the Pierre Auger Observatory, LAGO and GRID computing systems, among others -- demonstrate its collective ability to engage in multinational and multidisciplinary endeavours effectively.\bigskip

\noindent It is worth mentioning the existence of facilities in the region which could complement large scale projects for accelerators and experiments in particle, cosmology and astroparticle physics. For instance, the Brazilian Sirius Synchrotron Light Source and two particle accelerators dedicated to fundamental nuclear physics: one at the LAFN at the Institute of Physics of the University of Sao Paulo [I-4], and another called Tandar, at CNEA in Buenos Aires, Argentina. A strong collaboration with these technological facilities could be seminal for the realization of several scientific projects in the accelerator context in the region.\bigskip

\noindent Regarding \textbf{instrumentation}, the scientific key questions identified are the following:

\begin{itemize}
  \item What areas of instrumentation R\&D are ongoing in the region? And how do they meet the needs of future experimental programs? 
  \item Are there possibilities of collaborations to boost the efforts?
\end{itemize}

\noindent Next-generation experiments include those at colliders, fixed-target and beam-dump facilities, as well as dedicated projects for the search of elusive particles. Some of the proposed programmes exploit synergies between particle physics, astroparticle physics and cosmology (i.e. CTA or SAGO). Depending on the physics goals and experimental conditions, the technological challenges are varied and include: improved spatial resolutions (few $\mu$m per hit, low mass) and time resolutions (down to 10 ps per hit), high-performance photo-detectors, very high tolerance to radiation, extensive area coverage at low costs, a large number of channels, very high readout speed, enhanced performance for reconstruction and triggering, low threshold acquisition systems with low readout noise, increased exposure and/or single photon-electron resolution. \bigskip 

\noindent Regarding \textbf{computing and software} in high energy, cosmology and astroparticle physics, the landscape is shifting rapidly. New projects need to consider not only detectors and R\&D costs but also computing innovations, challenges and costs. Computing resources are nowadays becoming increasingly heterogeneous. Chip architectures and associated software and networking are under constant review and development for proper adaptation and optimizations.\bigskip

\noindent Long-term computing projects are moving more and more towards open source and open data models, and hybrid combinations of public and private (academic + volunteer + commercial) resources are being explored. No less relevant, the data management and preservation processes, public repositories, and protocols need substantial redefinition for future experiments, whiles keeping in mind the budget limitations, as also expressed in the updated European Strategy for Particle Physics Update~\cite{Ellis:2691414}.\bigskip

\noindent Taking those factors into account, the scientific key questions identified in the computing and software context are:
\begin{itemize}
  \item How the scientific software and data structure must be designed to take the best advantage of the current resources in the different institutions in the region, and associated partners in other places of the world? 
  \item How should computing evolve to support future scientific programs and their specific needs?
  \item How can current commercial solutions be a relevant player in more cost-effective usage of the funds allocated to computing power while keeping a pertinent role in the operations?
  \item What are the R\&D activities ongoing in the region? And how to boost them to enable the computing evolution while ensuring computation and science reproducibility as well as open access principles?
\end{itemize}

\noindent To try to address the identified scientific key questions, summaries of the technologies and their possible specific applications in the region (from the inputs received) are shown in sections~\ref{chapt:common} and \ref{chapt:computing}. These summaries highlight the diversity of  ongoing projects, as well as the need for coordinating bodies/panels in the region to maximize the scientific outcomes of these activities and make the most efficient use of resources. Such bodies/panels can oversee the creation of coherent strategies and collaboration in the region; opportunities for the development of, and access to common tools, infrastructures and services; development, deployment and access to an extensive network of information and worldwide expertise in different areas, wide dissemination of knowledge and results and a visible framework for institutions.\bigskip

\noindent The \textbf{timescales spanned by ongoing and future projects, range from a few years to decades}. There are short-term projects (< 5 years) like phase-I CMS, LHCb and ATLAS upgrades, ASTRID-mini array, AugerPrime and Reactor Neutrino experiments; and long-term (>5 years) experiments like CTA, DUNE, HL-LHC and the American Gravitational Wave Observatory, and big facilities like ANDES. In addition to the complexity and diversity of the required technologies, these timescales also constitute a challenge. The challenge is related to having funding stability over a long period of time and creating regional programs that facilitate regional collaborations.\bigskip


\subsection{Non-scientific drivers}

\noindent Non-scientific drivers are also important in this context. They include not only financial support, as stated before, but also:
\begin{itemize}
  \item Access to shared infrastructures and tools. Considering the regional landscape is essential, i.e. e-infrastructure and internet performance vary a lot in the region.
  \item The existence of effective collaborations and organizational networks, regional and international. It is important to build meaningful relations, and synergies with related strategy initiatives, like Snowmass\footnote{Snowmass 2021: \href{https://snowmass21.org}{https://snowmass21.org}} and the European Particle Physics Strategy\footnote{European Particle Physics Strategy: \href{https://europeanstrategy.cern}{https://europeanstrategy.cern}}.
  \item The improvement of existing structures and the creation of new mechanisms through which the community can build relationships with the industry—also leveraging the career opportunities of those interdisciplinary associations in both directions.
  \item Environmental and social impact: knowledge transfer, outreach \& communication, and the ecological impact of our activities.
  \item Training of the next generation of experts while understanding the natural flow of human resources between the scientific and private sectors.
  \item Appropriate support and recognition of the workforce engaged in accelerators, instrumentation and computing R\&D activities.
\end{itemize}
The last two points are particularly crucial in the context of Latin America. To effectively carry out the projects submitted to LASF4RI and to increase the visibility of the teams in the region, we need: 1) more skilled physicists, engineers and computer scientists, and 2) significant investment. The last section will be devoted to addressing this. Finally, it is crucial to keep in mind that planning for the next years implies considering the impact of unpredictable events (like the current pandemic) on the timelines of the activities and the institutions involved.\bigskip


\section{Topics within similar instrumentation drivers}\label{chapt:common}
In this section common developments are identified in terms of dedicated devices and technologies that are being implemented as part of different detectors, observatories and experiments.\bigskip

\subsection{FPGA Boards}
The recent developments in FPGA processor power and I/O bandwidths are enabling complex online feature extraction and increased readout flexibility. This also requires firmware development with increased complexity that demands higher level of expertise.

\noindent There is a general approach to use FPGAs boards, with projects in Argentina and Brazil, for a fast trigger decision and selection algorithms for the first level triggers, and back-end electronics of CMS [I-20] and ATLAS [I-2,I-19]. This approach is based on Calorimeter and Muon system information and includes efforts to implement challenging deep neural networks firmware algorithms. These projects include the development of new dedicated electronic laboratories in the region. The projects should be ready for the HL-LHC, which will start delivering collisions in 2026. 

\noindent FPGA Boards are the core of different general readout systems, which are included in the neutrino reactor instrumentation for the $\nu$-Angra [I-14] experiment in Brazil and in the new Skipper CCDs technologies [I-9,I-15] for a Low Threshold Acquisition Systems for Dark Matter searches [I-22] to be developed in Argentina.\bigskip

\subsection{Read Out systems}
In many projects that are proposed for the near future, which are being built now or are going through major upgrades, the readout systems are being implemented using new concepts for bridging the streams of data between various detector electronics and the endpoints of the data acquisition network. For the Run-3 upgrade of the ALICE experiment, groups in Brazil have greatly contributed with the full development of a new front-end readout chip (knwon as the SAMPA chip) for two ALICE detectors, the Time Projection Chamber (TPC) and the Muon Chamber (MCH), and the construction of the mechanical support of the new Muon Forward Tracker (MFT) [I-10].

\noindent In the context of the ALICE Upgrade for LHC Run-4, the FoCal is a high-granularity calorimeter located in the very high rapidity region of the experiment, adding new capabilities. The planned contribution in this project is the R\&D and construction of the FoCal Silicon Pad layers readout system. Given the large experience of the group with the development of the SAMPA chip, which will instrument the ALICE TPC and MCH detectors beyond Run-3, it is natural to expect that a significant contribution to this part of the project can be achieved [I-10]. Brazilian participation in the ALICE experiment for the next 6 or 7 years includes projects for the LHC Run 3, from May 2021 to the end of 2024, and the preparation for the LHC Run 4 with upgraded components to be installed during the Long Shutdown 3, from 2025 to mid-2027. 

\noindent Institutions in Colombia, collaborating with groups in Peru and Paraguay, are working directly on the design of the digitization boards of the DUNE experiment [I-1]. To read the SiPMs signals, digitization at room temperature will be performed by electronic boards, currently under design. The boards are known as DAPHNE (Detector Electronic for Acquiring Photons from Neutrinos) and will perform the digitization, initial processing and communication. Following this work at the beginning of the year 2020, the production of a small number of prototype boards is planned. The prototypes will be tested at different facilities around the world [I-7].

\noindent Different groups in Argentina have been involved in the development of the first and only existing electronics for the Skipper-CCD detectors [I-22]. The good performance has encouraged its use in more than 10 institutions in different countries. The newly developed readout system provides unique capabilities for low noise applications as required for the Skipper-CCD sensor.

\subsection{Small-area Photomultipliers (sPMTs)}
Small-area photomultipliers are designed to have a better single photon-electron resolution, with the ability to work in a "photon-counting mode".

The JUNO experiment consists of a set of 25000 of these photomultipliers in the so-called “counting mode". The Latin American activities, based in Brazil and Chile, are focused on the sPMT system with both hardware and software contributions [I-12]. The JUNO detector construction is well under way and test runs are expected to start in June 2022. Concerning the sPMT subsystem, 26000 sPMTs have already been tested and accepted well within schedule.

The Pierre Auger Observatory (PAO), located in Argentina, is going through a major Upgrade (AugerPrime). It includes the addition of a fourth SPMT in each Water Cherekov Detector (WCD), which will drastically reduce the number of events with saturated signals [I-17]. Brazilian groups will share efforts that involve the acquisition, characterization and assembly of the equipment in the PAO site in Argentina. They also foresee to operate these devices at least until the end of 2025.

\subsection{Silicon Photomultipliers (SiPMs)}
The Silicon Photomultiplier is a sensor that addresses the challenge of collecting, timing and quantifying low-light signals down to the single-photon level. It combines low-light detection capabilities with the benefits of a solid-state sensor.

\noindent For DUNE and SBND experiments, groups in Brazil, Colombia and Paraguay  [I-7,I-11] are collaborating on projects of cold electronics for the amplification of the SiPM and digitization of the signals with the DAQ system. To increase the effective collection area of SiPMs, the ARAPUCA concept has been developed in Brazil, based on the shifting and trapping of scintillation light in noble liquids.

\noindent The ASTRI MINI-ARRAY telescopes are characterized by an optical system based on a dual-mirror Schwarzschild-Couder design and a camera at the focal plane composed of silicon photomultiplier sensors managed by fast read-out electronics of custom design [I-5, I-6, I-24].  The ASTRI MINI-ARRAY is the precursor of CTA to be installed in Tenerife and with start-up date set to 2022. Brazilian engineers have been involved in the development of the prototype as well as in the construction of the camera. They will participate in the final testing of the telescope and foresee future manufacturing in Brazil of other telescopes for the CTA-South Array. The first telescopes are expected to be installed on site in 2020 in the north and in 2022 in the south. The completion of the first phase of operations, with a subset of telescopes, and the  start of a second phase, for the installation of the full array, is planned for the beginning of 2025.\bigskip

\subsection{Charge-Coupled Devices (CCDs and Skipper CCDs)}
CCDs and the new Skipper version are semiconductor device sensors that can achieve deep sub-electron readout noise levels and single electron counting capabilities.

\noindent There is a proposal to upgrade the CONNIE experiment with Skipper CCDs, which will allow it to achieve lower detection thresholds and greatly increase its sensitivity [I-9]. The plan is to install these new devices in 2020 in order to study their performance as low-energy neutrino coherent scattering detectors, as well as to characterise the backgrounds and perform noise and stability studies. Based on the results of this pilot program, further plans will be done on how to size up the detector.

\noindent There is a wider effort in Latin America to build the next big reactor neutrino experiment using Skipper CCDs [I-15]. One strong contender for the experiment site is the Atucha reactor in Argentina, which may allow placing the detector at a minimal distance of 12 m from the reactor core, inside its dome, thus increasing the neutrino flux and also profiting from the concrete shielding from the dome to decrease the cosmic muon background [I-9].

\noindent Efforts are ongoing to develop a new Smart Skipper-CCD camera to install it in new instruments [I-22]. This proposal seeks to develop technology and scientific forecasts for next-generation cosmic surveys probing the ‘dark sector’. Specific aims are: (1) to develop faster readout strategies for low-noise Skipper CCDs, (2) to characterize the optical performance of these detectors for their use in quantum imaging applications and cosmic surveys, (3) to demonstrate the first implementation of Skipper CCDs on a prototype astronomical instrument, and (4) to assess the sensitivity of future cosmic surveys to fundamental properties of dark matter. A 5-year plan has been developed to achieve these goals.

\noindent It is also worth mentioning that the TOROS experiment aims to perform optical follow-ups of gravitational-wave events conducted by unfiltered CCD observations.\bigskip

\subsection{Resistive Plate Chambers (RPC)}
Resistive Plate Chambers are gaseous parallel-plate detectors that combine precise spatial resolution with a good time resolution. They are, therefore, well suited for fast space-time particle tracking, in particular for muons.\bigskip

\noindent As part of the PAO upgrade that is currently ongoing, it is expected that the RPC assembled in Brazil [I-17] will be tested and operated on-site in the coming years.\bigskip



\noindent The Muon Upgrade project of CMS is an international effort with important contributions from institutions in Brazil, Colombia and Mexico. The upgrade includes the installation of a new Link System for the RPC system and new chambers in the high eta regions [I-1, I-20]. This project will be finished with the installation of the new RPC chambers and electronics during the Yearly Technical Stops at the end of 2022 and 2023. The collaboration, with participation of institutions in Brazil, proposes also updating the readout software for Run-3, and the data certification using Machine Learning (ML) techniques to evaluate performance.

\subsection{ARAPUCA Light Trap (Argon R\&D Advanced Program at UNICAMP)}
Liquid Argon Time Projection Chambers are a choice for the next generation of large neutrino detectors due to their optimal performance in particle tracking and calorimetry. ARAPUCA offers a light trap device to enhance Ar scintillation light collection (and thus the overall performance) based on a suitable combination of dichroic filters and wavelength shifters.

\noindent The far detector of the DUNE experiment will consist of a liquid Argon Time Projection Chamber (TPC). Dimmer light, coming from the most remote regions of the TPC, needs to be collected. For that reason, a large photon collection is required at the far end, made possible through the use of a new technology developed in Brazil [I-7,I-13]: the ARAPUCA concept.

\noindent The next-generation X-ARAPUCA concept currently represents the baseline choice for the photon detection system of the DUNE experiment. The X-ARAPUCA is a development of the traditional ARAPUCA. This concept was conceived to reduce losses on the internal surfaces of the ARAPUCA by diminishing the average number of reflections before detection [I-13].
These projects will follow the DUNE timeline and aim to be ready for operations starting in 2026, while planning for 10 years data taking in order to achieve ambitious physics results.

\subsection{Water Cherenkov Detectors}
Water Cherenkov Detectors are based on a very well established technique in which Cherenkov radiation, generated by charged particles going through water, is detected using photomultiplier tubes. The technique provides the possibility to cover huge areas of target mass at reasonable costs. Many countries are working on the development, construction and operation of Water Cherenkov detectors.

\noindent At present, for the PAO and for different sites for the LAGO project [I-1,I-17], work is ongoing to incorporate new developments of electronic readout, DAQ custom systems, and the modelling and simulation of signals.\bigskip

\noindent To be constructed in the Southern Hemisphere, the Southern Wide-field-of-view Gamma-ray Observatory (SWGO) is a next-generation WCD instrument with sensitivity to the very-high-energy band [I-23]. It will provide a unique view on gamma-ray and cosmic-ray emission from tens of GeV to hundreds of TeV. The facility will improve upon the success of the HAWC Gamma-ray Observatory in Mexico, which is surveying the northern gamma-ray sky. The recently formed SWGO Collaboration will conduct site selection and detector optimization studies prior to construction, with full operations foreseen to begin in 2026.\bigskip

\noindent The Neutrino Angra Experiment is a water-based Cherenkov detector located and oparational in the Angra dos Reis nuclear power plant [I-14]. In addition to the detector itself, a complete data acquisition system has been designed and integrated for the experiment. The long term goal of the experiment is to develop a reliable and cost-effective technology to routinely monitor the nuclear reactor power and possibly the neutrino spectral evolution.

\subsection{Laser Interferometer}
Laser interferometers detect gravitational waves that extend and contract the distance between mirrors.

After 2030, two new gravitational wave observatories will likely step in: the Einstein Telescope and the Cosmic Explorer. Both will be based on Laser Interferometer technology. They will be the first ones of the so-called third-generation (3G). They will need a third partner for triangulation of arrival times to determine more precisely the source position in the sky. Therefore, there will be an opportunity for the construction of a South American Gravitational wave Observatory (SAGO), if the Latin American community reaches, by that time, a critical mass of experimentalists already educated in the 3G technology.\par
\noindent The main goal is the construction and the subsequent operation of a 3G laser interferometer for gravitational wave observation in South America [16].  Currently, only two group in Latin America are involved in the LIGO Scientific Collaboration and one in Virgo. All three groups are in Brazil. And there are only a few Latin Americans involved in laser interferometer projects. Therefore, the challenges to overcome in order to propose a project for the construction and development of a gravitational wave observatory in South America are enormous.\bigskip


\section{Computing and software}
\label{chapt:computing}
Worldwide, the computing requirements of large experiments have been in a state of constant evolution. Innovative and flexible computing solutions and services have been developed to meet the numerous challenges of multiple experiments, use cases and scientific disciplines.

\noindent Regarding that evolution, several elements stand out: the hardware architectures and dispositions, the complexity and sizes of the datasets produced and the distributed nature of the researchers, engineers and students that form the human resource in those experiments.

\subsection{General remarks}
On the side of hardware -- computer processing power, networks, storage elements and so on -- one successful approach is the integration of already existing infrastructure to develop and consolidate linked macro-structures that perform as a single entity. Initiatives like the Worldwide LHC Computing Grid (WLCG)\footnote{Worldwide LHC Computing Grid (WLCG): \href{https://wlcg.web.cern.ch/}{https://wlcg.web.cern.ch/}} demonstrates how a very heterogeneous set of computer centres, universities and dedicated High-Performance Computing (HPC) facilities can join resources (including human expertise) to consolidate a global network. Such a system is currently serving multiple LHC and other experiments. More recently it was even used to collaborate in the relevant initiative far away from the HEP, like Folding@home\footnote{Folding@home: \href{https://foldingathome.org}{https://foldingathome.org}}, which is a distributed computing project to carry out complex simulations of protein molecules, how they fold, and how their movements are implicated in a variety of diseases. 

\noindent On the software side, constant changes and bigger collaborations have emerged inside the global scientific community. The sophistication needed to carry out the data acquisition, data handling and, of course, the data analysis of those datasets collected (and simulated) by large experiments forces us to develop a more global approach, keeping close contact with the appropriate enterprises, protocols and industrial standards.  In this context, scientific software and computational tools have earned a place by themselves as fields of study and careers to pursue.

\noindent The often-heard statement: “bad software is extremely expensive”, refers to the fact that researchers have been encouraged to enhance their computing and software skills so that their analysis tasks take good advantage of the new and legacy hardware. Also, by using leading programming languages and techniques available, researchers can optimize running time and CPU usage. The lack of such enhancements and optimizations could have a negative impact on the overall performance of the system and consequently on the ratio of results/resources or, on a more pragmatic view: result/budget.\par


\noindent In the Latin American region, synergies and collaborations have emerged thanks to the participation of multiple institutions in multinational experiments and organisations. This is intimately related to the capacity building that we will address in the next section.

\noindent Nowadays, ongoing scientific ventures and any new project need to consider the computational aspects as part of their R\&D initiatives.

\subsection{Large collaboration examples in the region}
A couple of concrete efforts shown at the LASF4RI-HECAP workshop come from Brazil:
\begin{itemize}
  \item At the CONNIE experiment, data is currently stored, processed and analysed in the CHE cluster at CBPF, which is the only machine allowed by Electronuclear to connect directly to the laboratory hosting CONNIE at Angra 2. The cluster has a professional storage system with currently 100TB dedicated to CONNIE and 280 cores available for CONNIE processing [I-9].
  \item In ALICE: several groups have been providing its fair share contribution to the experiment since 2007 through the SAMPA cluster (acronym from the Portuguese, Sistema de Analise e Multi-Processamento Avancado), hosted in the Physics Institute of Universidade de São Paulo. Currently, the cluster has 2408 CPU cores, totaling 18.7 kHS06 of processing power and 0.85 PB of storage. It is important to highlight that part of this processing capability, around six kHS06, benefits local users, outside the GRID [I-10].
\end{itemize}

\subsection{Training and knowledge transfer efforts}

More events, schools, workshops and trainee ships in the experiments looking to consolidate the flow of knowledge and expertise between students, scientists and industries are needed. They should address how to develop and deliver the right software infrastructure and tools, as well as how to effectively integrate experts and new members to the development of such common tools and protocols. An emblematic case is the HEP Software Foundation (HSF)\footnote{HEP Software Foundation (HSF): \href{https://hepsoftwarefoundation.org}{https://hepsoftwarefoundation.org}} that looks to facilitate cooperation and common efforts in High Energy Physics software and computing internationally.

\noindent Two relevant examples in the Latin American region that took place in 2019 are: the Latin American Workshop on Software and Computing (S\&C) Challenges in High-Energy Particle Physics (LAWSCHEP)\footnote{Latin American Workshop on Software and Computing (S\&C) Challenges in High-Energy Particle Physics (LAWSCHEP): \href{https://indico.cern.ch/event/813325/}{https://indico.cern.ch/event/813325/}} and the Symposium Artificial Intelligence for Science, Industry and Society (AISIS)\footnote{Symposium Artificial Intelligence for Science, Industry and Society (AISIS): \href{https://indico.cern.ch/event/781223/}{https://indico.cern.ch/event/781223/}} in Mexico. LAWSCHEP -under the auspices of the HSF and with support from the Institute for Research \& Innovation in Software for HEP (IRIS-HEP)- brought together the Latin American HEP groups to discuss opportunities for collaboration in topics of S\&C within the region and with the international community at large. In addition, AISIS looked to bringing together and creating synergies between different scientific research and industries using new S\&C technologies, including ML and other Artificial Intelligence (AI) branches, in fundamental research, ecology and also social disciplines.\bigskip

\noindent In general, the submitted proposals show that we have already multiple international collaborations working in the development of cutting-edge hardware, computing networks and sophisticated software. They also show the need to encourage communication between Latin American countries to strengthen collaborations in current and new projects. 
The next step is to wrap-up those efforts with a constant and coordinated flow of high-level training. This is essential in order to reach these goals of long-term scientific production and qualified human capital, which can later on integrate other scientific and economic activities in the region.

\section{Synergies with other chapters/scientific topics}\label{chapt:synergies}

Instrumentation and computing development for high energy, cosmology and astroparticle physics is both a driver for and a beneficiary of progress made within the topical sections presented before in this document. Some examples are: the development of silicon photomultiplier for neutrino experiments, pure water Cherenkov with photomultipliers and CCDs and Skipper CCDs.

A diversity of other areas benefit from and synergistically contribute to the advances in accelerators, instrumentation and computing. Some of the most significant examples mentioned in the inputs received are:

\begin{itemize}
\item Computer science: collaboration in the use of cloud computing, big data and external storage services. Usage and development of Machine Learning/Artificial intelligence [I-1, I-2, I-21].
\item Sharing of “generic” resources and trainings, including computing, software development and management, consultancy with external parties, like companies, is important for the computing and software development side [I-21].
\item The UniANDES group in Colombia participates in interdisciplinary applications of semiconductor and gaseous radiation detectors [I-1].
\item Geophysics: the work done by the UIS group in Colombia on cosmic ray detection has led to other projects such as the Muon
Telescope (MuTe) for muon volcanography [I-1].
\item Citizen science: the Red Ambiental CIudadana de MOnitoreo (RACIMO) initiative is a citizen network for environmental monitoring from the UIS group in Colombia [I-1].
\item Space weather: Paraguayan groups started working with space weather due to its location at the center of the South Atlantic Magnetic Anomaly. They are involved in the installation of magnetometers in order to be part of the EMBRACE net, which measures the magnetic field in South America and of a new muon detector that will help measuring not only the flux, but the incoming direction as well [I-11].
\item Nuclear physics: Argentina and its series of dedicated reactors and the collaboration between them due to a single overseeing agency.
\item Other known synergies were not specifically addressed in the inputs received but they are worth mentioning: medical physics, biology, photonics and neutronics, large civil, vacuum and cryogenic infrastructures, among others.

\end{itemize}

\noindent An important aspect of this exercise is to identify these synergies and support/encourage them through the  funding of interdisciplinary research. The joint developments with applied fields in academia and industry are also essential for advances in accelerators, instrumentation and computing. 


\section{Developing and preserving knowledge and expertise}\label{chapt:training}
Worldwide, one of the challenges faced by the particle physics and cosmology communities is to ensure an adequate level of development and preservation of expertise in instrumentation and computing-related R\&D activities. In Latin America, the situation is not different and it is deepened by episodic funding, and subcritical mass in instrumentation- and computing-trained human resources in the individual institutions.

\noindent Some efforts to develop and retain expertise in the region are ongoing. At the training level, enhancing fundamental physics, instrumentation and computing training at the university level by multiple members of the Latin American community (e.g. CEVALE2VE~\cite{SanchezPineda:2241903}\footnote{CEVALE2VE: \href{http://www.cevale2ve.org}{http://www.cevale2ve.org}}, LA-CoNGA physics [I-8]\footnote{Latin American Alliance for Capacity buildiNG in Advanced
Physics (LA-CoNGA physics): \href{https://laconga.redclara.net}{https://laconga.redclara.net}} and investing in specialized schools (e.g. CERN-Latinamerican school\footnote{CERN-Latinamerican school: \href{https://www.symmetrymagazine.org/article/building-the-future-two-weeks-at-a-time}{https://www.symmetrymagazine.org/article/building-the-future-two-weeks-at-a-time}}) is highly beneficial.

\noindent However, available economic and human resources mean that universities are  typically limited. Hence, creating regional training programs with clusters of institutions participating is an appealing option (e.g. LA-CoNGA physics [I-8], PPGCosmo experience [I-31]). Pursuing these ideas will need investment in TechEd and digital education skills in the region. It will be a great opportunity to invite other countries in the region not currently actively working in the field (and not in the LASF4RI-HECAP network yet) to include such topics in their university programs.

\noindent Regarding expertise preservation and career opportunities, even the visibility given by academia often cannot compensate for the higher salaries and attractive challenges in industry for physicists and engineers with instrumentation and/or computing skills. Ways for academic institutions to provide stable employment and more competitive wages for these crucial professionals must be found with specialized and attractive grants, as well as career prizes for instrumentation and computing R\&D. Exploiting synergies with engineering and computing science departments is also important (as stated in several of the inputs received).
Efforts in outreach and communication are welcome, not only to attract more students but also to increase the positive view at the society and governmental level of particle physics and cosmology like a playground for innovation with far-reaching impact on industry and society.


\section{Conclusions}\label{chapt:conclusion}

The presence and construction of new local facilities and infrastructures to develop, build, test and characterise new instruments, technologies and detectors are growing significantly in the region. The review presented in this chapter is based on the different inputs that were submitted to LASF4RI and show the large diversity of experience and skills developed in the different groups and institutions. They already show a large number of projects and involvement in international collaborations. Many of them are now at the point where collaborative efforts, knowledge sharing and facilities will be boosting the local efforts towards significant contributions to sciences with great international implications.

\noindent It is crucial for the community to maintain a strong focus on instrumentation and computing R\&D and to foster an environment that stimulates innovation. Creating this environment depends on several factors, including: stable funding, access to shared infrastructures and tools, existence of networks and consortia and the training, recognition and retention of human resources. 

\noindent Capacity building must be a priority in the region. It is important to highlight that those high-level knowledge transfer and educational efforts must take into account the local situations, and also aim to generate a critical mass of human resources that remains in the region so that a long-term and progressive development of such projects can be a sustained reality.

\noindent Finally, LASF4RI-HECAP is the first overall regional effort to bring together the Latin American particle, astroparticle physics and cosmology community. It offers an invaluable opportunity to perform surveys and create working panels to assess the global status of the community in terms of demographics, resources, early-career scientists status, career recognition and perspectives, and diversity, among others.




\pagebreak




\chapter{Appendix}
\section{List of White Papers}\label{wp}

\begin{table}[h!]
  \centering
  \caption{Details of the Inputs of White Papers for this Section}
  \label{tab:iwp}

  \resizebox{\textwidth}{!}{
  \begin{tabular}{ll}
    \hline
    Input ID & Title \\
    \hline
    \href{https://zenodo.org/record/4035221}{I-1}&Colombian Network on High Energy Physics - Input on Experimental HEP \\
    \href{https://zenodo.org/record/4060410}{I-2}&Argentina Experimental HEP Input \\
    \href{https://zenodo.org/record/4065532}{I-3}&Brazilian Participation in the Next-Generation Collider Experiments (Young Scientists) \\
    \href{https://zenodo.org/record/4313952}{I-4}&White Paper on Nuclear Science in Brazil. Contribution to the Latin American Strategy Forum for Research Infrastructure\\
    \href{https://zenodo.org/record/4035193}{I-5}&The ASTRI MINI-ARRAY: a Precursor for the Cherenkov Telescope Array (CTA)\\
    \href{https://zenodo.org/record/4060346}{I-6}&The Cherenkov Telescope Array (CTA)\\
    \href{https://zenodo.org/record/4060350}{I-7}&DUNE in the context of LASF4RI. The Colombian Case\\
    \href{https://zenodo.org/record/4060398}{I-8}&LA-CoNGA Physics perspectives for the Latin America Strategy Forum for Research Infrastructure\\
    \href{https://zenodo.org/record/4065544}{I-9}&Coherent Neutrino-Nucleus Scattering Experiment (CONNIE)\\
    \href{https://zenodo.org/record/4060405}{I-10}&The Study of the Quark-Gluon Plasma with the ALICE-LHC Experiment\\
    \href{https://zenodo.org/record/4313916}{I-11}&Letter of Intent of the Paraguayan Group\\
    \href{https://zenodo.org/record/4313914}{I-12}&Latin American Contribution to JUNO\\
    \href{https://zenodo.org/record/4313948}{I-13}&Brazilian Community Report on Neutrino Physics\\
    \href{https://zenodo.org/record/4313930}{I-14}&Brazilian Report on Safeguards Application of Reactor Neutrinos\\
    \href{https://zenodo.org/record/4065538}{I-15}&Short baseline neutrino experiment in nuclear reactors in Argentina\\
    \href{https://zenodo.org/record/4313941}{I-16}&The South American Gravitational wave Observatory (SAGO) White Paper\\
    \href{https://zenodo.org/record/4313971}{I-17}&Unravelling the Mysteries of Ultraenergetic Cosmic Rays with AugerPrime\\
    \href{https://zenodo.org/record/4313962}{I-18}&Physics exploration with the LHCb experiment. LHCb Group in Brazil\\
    \href{https://zenodo.org/record/4313958}{I-19}&The ATLAS/Brazil Cluster: Current Status and Perspectives from the ATLAS Upgrade Programme\\
    \href{https://zenodo.org/record/4313964}{I-20}& CMS Group - Universidade do Estado do Rio de Janeiro (UERJ)-- This contribution was retracted.\\
    \href{https://zenodo.org/record/3614109}{I-21}&A proposal for Transversal Computer-related Strategies and Services for Scientific and Training efforts\\
    \href{https://zenodo.org/record/4313934}{I-22}&Developing the first astronomical and quantum imaging instrument using the Smart Skipper-CCD technique\\
    \href{https://zenodo.org/record/4313936}{I-23}&Southern Wide-field-of-view Gamma-ray Observatory (SWGO)\\
    \href{https://zenodo.org/record/4313950}{I-24}&The Cherenkov Telescopes Array: fundamental physics and instrumentation\\
    \href{https://zenodo.org/record/4060394}{I-25}& GRAND: Giant Radio Array for Neutrino Detection\\
    \href{https://zenodo.org/record/3614096}{I-26} & A Venezuelan input to the Latin American Strategy for Research Infrastructures (LASF4RI)\\
    \href{https://zenodo.org/record/4035226}{I-27} & Colombian Network on High Energy Physics Input on Theoretical HEP\\
    \href{https://zenodo.org/record/4065546}{I-28} & Hyper-Kamiokande: Possible Contributions from Latin America\\
    \href{https://zenodo.org/record/4060402}{I-29} & The Latin American Giant Observatory \\
    \href{https://zenodo.org/record/4060380}{I-30} & Ecuadorian HECAP Groups Input to the Latin American Strategy Forum for Research Infrastructure\\
    \href{https://zenodo.org/record/4313969}{I-31} & A Latin American graduate school - The PPGCosmo experience\\
    \href{https://zenodo.org/record/4031864}{I-32} & The ANDES Deep Underground Laboratory\\
    \href{https://zenodo.org/record/4036118}{I-33} & Mac\'on Ridge Astronomical Site\\
    \href{https://zenodo.org/record/4035233}{I-34} & Brazilian Community Report on Dark Matter \\
    \href{https://zenodo.org/record/4065534}{I-35} & Dark Matter and Neutrino Physics - Contribution from Buenos Aires to the LASF4RI\\
    \href{https://zenodo.org/record/4313924}{I-36} & An Andean Deep-Valley Detector for High-Energy Tau Neutrinos\\
    \href{https://zenodo.org/record/4313928}{I-37} & QUBIC: Q\&U Bolometric Interferometer for Cosmology\\
    \href{https://zenodo.org/record/4313918}{I-38} & Cosmology with the Large Synoptic Survey Telescope (LSST)\\
    \href{https://zenodo.org/record/4035205}{I-39} & The BINGO Radio Telescope: an instrument to investigate the Universe through the 21 cm neutral hydrogen line\\
    \hline
 \end{tabular}}
\end{table}

\clearpage
\newpage

\section{Glossary of Experiments}
\noindent ABRAS: Argentina-Brasil Astronomical Center\\
ALICE: A Large Ion Collider Experiment\\
ALPACA: Andes Large-area PArticle detector for 
Cosmic-ray physics and Astronomy\\
ANDES: Agua Negra Deep Experiment Site\\
ANGRA: Angra Dos Reis Nuclear complex\\
ASTRI: Astrofisica a Specchi con Tecnologia Replicante Italiana\\
ATLAS: A Toroidal LHC ApparatuS\\
AUGER: Pierre Auger Observatory\\
BELLE 2: B detector at KEK \\
BINGO: BAO from Integrated Neutral Gas Observations\\
CEPC: Circular Electron-Positron Collider\\
CLAS: CEBAF Large Acceptance Spectrometer\\
CMS: Compact Muon Solenoid\\
CNEA TANDAR: Acelerador TANDAR, Comisi\'on Nacional de Energ\'{\i}a At\'omica\\
CONNIE: Coherent Neutrino Nucleus Interaction Experiment\\
CTA: Cherenkov Telescope Array\\
DARKSIDE: Two phase TPC for Dark Matter Direct Detection\\
DUNE: Deep Underground Neutrino Experiment\\
FCC: Future Circular Collider\\
GRAND: Giant Radio Array for Neutrino Detection\\
HAWC: High Altitude Water Cherenkov Gamma-ray Observatory\\
HL-LHC: High Luminosity Large Hadron Collider\\
Hyper-K: Hyper Kamiokande\\
ILC: International Linear Collider\\
JUNO: Jiangmen Underground Neutrino Observatory\\
KM3NeT: Cubic Kilometre Neutrino Telescope\\
LAGO: Latin American Giant Observatory\\
LAFN Tandem: Tandem Accelerator at Laboratorio Aberto de Fisica Nuclear\\
LHCb:Large Hadron Collider beauty\\
LLAMA: Large Latin American Millimetre Array\\
LSST: Legacy Survey of Space and Time\\
No$\nu$A: NuMI Off-axis $\nu$e Appearance\\
PHENIX:Pioneering High-Energy Nuclear Interaction eXperiment\\
PPS: Precision Proton Spectrometer\\
QUBIC: Q$\&$U Bolometric Interferometer for Cosmology\\
RHIC: Relativistic Heavy Ion Collider\\
SAGO: South American Gravitation Wave Observatory\\
SBND: Short Baseline Neutrino Detector\\
SENSEI: Sub-Electron-Noise Skipper Experimental Instrument\\
STAR:Solenoidal Tracker at RHIC\\
SWGO:Southern Wide-field-of-view Gamma-ray Observatory\\
TAMBO:Tau Air Shower Mountain-Based Observatory\\
TOROS: The Transient Optical Robotic Observatory of the South\\
$\nu$IOLETA:Neutrino Interaction Observation with a Low Energy Threshold Array\\

\end{document}